\documentclass[useAMS,usenatbib]{mn2e}
\usepackage{epsfig, natbib, enumerate, url}
\usepackage{amssymb, amsmath}
\usepackage{graphicx, xcolor}
\usepackage{placeins, float}
\usepackage{times}

\newcommand\ion[2]{#1\,{\sc #2}}
\newcommand{\NHI}{\ensuremath N_{\rm H\,\textsc{i}}}

\newcommand{\FeII}{\ion{Fe}{ii}}
\newcommand{\ZnII}{\ion{Zn}{ii}}

\newcommand{\CrII}{\ion{Cr}{ii}}
\newcommand{\SiII}{\ion{Si}{ii}}
\newcommand{\MgI}{\ion{Mg}{i}}

\newcommand{\HI}{\ion{H}{i}}

\newcommand{\zDLA}{z_{\textsc{dla}}}
\newcommand{\lya}{Ly$\alpha$}
\newcommand{\logNHI}{\log (N_{\rm H\,\textsc{i}} / {\rm cm}^{-2})}

\usepackage{booktabs}

\title[The metallicity--luminosity relation for DLAs]
{Consensus report on 25 years of searches for damped Ly$\alpha$ galaxies in emission: Confirming their metallicity--luminosity relation at \boldmath{$z\gtrsim2$}}

\author[Krogager et al.]{
J.-K. Krogager$^{1,2}$\thanks{E-mail:
jens-kristian.krogager@iap.fr},
P. M\o ller$^{3}$,
J.~P.~U. Fynbo$^{2}$, \&
P. Noterdaeme$^{1}$\\
$^{1}$Institut d'Astrophysique de Paris, CNRS-UPMC, UMR7095, 98bis bd Arago, 75014 Paris, France\\
$^{2}$Dark Cosmology Centre, Niels Bohr Institute, Copenhagen University, Juliane Maries Vej 30, 2100 Copenhagen \O, Denmark\\
$^{3}$European Southern Observatory, Karl-Schwarzschild-Stra{\ss}e 2,
85748 Garching bei M\"unchen, Germany\\
}

\begin{document}


\pagerange{\pageref{firstpage}--\pageref{lastpage}} \pubyear{2016}

\maketitle

\label{firstpage}

\begin{abstract}
Starting from a summary of detection statistics of our recent X-shooter campaign, we review the major surveys, both space and ground based, for emission counterparts of high-redshift damped \lya\ absorbers (DLAs) carried out since the first detection 25 years ago. We show that the detection rates of all surveys are precisely reproduced by a simple model in which the metallicity and luminosity of the galaxy associated to the DLA follow a relation of the form,
${\rm M_{UV}} = -5 \times \left(\,[{\rm M/H}] + 0.3\, \right) - 20.8$,
and the DLA cross-section follows a relation of the form $\sigma_{\textsc{dla}} \propto L^{0.8}$.
Specifically, our spectroscopic campaign consists of 11 DLAs preselected based on their equivalent width of \ion{Si}{ii}\,$\lambda1526$ to have a metallicity higher than $[{\rm Si/H}]>-1$. The targets have been observed with the X-shooter spectrograph at the Very Large Telescope to search for emission lines around the quasars. We observe a high detection rate of 64\% (7/11), significantly higher than the typical $\sim$10\% for random, \HI-selected DLA samples. We use the aforementioned model, to simulate the results of our survey together with a range of previous surveys: spectral stacking, direct imaging (using the `double DLA' technique), long-slit spectroscopy, and integral field spectroscopy. Based on our model results, we are able to reconcile all results. Some tension is observed between model and data when looking at predictions of \lya\ emission for individual targets. However, the object to object variations are most likely a result of the significant scatter in the underlying scaling relations as well as uncertainties in the amount of dust which affects the emission.

\end{abstract}


\section{Introduction}

One of the main limitations when studying galaxies at high redshift is the rapid decrease in flux with increasing lookback time. Hence, only the very brightest part of the galaxy population is directly observable in large scale surveys.
However, by using the imprint of neutral hydrogen observed in the spectra of bright background sources, we are able to study the gas in and around galaxies at high redshift.
The various strengths of absorption systems are thought to probe different parts of the galaxy environments with an observed anti-correlation between the column density of neutral hydrogen ($\NHI$) and the impact parameter \citep{Katz1996, Gardner2001, Zwaan2005, Monier2009, Peroux2011a, Rahmati2014, Rubin2015}.
Thus, the higher column densities typically trace the medium in (or very nearby) galaxies whereas the lower column densities trace the surrounding medium and the intergalactic gas clouds.
A specific class of such neutral hydrogen absorbers is the so-called damped \lya\ absorbers \citep[][DLAs]{Wolfe1986}, whose large column density of \HI\ ($\NHI > 2\times 10^{20}$~cm$^{-2}$) makes these absorbers great probes of the gas on scales up to $\sim$30~kpc \citep{Rahmati2014}.
DLAs might therefore serve as direct tracers of galaxies irrespective of their luminosities.

While it is possible to study the metal abundances in DLAs in great detail \citep[e.g.,][]{Kulkarni2002, Prochaska2003b, Dessauges-Zavadsky2006, Ledoux2006, Rafelski2014}, we still do not have a good understanding of the underlying physical origin of the systems hosting DLAs. 
Some insights can be obtained through the study of kinematics of the absorption lines. For this purpose, the velocity width, $\Delta V_{90}$ \citep{Prochaska1997}, has been widely used to quantify the kinematics of the absorbing medium \citep[e.g.,][]{Prochaska1998}. Using the velocity width as a proxy for the mass of the dark matter halo, several authors have used $\Delta V_{90}$ to decipher the underlying host properties of DLAs \citep[e.g.,][]{Haehnelt1998, Haehnelt2000, Ledoux2006, Bird2015}.
However, a more direct method to study the host of the absorption is to search for the emission associated with the host (here we use the terms `DLA galaxy', `counterpart' or `host' to refer to the galaxy associated with the absorption).
Direct detections have been sparse in the past; since the first study of DLAs in 1986 until 2010 only 3 counterparts of high-redshift DLAs had been identified \citep{Moller2004}.
At lower redshifts, however, the detections of counterparts have been more frequent \citep[e.g.,][]{Chen2003, Rao2011, Straka2016, Rahmani2016}. Various techniques to search for DLA galaxies have been utilized: narrow-band imaging of the field around the quasar can reveal the associated emission \citep[e.g.,][]{Smith1989, Moller1993, Moller1998b, Kulkarni2006, Fumagalli2010, Rahmani2016}, long-slit spectroscopy has been used to search for emission lines from the DLA galaxy \citep[e.g.,][]{Warren1996, Moller2002, Moller2004, Fynbo2010, Fynbo2011, Noterdaeme2012a, Srianand2016}, and integral field spectroscopy combines the power of these two approaches allowing an extended search for emission lines around the quasar \citep[e.g.,][]{Peroux2011a, Bouche2012a, Wang2015}.
Moreover, stacking of spectra from the Sloan Digital Sky Survey \citep[SDSS;][]{York2000} can constrain the average properties of \lya\ emission from DLAs at high redshift \citep{Rahmani2010, Joshi2017}.
Although the number of detections has increased \citep[see also \citealt{Christensen2014}]{Krogager2012}, the detection rate of emission counterparts in blindly selected samples remains very low \citep{Fumagalli2015}.  

This low detection rate can be understood as a consequence of the way DLAs are selected \citep{Fynbo1999}.
Due to the selection against background sources, DLAs are selected based on the cross-section of neutral gas, $\sigma_{\rm DLA}$, which (in the CDM cosmology) scales with the mass of the host halo \citep[e.g.,][]{Gardner2001, Pontzen2008, Bird2013}. Moreover, there is evidence that $\sigma_{\rm DLA}$ scales with the luminosity of the host galaxy in the local universe \citep{Chen2003}. Assuming that a similar relation holds at higher redshifts, the weighting of the luminosity function by $\sigma_{\rm DLA}\propto L^{2\beta}$ leads to a flattening of the faint-end slope (for observationally motivated values of $\beta\sim 0.4$). This means that DLAs sample the luminosity function over a wide range of luminosities, both the bright and faint ends.
The underlying assumption that high-redshift DLA galaxies are regular star-forming galaxies is supported by observations of \lya\ emitters \citep{Fynbo2001, Fynbo2003, Rauch2008, Barnes2009, Grove2009}.
\citeauthor{Rauch2008} propose that the counterparts of neutral hydrogen absorbers seen in quasar spectra have emission properties similar to those of \lya\ emitting galaxies \citep[see also][]{Krogager2013, Noterdaeme2014}.
Furthermore, \citet{Fynbo2001} and \citet{Verhamme2008} argue that luminous LAEs overlap with the population of bright star-forming galaxies selected as Lyman break galaxies (LBGs).
\citet{Moller2002} also established that DLA galaxies found in emission have properties overlapping those of LBGs at similar redshifts.
Several studies of the nature of LAEs have shown that the galaxies associated to \lya\ emission probe a mix of different galaxy properties \citep{Finkelstein2007, Finkelstein2009, Nilsson2007, Kornei2010, Shapley2011}, possibly with a dependence on redshift \citep{Nilsson2009, NilssonMoller2009}. This is in good agreement with a scenario in which DLAs trace star-forming galaxies with a large span of masses, luminosities and star-formation rates.
In this way, DLAs reveal complementary information to the population of luminosity selected galaxies, for which we can directly infer star formation rates, stellar masses, morphologies and sizes \citep{Kauffmann2003, Shen2003, Ouchi08, Reddy09, Cassata2011, Alavi2014}. Moreover, for DLAs we are able to obtain precise metallicity measurements allowing us to determine metallicity scaling relations and their evolution out to large redshifts.

Numerical simulations provide an important tool to reveal the physical nature of the galaxies associated to DLAs, and recent simulations are starting to match the observed absorption properties very well, e.g., velocity widths ($\Delta V_{90}$), metallicities, and column densities of \HI\ \citep{Fumagalli2011, Rahmati2014, Bird2014, Bird2015}. Such numerical studies also reveal a mixed population of galaxies associated to DLAs spanning many orders of magnitude in stellar mass and star formation rate \citep{Berry2016}.

Although the numerical simulations are powerful and allow detailed studies of individual galaxies, a simpler approach using well-established scaling relations enables us to easily gauge the galaxy population responsible for DLA absorption as a whole. Using this approach, \citet{Fynbo2008} have tested the hypothesis that DLAs are drawn from the same parent population of star-forming galaxies that give rise to LBGs. The two observed phenomena (either a DLA or a bright LBG) result from two different ways of sampling the same luminosity function; as mentioned previously, the DLAs probe a large span of luminosities. In order to further examine this hypothesis and to increase the sample of spectroscopically identified DLA emission counterparts at high redshift, a spectroscopic campaign was initiated targeting high-metallicity DLAs \citep{Fynbo2010, Fynbo2011}. The focus on high-metallicity was based by the hypothesis that DLAs follow a mass--metallicity relation, which is motivated by the observed metallicity--velocity relation \citep{Ledoux2006, Moller2013, Neeleman2013}.
Assuming that luminosity scales with stellar mass, one would then expect that high-metallicity DLAs have brighter counterparts \citep[see also][]{Moller2004}.

In this paper, we summarise the efforts of this spectroscopic campaign for counterparts of metal-rich DLAs. In total, we have observed 12 sightlines, 6 of which have been published previously \citep{Fynbo2010, Fynbo2011, Krogager2012, Fynbo2013b, Hartoog2015}. The remaining data (from the observing runs 086.A-0074 and 089.A-0068) have been reduced and analysed in this work. However, we restrict our analysis of the X-shooter data to the rest-frame UV properties derived from \lya, so as to keep the analysis and modelling as concise as possible. The analysis and modelling of the near-infrared data will be presented in a forthcoming paper (Fynbo et al. in preparation). Using the entire sample of metal-rich DLAs, we test the expectations from the model by \citet{Fynbo2008} and find an excellent agreement between the data and our model. Moreover, we apply our model to all major past surveys of high-redshift DLAs and find that all the previous results are in agreement with our model expectation.

The paper is structured as follows:
In Section~\ref{data}, we summarize the X-shooter sample selection and the observations;
in Section~\ref{analysis}, we present the analysis of absorption and emission properties;
in Section~\ref{model}, we briefly describe the model from \citet{Fynbo2008} and present our comparison of this model to the \lya\ detections from our campaign;
in Section~\ref{comparison}, we apply our model framework to various samples from the literaute,
and in Section~\ref{discussion}, we discuss the limitations and implications of our results.

\defcitealias{Planck2014}{Planck Collaboration 2014}
Throughout this paper we assume a standard $\Lambda$CDM cosmology
with $H_0=67.8\, {\rm km\ s^{-1}\ Mpc}^{-1}$, $\Omega_{\Lambda}=0.69$ and
$\Omega_{\mathrm{M}} = 0.31$ (\citetalias{Planck2014}). We use the standard notation of $[{\rm X/Y}] \equiv \log N({\rm X})/N({\rm Y}) - \log N({\rm X})_{\odot}/N({\rm Y})_{\odot}$, where $N({\rm X})$ and $N({\rm Y})$ refer to the column densities of elements ${\rm X}$ and ${\rm Y}$. We use the photospheric Solar values from \citet{Asplund09}. The notation $[{\rm M/H}]$ refers to the metallicity of any volatile element, typically zinc.
When referring to the quasars in our sample, we use the following shorthand notation based on the J\,2000 epoch coordinates: Qhhmm$\pm$ddmm. However, for two targets, which have been published previously, we use their original names: Q2348$-$011 and PKS0458$-$020.

\section{X-shooter Sample Selection and Observations}
\label{data}
The targets are selected from the Sloan Digital Sky Survey (SDSS, \citealt{Richards2001}) using the rest-frame equivalent width ($W_{\rm rest}$) of \ion{Si}{ii}$\,\lambda1526$ as a proxy for metallicity. We require that $W_{\rm rest}$, measured for the DLA in the SDSS spectrum \citep{Noterdaeme2009b}, be larger than $1$~\AA\ as this is a good indication that the metallicity of the DLA is higher than $[{\rm M/H}]>-1$ \citep[see figure~6 of][]{Prochaska2008}. From the initial candidates, we select targets that have suitable redshifts to allow us to look for nebular emission lines in the near-infrared outside the strong telluric absorption bands between the $J$, $H$, and $K$ bands (i.e., $\zDLA\sim 2.2 - 2.5$). Lastly, we give priority to targets that also exhibit strong iron lines, specifically the \ion{Fe}{ii} lines at 2344, 2374, 2382.

\begin{figure}
	\includegraphics[width=0.48\textwidth]{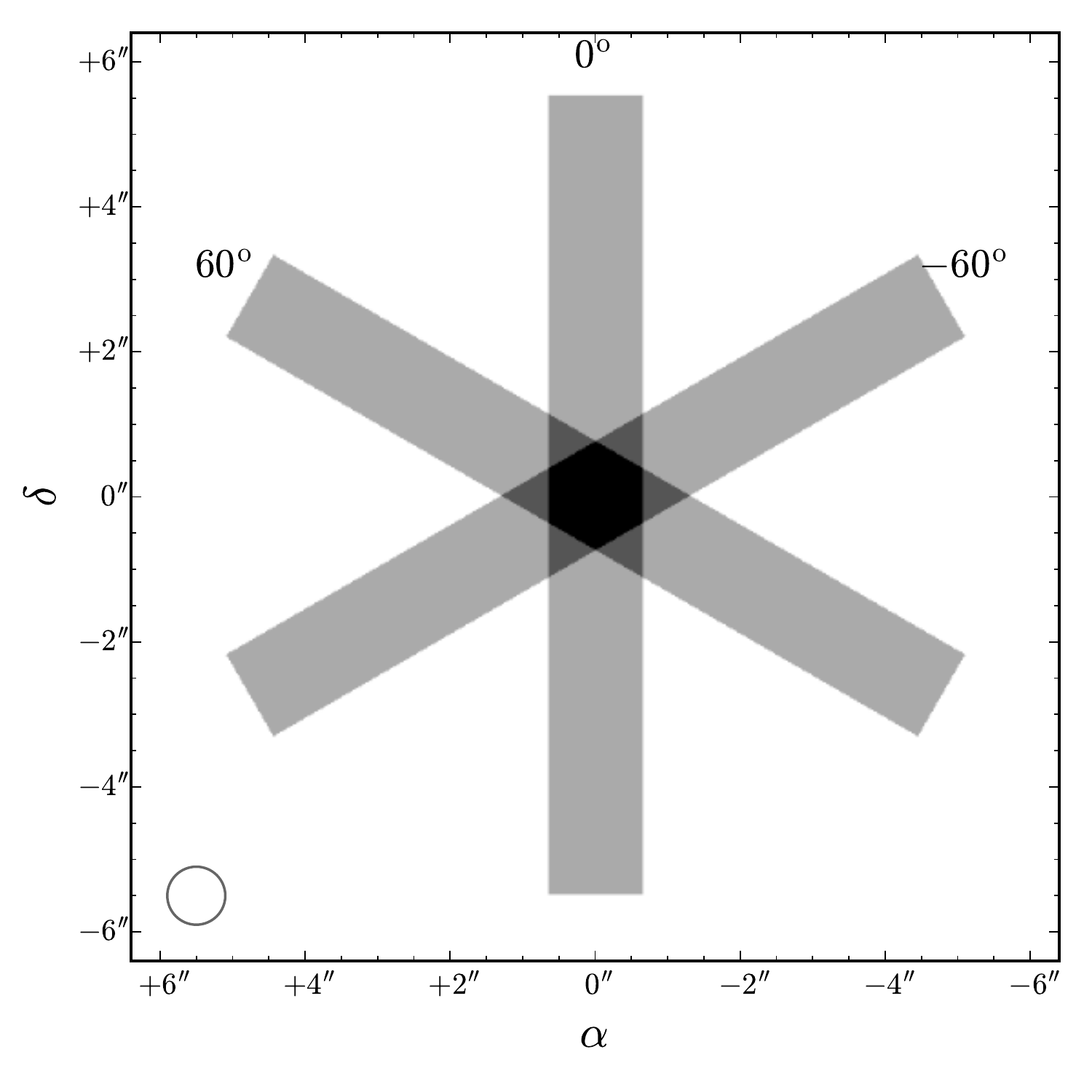}
	\caption{Effective exposure time map on the sky. The three individual slits are shown. Darker colour corresponds to higher effective exposure time. The circle in the lower left corner shows the average seeing disk of 0\farcs8. \label{fig:slit_image}}
\end{figure}

\subsection{Compilation of the Statistical Sample}
\label{data:sample}
The emission counterpart of PKS0458$-$020 was known from previous work \citep{Moller2004} and included in our campaign as a sanity check in order to ensure that our spectroscopic setup is sensitive enough. Moreover, the new observations allow us to measure the \lya\ flux at the position angle reported by \citet{Moller2004}. Furthermore, the target Q0030$-$5129 was observed as a back-up target during one of the observing runs (088.A-0101). The DLA towards Q0030--5129 does not meet the line-strength criterion and consequently does not meet the metallicity requirement \citep{Hartoog2015}. Lastly, we notice that there are two DLAs towards Q2348--011. The second of these two DLAs (at $z_{\rm DLA}=2.614$) does not meet the line-strength criterion (with a $W_{\rm rest}$ of \ion{Si}{ii}$\,\lambda1526$ of only 0.4~\AA). 
Since those three DLAs have not been selected in the same way as the rest of our sample (i.e., one was known already and the other two did not meet our metal line-strength criterion), these targets are excluded from our statistical analyses. We thus have a final {\it statistical} sample of 11 DLAs.
The three targets mentioned above are included in this work for completeness.

\subsection{Spectroscopic Observations}
For the spectroscopic observations, we have used the X-shooter instrument, which is mounted on unit 2 of the Very Large Telescope at Paranal observatory in Chile operated by the European Southern Observatory. The spectrograph covers the observed wavelength range from $3000$~{\AA} to $2.5~\mu$m simultaneously by splitting the light into three separate spectrographs, the so-called `arms': UVB ($3000-5500$~\AA), VIS ($5500-10000$~\AA), and NIR ($10000-25000$~\AA). 

The main strategy of the campaign is presented in \citet{Fynbo2010}, and we will only briefly summarize the main points of the spectroscopic setup. Each quasar is observed using long-slit spectroscopy at three position angles (PA1=$0$\degr, PA2=$+60$\degr, and PA3=$-60$\degr east of north) in order to cover as much of the region around the quasar as possible. These three position angles are referred to as PA1, PA2, and PA3, respectively.
All observations are carried out using the same slit widths of 1\farcs3, 1\farcs2 and 1\farcs2 for UVB, VIS and NIR, respectively. All slits have the same length of 11\arcsec. The effective slit configuration is shown schematically in Fig.~\ref{fig:slit_image}. An overview of the sample, including information regarding the observations, is provided in Table~\ref{tab:log}.

\subsection{Data Reduction} 
For seven quasars, PKS0458--020, Q0316+0040, Q0338$-$0005\footnote{The initial detection of emission for Q0338$-$0005 is reported in \citet{Krogager2012}.}, Q2348$-$011, Q0845+2008, Q1435+0354, and Q1313+1441, the data are published here for the first time. The data processing of these seven quasars is described in the following section. 
The raw data frames are first corrected for cosmic ray hits using the code {\sc DCR} \citep{DCR}. The spectra are subsequently reduced using the official X-shooter pipeline version 2.5 for `stare mode'. The pipeline performs the following steps for each arm independently: First, the raw frames are corrected for the bias level (UVB and VIS) and dark current (NIR). Then the background is subtracted followed by a subtraction of the sky emission lines using the method laid out by \citet{Kelson03}. After division by the spectral flat-field, the individual orders are extracted and rectified in wavelength space. The individual orders are then merged using error weighting in the overlapping regions. The resulting spectrum is a merged 2-dimensional spectrum and its error spectrum.
Intermediate products such as the sky spectrum and individual echelle orders (with errors and bad-pixel maps) are also produced. From the 2-dimensional spectrum, we extract a 1-dimensional spectrum using our own python implementation of the optimal extraction algorithm \citep{Horne1986}. The 1-dimensional spectrum is subsequently converted to vacuum wavelengths and shifted to the helio-centric rest-frame. No correction of telluric absorption has been performed.

The relative flux calibration performed by the X-shooter pipeline provides a robust recovery of the spectral shape (to within 5\%, measured from our spectra by comparing to photometry); however, in order to improve the absolute flux calibration\footnote{Small offsets in the fluxes are observed between the different arms}, we have scaled our spectra to their corresponding photometry from SDSS. We scale the UVB arm in order to obtain the most precise flux calibration for the \lya\ emission detections and match the VIS arm to the calibration achieved in the UVB. For this purpose, we use the $g$-band from SDSS as this band is more robust than the $u$-band\footnote{We note that consistent scaling factors were derived for the $u$-band, and for the $r$, $i$ and $z$ bands in the VIS arm.}. Since the quasars could have undergone intrinsic variations in their luminosities between our spectroscopic observations and the epoch of observation by the SDSS (of the order $\sim$10 to 15\%; \citealt{Giveon1999}), we assign a conservative uncertainty on the flux calibration of 15\%.
One target is not covered by the SDSS footprint, namely PKS0458--020. For this target, we use instead observations from \citet{Souchay2012} in the Johnson $B$-band ($B_J = 19.1\pm0.1$) to calibrate the UVB arm. The flux calibration for this target is much less reliable due to the worse quality of photometric data available. The large uncertainty has been taken into account in the analysis of this target.\\

In order to study the absorption lines from the DLAs in greater detail, we combine the three 1-dimensional spectra for each target (corresponding to each position angle) using the error spectra as weights for the combination and masking bad pixels in individual spectra. For Q1313+1441, we observe a small shift in wavelengths between the UVB and VIS arms of 0.6~\AA\ (3 pixels in the UVB arm, corresponding to roughly one third of the used slit-width) due to uncertainties in the wavelength calibration and centring of the object in the three slits. Similar offsets have been noted previously for X-shooter\footnote{An in-depth description of the shifts is available on the instrument webpage: \url{https://www.eso.org/sci/facilities/paranal/instruments/xshooter/doc.html}}. We have subsequently shifted the UVB spectrum to match the VIS wavelength calibration.\\

As mentioned, the seeing was smaller than the used slit-widths. This affects not only the wavelength calibration but also the determination of the resolution, $\mathcal{R}$, as the instrument specific values are no longer valid. To overcome this, we infer the resolving power of each spectrum by convolving a telluric absorption template with a Gaussian kernel to match the observed telluric profiles. In order not to blur the telluric lines, the resolving power was inferred from a separate combination of the spectra before applying the air-to-vacuum conversion and the correction for the relative motion of the observatory relative to the helio-centric frame. Since we mainly fit absorption lines in the VIS arm, we only report the spectral resolution for this arm, the obtained values are given in Appendix~\ref{app:fits}. For one case (Q1313+1441) we also fit transitions in the UVB arm. We therefore determine the resolving power by using the seeing as an estimate of the effective slit width. We then interpolate between the tabulated values of resolution for given slit widths (assuming an inverse proportionality between $\mathcal{R}$ and slit width). Using the average seeing in the $V$-band, we infer a resolution in the UVB arm of 8000.

\begin{table*}
\caption{X-shooter observing log \label{tab:log}}
\begin{center}
\begin{tabular}{lccccccccr}
\hline
    Target  &      R.A.    &      Decl.      &   P.A.$^{(a)}$ & Exp. time &   Date    &  Airmass$^{(b)}$ & Seeing$^{(b)}$ &   Prog. ID  &  Reference \\
            &              &                 &    & (sec) &  &  & (arcsec) &  &  \\
\hline
Q0030--5129 & 00:30:34.37  &  $-$51:29:46.3  &   $0$\degr & 3600  &  2011-10-21  &  1.13  &  0.78  &  088.A-0601 & (8) \\
            &              &                 & $+60$\degr & 3600  &  2011-10-21  &  1.21  &  1.00  &  088.A-0601 &  \\
            &              &                 & $-60$\degr & 3600  &  2011-10-21  &  1.37  &  0.90  &  088.A-0601 &  \\[1mm]
Q0316+0040  & 03:16:09.75  &  $+$00:40:42.6  & $0$\degr & 3200  &  2010-11-09  &  1.17  &  0.53  &  086.A-0074 &  \\
            &              &                 & $+60$\degr & 3200  &  2010-11-09  &  1.11  &  0.78  &  086.A-0074 &  \\
            &              &                 & $-60$\degr & 3200  &  2010-11-09  &  1.13  &  0.56  &  086.A-0074 &  \\[1mm]
Q0338--0005 & 03:38:54.74  &  $-$00:05:21.3  & $0$\degr & 3200  &  2010-11-09  &  1.18  &  0.71  &  086.A-0074 & (5) \\
            &              &                 & $+60$\degr & 3200  &  2010-11-09  &  1.36  &  0.53  &  086.A-0074 &  \\
            &              &                 & $-60$\degr & 3200  &  2010-11-09  &  1.75  &  0.56  &  086.A-0074 &  \\[1mm]
PKS0458--020& 05:01:12.77  &  $-$01:59:14.8  & $-60.4$\degr & 3600  &  2010-02-16  &  1.40  &  1.86  &  084.A-0303 & (1,5,9) \\[1mm]
Q0845+2008  & 08:45:02.85  &  $+$20:08:50.7  & $0$\degr & 3600  &  2012-04-19  &  1.48  &  0.68  &  089.A-0068 &  \\
            &              &                 & $+60$\degr & 3600  &  2012-04-20  &  1.75  &  0.69  &  089.A-0068 &  \\
            &              &                 & $-60$\degr & 3600  &  2012-04-21  &  1.64  &  0.72  &  089.A-0068 &  \\[1mm]
Q0918+1636  & 09:18:26.16  &  $+$16:36:09.0  & $0$\degr & 3600  &  2010-02-16  &  1.43  &  0.70  &  084.A-0303 & (4,7) \\
            &              &                 & $+60$\degr & 3600  &  2010-02-16  &  1.34  &  0.71  &  084.A-0303 &  \\
            &              &                 & $-60$\degr & 3600  &  2010-02-16  &  1.40  &  0.65  &  084.A-0303 &  \\[1mm]
Q1057+0629  & 10:57:44.45  &  $+$06:29:14.5  & $0$\degr & 3600  &  2010-03-19  &  1.35  &  0.71  &  084.A-0524 & (8) \\
            &              &                 & $+60$\degr & 3600  &  2010-03-19  &  1.20  &  0.57  &  084.A-0524 &  \\
            &              &                 & $-60$\degr & 3600  &  2010-03-19  &  1.20  &  0.50  &  084.A-0524 &  \\[1mm]
Q1313+1441  & 13:13:41.17  &  $+$14:41:40.4  & $0$\degr & 3600  &  2012-04-20  &  1.34  &  0.61  &  089.A-0068 &  \\
            &              &                 & $+60$\degr & 3600  &  2012-04-20  &  1.29  &  0.74  &  089.A-0068 &  \\
            &              &                 & $-60$\degr & 3600  &  2012-04-21  &  1.41  &  0.86  &  089.A-0068 &  \\[1mm]
Q1435+0354  & 14:35:00.22  &  $+$03:54:03.7  & $0$\degr & 3600  &  2012-04-21  &  1.24  &  0.91  &  089.A-0068 &  \\
            &              &                 & $+60$\degr & 3600  &  2012-04-21  &  1.15  &  0.97  &  089.A-0068 &  \\
            &              &                 & $-60$\degr & 1100  &  2012-04-20  &  1.16  &  0.64  &  089.A-0068 &  \\[1mm]
Q2059--0528 & 20:59:22.43  &  $-$05:28:42.8  & $0$\degr & 3600  &  2011-10-20  &  1.09  &  0.74  &  088.A-0601 & (8) \\
            &              &                 & $+60$\degr & 3600  &  2011-10-21  &  1.21  &  1.24  &  088.A-0601 &  \\
            &              &                 & $-60$\degr & 3600  &  2011-10-21  &  1.50  &  1.24  &  088.A-0601 &  \\[1mm]
Q2222--0946$^{(c)}$ & 22:22:56.11  &  $-$09:46:36.2  & $0$\degr & 3600  &  2009-10-21  &  1.06  &  1.05  &  084.A-0303 & (3,6) \\
            &              &                 & $0$\degr & 3600  &  2009-10-22  &  1.05  &  1.15  &  084.A-0303 &  \\
            &              &                 & $-60$\degr & 3600  &  2009-10-22  &  1.16  &  1.35  &  084.A-0303 &  \\[1mm]
Q2348--011   & 23:50:57.82  &  $-$00:52:09.8  & $0$\degr & 3200  &  2010-11-09  &  1.09  &  0.84  &  086.A-0074 & (2) \\
            &              &                 & $+60$\degr & 3200  &  2010-11-09  &  1.15  &  0.58  &  086.A-0074 &  \\
            &              &                 & $-60$\degr & 3600  &  2010-11-10  &  1.09  &  0.93  &  086.A-0074 &  \\
\hline
\end{tabular}
\end{center}

{\flushleft
$^{(a)}$ Position angle of the slit measured East of North.\\
$^{(b)}$ Airmass and seeing is averaged over the exposure.\\
$^{(c)}$ Due to an error in the execution of our observations at the telescope, the object was observed twice at ${\rm PA} = 0$\degr\ and once at ${\rm PA} = -60$\degr.
\newline\noindent
\hphantom{$^{(c)}$ }Hence, we did not get a spectrum at ${\rm PA} = +60$\degr.\\[2mm]

References:
(1) \citet{Moller2004};
(2) \citet{Noterdaeme2007};
(3) \citet{Fynbo2010};
(4) \citet{Fynbo2011};
(5) \citet{Krogager2012};
\newline\noindent
\hphantom{References: }(6) \citet{Krogager2013};
(7) \citet{Fynbo2013b};
(8) \citet{Hartoog2015};
(9) \citet{Ledoux2006}.

}

\end{table*}

\section{Analysis of X-shooter data}
\label{analysis}

\subsection{Absorption Lines}
\label{absorption}
\begin{figure}
	\includegraphics[width=0.48\textwidth]{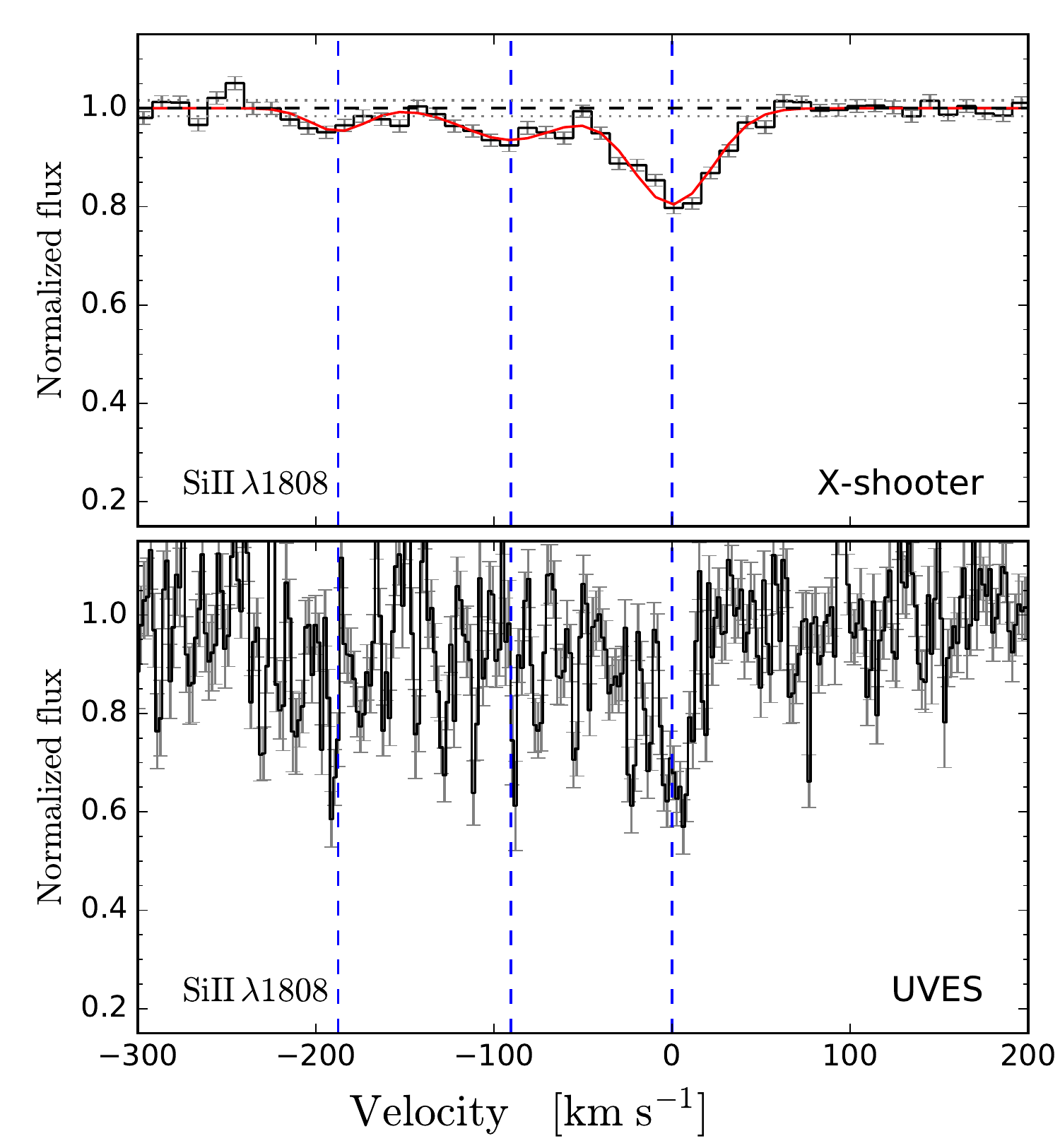}
	\caption{Comparison of the \ion{Si}{ii}$\lambda\,1808$ line for the DLA toward the quasar
	Q0338$-$0005 presented in this work (top panel)
	and the UVES data of the same transition (lower panel). The red solid line
	in the top panel indicates the best-fit model from our Voigt-profile analysis.
	The improved signal-to-noise ratio in the top panel is evident and the broad velocity structure
	is clearly seen in the X-shooter data. The UVES data shown here have been obtained from the ESO
	archive of science-grade, pipeline-processed data products.
	\label{fig:comparison}}
\end{figure}

The column densities of \ion{H}{i} and low-ionization metal lines have been obtained through Voigt-profile fitting using our own python code \citep[see Appendix A of][]{Krogager_PhD}. We search the spectra for suitable transitions of \ion{Fe}{ii}, \ion{Si}{ii}, \ion{Zn}{ii}, \ion{Cr}{ii}, and \ion{S}{ii}, however, not all species are available for all the DLAs. Under the assumption that the low-ionization lines arise from similar conditions in the absorbing medium, we fit all the lines from the singly ionized state using the same velocity structure, i.e., the number of components, relative velocities, and line broadening parameters are tied for all species. The DLA at $\zDLA=2.425$ toward Q2348$-$011 has been analysed previously by \citet{Noterdaeme2007}. For the species covered by the analysis of Noterdaeme et al. (\ion{Fe}{ii}, \ion{Si}{ii}, and \ion{S}{ii}), we use their measured values as the high-resolution data from UVES\footnote{The UV-visual echelle spectrograph (UVES) is mounted on unit 2 of the Very Large Telescope at Paranal observatory in Chile operated by the European Southern Observatory.} provide a better fit (though we obtain consistent values from our fits). The metallicities obtained from the absorption line analysis are listed in Table~\ref{tab:abs} together with values for the previously analysed DLAs in the sample (these are marked with a number pointing to the reference, from which the measurement was taken). The fitted transitions and the best-fit profiles are shown in Appendix~\ref{app:fits}. 
For the DLA at $\zDLA=2.229$ towards Q0338$-$0005, we note that a previous measurement of [Si/H] $=-1.22\pm0.11$ has been published using high-resolution data from UVES \citep{Jorgenson2013}. Although the UVES data have higher spectral resolution, the X-shooter data presented here have a much higher signal-to-noise ratio. We therefore use the measured quantity from this work over the one measured from the UVES data. The comparison of the two datasets is shown in Fig.~\ref{fig:comparison}.

For the target Q1313+1441, we fitted the available transitions (\SiII, \ZnII, \FeII, \CrII, and \MgI) in the VIS and UVB arms separately. Since the velocity structure of \MgI\ is consistent with the observed structure for the singly ionized species, we tied the relative velocities and broadening parameters of the \MgI\ line to the singly ionized species (\FeII, \ZnII, and \CrII). For the \SiII\,$\lambda1808$ line in the UVB arm, we then used the same velocity structure derived from the higher resolution data in the VIS arm to fit the \SiII\ line while only allowing the column density to vary.

We measure the velocity width of the absorption lines, $\Delta V_{90}$, following the definition by \citet{Prochaska1997}. For this purpose, we select weak, unblended low-ionization transitions in the VIS spectra. We deconvolve the measured $\Delta V_{90}$ using equation 1 from \citet{Arabsalmani2015}. The deconvolved velocity widths are given in Table~\ref{tab:abs}.

\begin{table*}
\caption{Absorption properties of the X-shooter campaign. \label{tab:abs}}
\begin{center}
\begin{tabular}{@{}lcccccccc@{}}
\hline
Target  &  $\zDLA$  &  $\logNHI$  &  [Zn/H] & [Si/H] & [Fe/H] & [Cr/H] & [S/H] & $\Delta V_{90}\, ^{(a)}$ \\
\hline
Q0030--5129	 &	2.452  & $20.8 \pm0.2^{(7)}$  & $-1.48\pm0.34^{(7)}$  &  --  &  $-1.55\pm0.20^{(7)}$  &  $-1.57\pm0.27^{(7)}$  &  --   & 41 \\
Q0316+0040	 &	2.179  & $21.04\pm0.05$  & $-1.02\pm0.07$  &  $-0.95\pm0.06$  &  $-1.45 \pm0.06 $  &  $-1.17\pm0.06$  &  --   & 69 \\
Q0338--0005	 &	2.229  & $21.12\pm0.05$  & $-1.36\pm0.07$  &  $-1.36\pm0.06$  &  --  &  $-1.46\pm0.06$  &  --   & 221 \\
PKS0458--020 &	2.040  & $21.70\pm0.10^{(1)}$  & $-1.22\pm0.10^{(1)}$  &  --  &  --  &  --  &  --   & 87$^{(1)}$ \\
Q0845+2008	 &	2.237  & $20.41\pm0.06$  & $+0.05\pm0.08$  &  $-0.29\pm0.07$  &  $<-0.8\pm0.1$  &  $-0.79\pm0.07$  &  --   & 155 \\
Q0918+1636-1 &	2.412  & $21.26\pm0.06^{(5)}$  & $-0.6 \pm0.2^{(5)}$  &  $-0.6 \pm0.2^{(5)}$  &  $-1.2 \pm0.2^{(5)}$  &  $-1.2 \pm0.2^{(5)}$  &  --   & 350$^{(5)}$ \\
Q0918+1636-2 &	2.583  & $20.96\pm0.05^{(4)}$  & $-0.12\pm0.05^{(4)}$  &  $-0.46\pm0.05^{(4)}$  &  $-1.03\pm0.05^{(4)}$  &  $-0.88\pm0.05^{(4)}$  &  $-0.26\pm0.05^{(4)}$   & 293$^{(4)}$ \\
Q1057+0629	 &	2.499  & $20.51\pm0.03^{(7)}$  & $-0.24\pm0.11^{(7)}$  &  $-0.37\pm0.05^{(7)}$  &  $-0.98\pm0.03^{(7)}$  &  --  &  $-0.15\pm0.06^{(7)}$   & 328 \\
Q1313+1441	 &	1.794  & $21.3 \pm0.1 $  & $-0.7 \pm0.1 $  &  $-0.8\pm0.1$  &  $-1.5 \pm0.1 $  &  $-1.2 \pm0.1 $  &  --   & 164 \\
Q1435+0354	 &	2.269  & $20.42\pm0.08$  & $-0.4 \pm0.1 $  &  $-0.6\pm0.1$  &  $-1.1\pm0.1$  &  $-0.7 \pm0.1 $  &  --   & 183 \\
Q2059--0528	 &	2.210  & $21.00\pm0.05^{(7)}$  & $-0.96\pm0.06^{(7)}$  &  $-0.99\pm0.05^{(7)}$  &  $-1.41\pm0.05^{(7)}$  &  $-1.19\pm0.05^{(7)}$  &  $-0.91\pm0.06^{(7)}$   & 114$^{(7)}$ \\
Q2222--0946	 &	2.354  & $20.65\pm0.05^{(6)}$  & $-0.38\pm0.05^{(6)}$  &  $-0.54\pm0.05^{(6)}$  &  $-1.02\pm0.05^{(6)}$  &  --  &  $-0.49\pm0.05^{(6)}$   & 181$^{(3)}$ \\
Q2348--011-1 &	2.425  & $20.53\pm0.06$  & $-0.33\pm0.08$  &  $-0.80\pm0.10^{(2)}$  &  $-1.17\pm0.10^{(2)}$  &  $-1.25\pm0.07$  &  $-0.62\pm0.10^{(2)}$   & 240 \\
Q2348--011-2 &	2.614  & $21.34\pm0.06$  & --  &  $-1.98\pm0.08$  &  $-2.46\pm0.08$  &  $-2.13\pm0.08$  &  --   & 63 \\
\hline
\end{tabular}
\end{center}

{\flushleft
$^{(a)}$ $\Delta V_{90}$ in units of km~s$^{-1}$, corrected for resolution effects following \citet{Arabsalmani2015}.
\newline\noindent
\hphantom{$^{(a)}$ }Typical uncertainties on $\Delta V_{90}$ are of the order 10--20~km~s$^{-1}$. \ion{Si}{ii} $\lambda1808$ was used in all cases except for Q1313+1441, where \ion{Cr}{ii} $\lambda2056$ was used.\\[1mm]

References:
(1) \citet{Ledoux2006};
(2) \citet{Noterdaeme2007};
(3) \citet{Fynbo2010};
(4) \citet{Fynbo2011};
(5) \citet{Fynbo2013b};
\newline\noindent
\hphantom{References: }(6) \citet{Krogager2013};
(7) \citet{Hartoog2015}.

}
\end{table*}

\subsection{Emission Counterparts}
\label{emission}

For every quasar, we search for \lya\ emission at the redshift of the DLA in the individual 2D
spectra for each position angle separately. We detect \lya\ emission associated with three out
of the seven DLAs studied. This includes the previously published DLA towards Q0338$-$0005.
The remaining two systems are the $\zDLA=2.425$ DLA towards Q2348$-$011 (hereafter Q2348$-$011-1)
and the $\zDLA=1.794$ DLA towards Q1313+1441. The emission from Q2348$-$011-1 is only detected in
one spectrum (for ${\rm PA}=0$\degr) at an impact parameter of $b=0.7$~arcsec. We are therefore
not able to firmly constrain the position angle; however, given the fact that the emission is not
detected in the two other slits, the most likely position angle would be $\sim$180\degr\ east of north,
since this angle minimizes the overlap between the source and the two other slits.

The emission for Q1313+1441 is detected in two slit positions. For PA2 ($+60$\degr), we
measure an impact parameter of $b_{+60}=-0.5\pm0.2$ arcsec, and for PA3 ($-60$\degr), we
measure an impact parameter of $b_{-60}=-1.3\pm0.2$ arcsec.
The detection in PA2 at smaller impact parameter cannot be the same object as observed
in PA3 due to the relative orientation of the slits. We therefore propose that the
detection in PA2 could be a neighbouring member of a small group, however, this will
require further follow up to confirm.
In the sample by \citet{Christensen2014} there are two DLAs with more than one clear
counterpart. Following their work, we choose the brightest galaxy as the main
counterpart of the group as this galaxy will dominate the scaling relations of
the environment in terms of metallicity and luminosity. Moreover, the metallicity
gradients used in this work have been defined following this definition by
\citet{Christensen2014} and for consistency we apply the same definition here.
The detection in PA3 is the most significant detection and in the following we
thus quote this as the main counterpart for the DLA towards Q1313+1441. 
For the main counterpart of this DLA, we infer an impact parameter of
$b=1.3\pm0.2$~arcsec and a position angle of 120$\pm27$\degr\ east of north.

For the rest of the sample we do not detect any emission from \lya. Instead, we place upper limits on the \lya\ flux by estimating the noise in a square aperture of 20 by 20 pixels (corresponding to an extent of 3.2~arcsec in the spatial direction and $\sim750$~km~s$^{-1}$ in velocity space). For each of the three PAs, we evaluate the noise in various apertures placed at different impact parameters to gauge any variations in the background in the spectra. The noise estimates for each of the individual PA spectra are all consistent, except for the quasar Q1435+0354 where the spectrum of PA3 was exposed for only 1100 sec instead of 3600 sec.

In Table~\ref{tab:lya}, we quote the average of the individual limits derived for each slit as the 3-$\sigma$ upper limit together with the measured fluxes and impact parameters. We also provide an estimated star formation rate based on \lya\ assuming case B recombination and the \citet{Kennicutt1998} conversion; however, these should be regarded as lower limits due to the unknown attenuation from dust and multiple scattering of the resonant \lya\ photons. The individual 2-dimensional spectra are shown in Appendix~\ref{app:lya}.

None of the new detections was observed in the continuum due
to the combination of faintness and small impact parameter.

\begin{table*}
\caption{Emission properties of the X-shooter campaign. \label{tab:lya}}
\begin{center}
\begin{tabular}{lcccccc}
\hline
Target           &  $\zDLA$  &      $b$      &      $b$      &      P.A.     &  $F({\rm Ly}\alpha)$                   & SFR$^{(a)}$                \\
                 &           &   (arcsec)    &     (kpc)     & (Deg. East of North) & $\left( 10^{-17}~{\rm erg\ s^{-1}\ cm^{-2}} \right)$  & $\left({\rm M}_{\odot}~{\rm yr}^{-1} \right)$    \\
\hline
Q0030--5129      &   2.452   &       --      &        --     &     --     & $<0.8$        &  --      \\
Q0316+0040       &   2.179   &       --      &        --     &     --     & $<0.7$        &  --      \\
Q0338--0005      &   2.229   & $0.49\pm0.12$ &   $4.2\pm1.0$ & $-58\pm20$ & $1.3\pm0.2$   & $>0.3$  \\
PKS0458--020     &   2.040   & $0.31\pm0.04$ &   $2.7\pm0.3$ &$300\pm65$ & $6.4\pm1.3$  & $>1.1$  \\
Q0845+2008       &   2.237   &       --      &        --     &     --     & $<0.8$        &  --      \\
Q0918+1636-1$^{(b)}$ & 2.412 &     $<0.3$    &       $<2$    &     --     & $<0.5$        &  --      \\
Q0918+1636-2$^{(b)}$ & 2.583 &  $1.98\pm0.02$  &  $16.2\pm0.2$ & $245\pm1$ & $<0.5$        & $22\pm7$   \\
Q1057+0629       &   2.499   &       --      &        --     &     --     & $<0.9$        &  --      \\
Q1313+1441       &   1.794   &  $1.3\pm0.2$  &  $11.3\pm1.7$ & $120\pm27$ & $2.5\pm0.7$   & $>0.3$  \\
Q1435+0354       &   2.269   &       --      &        --     &     --     & $<0.8$        &  --      \\
Q2059--0528      &   2.210   &     $<0.8$    &       $<6.3$  &     --     & $1.02\pm0.17$ & $>0.2$  \\
Q2222--0946      &   2.354   &  $0.75\pm0.03$  &   $6.3\pm0.3$ &  $44\pm3$ & $14.3\pm0.3$  & $13\pm1$  \\
Q2348--011-1     &   2.425   &  $0.7\pm0.2$  &   $5.9\pm1.4$ & $180\pm42$ & $0.55\pm0.15$ & $>0.2$  \\
Q2348--011-2     &   2.614   &       --      &        --     &      --    & $<0.3$        &  --      \\
\hline
\end{tabular}
\end{center}

{\flushleft
$^{(a)}$ Star formation rates inferred from \lya\ assuming standard case B recombination theory (\lya\ / H$\alpha$ = 8.7)
and using \citet{Kennicutt1998} converted to the initial mass function of \citet{Chabrier2003}.
For Q0918+1636-2 and Q2222--0946, we give the more precise measurements derived by \citet{Fynbo2013b} and \citet{Krogager2013}, respectively.\\[1mm]

$^{(b)}$ Emission detected from near-infrared lines \citep{Fynbo2011, Fynbo2013b}.\\[1mm]

{\sc Note} --- All fluxes except for PKS0458--020 and Q2222--0946 should be considered lower limits due to unknown slit-losses.}

\end{table*}

\section{Model Comparison}
\label{model}

We compare the \lya\ detections in our high-metallicity sample of DLAs with predictions from a model in
which DLAs (at redshift 2 to 3) arise in gas distributed around star-forming galaxies \citep{Fynbo2008}.
In short, the model assumes that the galaxies hosting DLAs are drawn from the population of UV-selected,
star-forming galaxies, however, instead of being selected based on their luminosity, the galaxies are
selected based on their \ion{H}{i} absorption cross-section, $\sigma_{\rm H\,\textsc{i}}$. The model
assumes simple, observationally motivated scaling relations between luminosity, metallicity, and
$\sigma_{\rm H\,\textsc{i}}$. For simplicity, the \HI-extent of the galaxies is approximated by uniform,
circular, thin discs with random inclinations. For a given galaxy, a random impact parameter for the DLA
is subsequently drawn from the projected area of \HI\ on the sky, and an absorber metallicity is assigned
by assuming a metallicity gradient as a function of luminosity.
High redshift galaxies are clearly not simple, homogeneous
systems, and the model is not to be seen as an actual one-to-one
description of individual galaxies. Instead, the model should be understood
as a method to statistically predict the expected distribution of the
individual parameters based on the underlying scaling relations.
For further details, see \citet{Fynbo2008}. The output from the model, which we use
as the basis for our model comparison, is a table of metallicity and corresponding
impact parameter for each model realization.

\subsection{Modelling Ly\boldmath{$\alpha$} Emission}
In order to directly compare our \lya\ emission statistics with the model by
\citet[][hereafter the F08 model]{Fynbo2008}, we simulate the observation of the
11 DLAs in our statistical sample (Sect.~\ref{data:sample}) within the framework
of this model. Since the original model did not include information about the
\lya\ flux, we extend the model to calculate the expected \lya\ line flux given
the galaxy's luminosity. This line flux is subsequently `observed' through our
spectroscopic setup. The details are explained below.
\defcitealias{Fynbo2008}{F08}

First, we establish the probability distribution of impact parameter (in units of kpc), $b$,
as a function of metallicity, [M/H], from the \citetalias{Fynbo2008} model realizations.
We denote this distribution: $P\,(b\,|\,[{\rm M/H}])$.
We quantified the probability distribution in terms of its percentiles as a function of
metallicity in order to speed up this calculation, and to account for the limited realizations
of the F08 model at high metallicity, see Appendix~\ref{app:PDF} for details.

The original model by \citetalias{Fynbo2008} employed a complex description of the metallicity
gradient which was a function of the luminosity of the galaxy. However, recent theoretical work
has failed to reproduce this trend and conclude that `low-mass galaxies tend to have flat gradients'
\citep{Ma2017} in contrast with the original gradient implemented by \citetalias{Fynbo2008}. 
In the following, we therefore revert to a much simpler assumption of a single, constant metallicity
gradient. This assumption is in agreement with both recent theoretical work \citep[e.g.,][]{Ma2017}
and observations \citep{Christensen2014, Peroux2014, Stott2014, Wuyts2016, Kaplan2016}.
For the metallicity gradient we adopt the value $0.022 \pm 0.004$~dex~kpc$^{-1}$ reported by \citet{Christensen2014}.
This gradient is determined purely from absorption metallicities in DLA galaxies with known impact parameters,
and is therefore not subject to systematic uncertainties between
absorption and emission based metallicities, which could be the case
for gradients in absorption-emission pairs. We refer to this simplified model as the F08 model with constant gradient (hereafter F08-CG).

For each given DLA in our sample, we then use the observed [M/H] and $\zDLA$ to simulate the
expected \lya\ emission from the host galaxy in the F08-CG model framework. We generate 2000
model realizations of the \lya\ emission for each DLA in the statistical sample.
For each model realization, we go through the following steps:

\begin{enumerate}[I]
	\item {Assign a random impact parameter, $b$, at the given observed absorption metallicity from the distribution, $P\,(b\,|\,[{\rm M/H}])$, and invert the metallicity--luminosity relation from \citetalias{Fynbo2008} to find a continuum rest-frame UV absolute magnitude (at 1700~\AA) assuming the metallicity gradient of $-$0.02~dex~kpc$^{-1}$ \citep{Christensen2014}:
\begin{equation}
	{\rm M_{UV}} = -5\times([{\rm M/H}]_{0}+0.3) - 20.8\ ,
\end{equation}

\noindent where ${\rm [M/H]_{0}}$ is the central metallicity given the randomly drawn impact parameter, $b$\,:
\begin{equation}
	{\rm [M/H]_{0}} = [{\rm M/H}] + 0.02\times b\ .
\end{equation}
	}
	
	\item {Absolute magnitude, ${\rm M_{UV}}$, is converted to continuum flux density, $F_{\rm UV}$, at 1700~\AA\ in terms of $F_{\lambda}$, where the applied distance modulus is calculated for $\zDLA$.\\
	}
	
	\item {Assign a random rest-frame equivalent width of \lya, $W_{\rm Ly\alpha}$,
	from an exponential distribution:
	\begin{equation}
		P(W_{\rm Ly\alpha}) = \frac{1}{w_0} e^{-W_{\rm Ly\alpha}/w_0}\ ,
	\end{equation}
	 where the width of the exponential distribution, $w_0$, is determined from the redshift dependent, observed relation by \citet{Zheng2014} evaluated at $\zDLA$: $w_0 = 14.0~{\rm \AA} \times (1+z_{\rm DLA})^{1.1}$.
	$W_{\rm Ly\alpha}$ is subsequently converted to a \lya\ line flux, $F_{\rm Ly\alpha}$, given $F_{\rm UV}$.
	}\\
	
	\item {Convert the impact parameter, $b$, from proper distance units at $\zDLA$ to angular separation in arcsec, $\beta$, and assign a random position angle on the sky, $\phi$, uniformly distributed between $0$ and $2\pi$. A sky position is then calculated:
	\begin{equation}
		\alpha = \beta \cos(\phi)\ \ {\rm and}\ \ \delta = \beta \sin(\phi)\ .
	\end{equation}}\\
	
	\item {Create a \lya\ emission profile at the sky location ($\alpha$, $\delta$). For this purpose, we use a 2-dimensional S\'ersic profile with a fixed index ($n_s = 1$), i.e., an exponential profile.
\citet{Wisotzki2016} find an average scale length for the \lya\ emission profile of $r_{\rm Ly\alpha} = 4\pm2$~kpc at redshift $z \approx 3.1$ (from their tables 1 and 2). Moreover, the authors report a redshift evolution of roughly a factor of two increase from $z=5.1$ to $z=3.7$. Assuming a similar increase from $z=3.7$ to $z=2.3$, we find an average $r_{\rm Ly\alpha}$ of $8\pm4$~kpc at $z\approx 2.3$. This is furthermore consistent with the stacking results from \citet{Momose2014}, who find $\langle r_{\rm Ly\alpha} \rangle = 7.9$~kpc for $z=2.2$. We then draw a random scale length from a Gaussian distribution centred at 8~kpc with a standard deviation of 4~kpc, however, we truncate the distribution at the high end ($r_{\rm Lya}<20$~kpc) motivated by the observed sample variance reported by \citet{Wisotzki2016}, and we require that the scale length be larger than zero.}\\
	
	\item {Convolve the \lya\ emission profile with a Gaussian point spread function (PSF) with a full width at half maximum of 0\farcs8. This corresponds to the average seeing for all the observations. The convolved emission profile is then scaled to yield a total flux of $F_{\rm Ly\alpha}$.\\}

	\item {For each of the three X-shooter slits, calculate the amount of the total flux which is covered by the slit at a given position angle: ${\rm PA} = -60\degr, 0\degr, 60\degr$.
If the flux in a slit is larger than the detection limit for the given target, then we mark the emission line as detected in this slit. We take into account the higher flux limit derived for PA3 of Q1435+0354.\\}

	\item {Lastly calculate the amount of flux in the overlap of the three slits.
If this flux is larger than the detection limit divided by $\sqrt{3}$, then the emission line is marked as detected in the stacked region.
If the emission line is detected in any of the three slits {\it or} in the combined central region, then the emission line is detected for this DLA.}

\end{enumerate}

Integrating the number of detections over a full set of realizations for the 11 DLAs gives us a directly comparable measure to the number of detections of \lya\ obtained in our campaign. Doing so for the 2000 realizations yields an expected number of \lya\ detections of $N_{\rm Ly\alpha} = 5.7\pm1.1$ out of 11 DLAs, see Fig.~\ref{fig:simulations}. Compared to the 5 detections out of 11 in this work, our detection rate is consistent with the model within 1\,$\sigma$.

\subsection{Simulating a Control Sample of DLAs}
\label{model:control_sample}
\begin{figure}
	\includegraphics[width=0.48\textwidth]{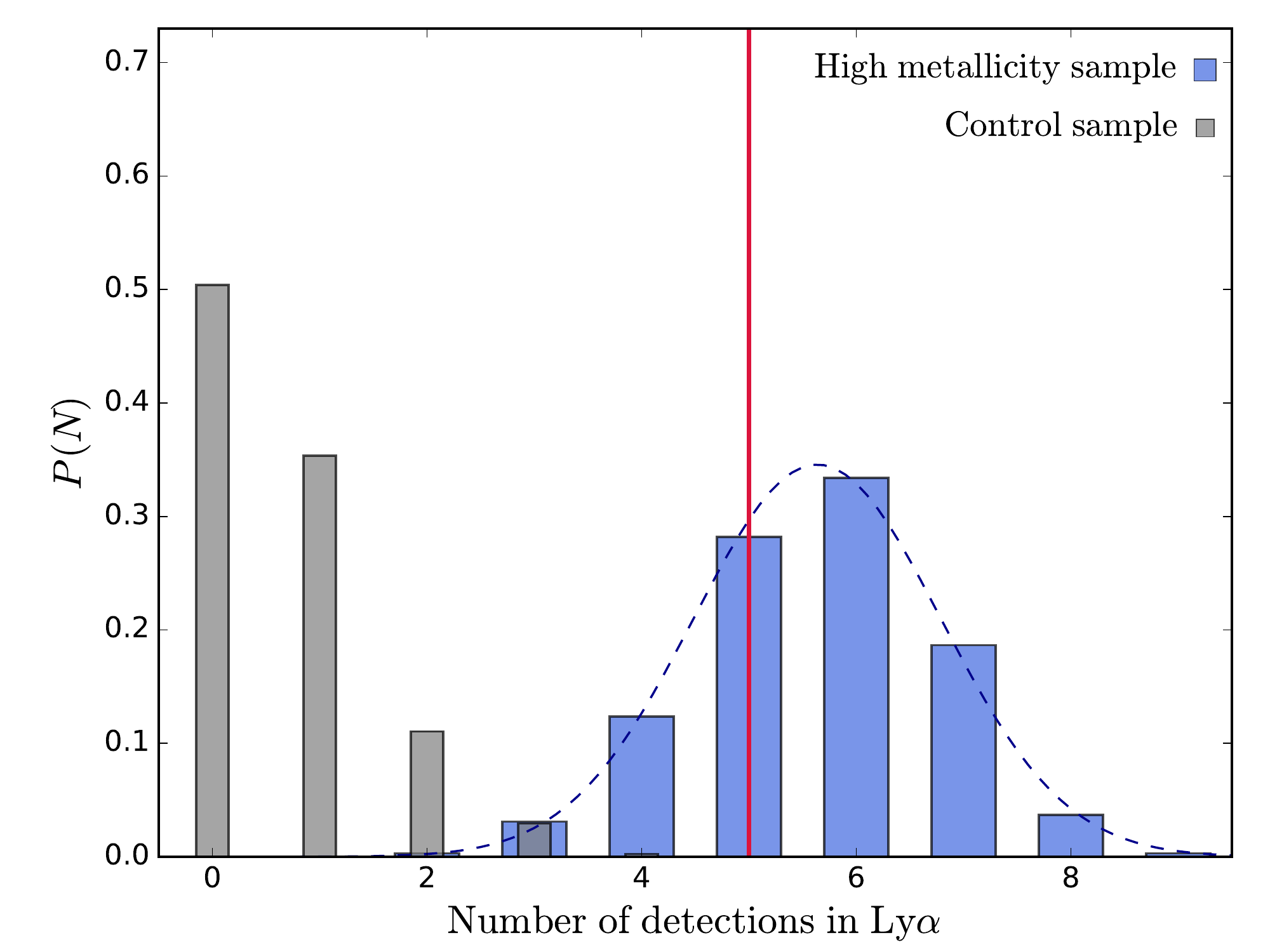}
	\caption{Number of detected \lya\ emission lines out of 11 DLAs. The blue distribution with wide bins shows the results from the simulated datasets given the model presented in the text. The distribution is well approximated by a Gaussian distribution (the dashed line). The red, vertical line marks the actual number of detections of \lya\ in our statistical sample (5 out of 11). The grey distribution with narrow bins shows the results from the simulation of 11 DLAs drawn randomly from the overall metallicity distribution for $z\approx2$ DLAs.\label{fig:simulations}}
\end{figure}

In order to compare our selection of metal-rich DLAs to a random sample with no prior selection on metallicity, we run the same simulation as above for a sample with randomly drawn metallicities.
We assume a fixed redshift of $z = 2.3$, which corresponds to the median redshift of the statistical sample.
For every model realization we draw a set of 11 metallicities from a Gaussian distribution ($\mu, \sigma) = (-1.51, 0.57)$, motivated by observations of DLAs at $z\approx2$ \citep{Rafelski2014}. The number of expected \lya\ detections for random DLAs (observed with the 3-slit X-shooter setup) within the F08-CG model is found to be well-represented by a Poissonian distribution with an average number of detections of $\langle N_{\rm Ly\alpha} \rangle = 0.7$.
The results for the simulated random sample are shown as the grey distribution with narrow bins in Fig.~\ref{fig:simulations}.

\subsection{Summary of X-Shooter Campaign}
Overall we observe good agreement between the observed detection rate presented in this work and the model predictions from \citetalias{Fynbo2008}-CG. We find a high detection rate of \lya\ of 45\% (5/11) when considering only the statistical sample. This is fully consistent with the predicted 52\% from the model. When including the two targets, which are detected only in the near-infrared (the two DLAs toward Q0918+1636), this yields a total detection rate of 64\% (7/11). 
The overall detection rate (64\%) presented here is significantly higher than what has been reported from previous surveys with no preselection on metallicity where a detection rate of roughly 10\% at $z\sim2$ is recovered \citep[e.g.,][]{Warren2001, Moller2002}. Similar low detection rates are inferred from integral field spectroscopy of H$\alpha$ \citep{Bouche2012a, Peroux2012}. This low detection rate from a purely \HI-defined sample is consistent with the low average number of detections found in our modelled control sample of $6.4_{-5.8}^{+19.8}$\%. Thus when looking at the average for our sample, we find that the model provides a good agreement with the observations, even in terms of the previous low detection rates. One thing that has not explicitly been addressed in the modelling is the effect of dust. Since the observed distribution of equivalent widths of \lya\ already takes the average attenuation into account, this is in large part included indirectly in the modelling. The effect of dust is discussed in more detailed in Sect.~\ref{discussion:dust}.

\begin{figure*}
	\includegraphics[width=0.98\textwidth]{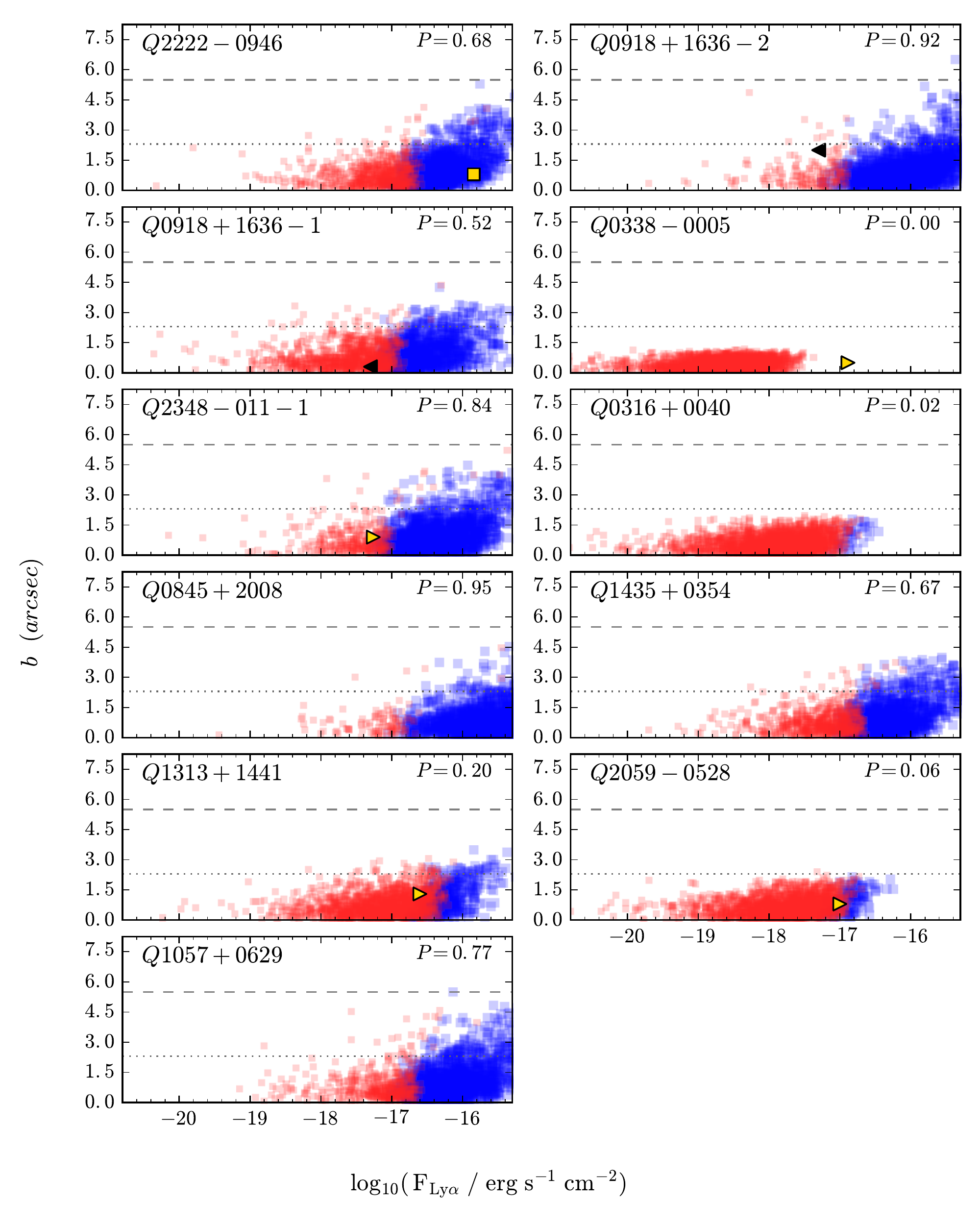}
	\caption{Individual predictions of impact parameter and \lya\ flux for each target in our statistical sample. Each point corresponds to one of the 2000 model realizations performed for each DLA. Small, red points indicate cases where the given modelled flux would {\it not} have been detected in our data. Conversely, large, blue points indicate cases where the modelled flux would have been detected. Actual detections of \lya\ from this work are marked by yellow squares and triangles. The yellow triangles mark lower limits to the \lya\ flux in cases where we cannot account for slit loss. The black triangles mark upper limits on the \lya\ flux, where the counterpart is detected through other emission lines at the indicated impact parameter. The dashed line shows the extent of the area covered by the slit (i.e., half the slit-length, 11 arcsec). The dotted line marks the extent out to which our slit configuration has a coverage of 50\% or more. \label{fig:individual_sim}}
\end{figure*}

While the predictions from the model are mainly valid as a statistical average over the entire sample, we can gain some additional insight by looking at the model predictions for individual targets, see Fig.~\ref{fig:individual_sim}. In this figure, we show the modelled impact parameter as a function of line flux for each of the DLAs in the statistical sample. In each panel, we give the probability of detecting \lya\ for this DLA, $P$, given the observed noise level of the data. Each detection is marked by an orange symbol, and for the two DLAs toward Q0918+1636, we have detections from other emission lines but no detection of \lya\ \citep{Fynbo2013b}. These two DLAs are marked by black triangles indicating the observed impact parameter. When interpreting the results for the individual targets, we observe some tension between model and data. This will be discussed further in Sect.~\ref{discussion:individual}.

\section{Application to Literature Samples}
\label{comparison}

Many surveys for emission counterparts of high-redshift DLAs have been carried out in the past, and a low detection rate ($\sim10$\%) has been reported in all of these surveys. In order to test how these numerous non-detections fit into our model framework, we have adapted our model to mimic the various strategies applied in these surveys ranging from far UV continuum emission to H$\alpha$ line emission. In the following, we will compare our model expectations to the results of these surveys and explain the slight modifications to our model in each case.
Additionally, we will compare our model expectations to the spectral stacking analyses performed on the SDSS and BOSS data \citep{Rahmani2010, Noterdaeme2014, Joshi2017}.

\subsection{The HST NICMOS/STIS Survey}
\label{comparison:warrenmoller}

The observing strategy for the X-shooter survey we have reported on
here was developed based on the successes, and failures, of the large
{\it HST}--NICMOS/STIS and VLT--FORS/ISAAC survey of 24 DLAs and sub-DLAs
\citep{Warren2001, Moller2002, Weatherley2005}.
The sample definition of that survey was aimed at spanning a wide
parameter space, and was therefore not optimized to target objects with
high metallicity. In fact, at the time the metallicity of most of the
target DLAs was unknown. In hindsight, we therefore now understand why
the success rate was correspondingly low \citep{Moller2004}.

The strategy of the original survey was markedly different from the
current X-shooter survey. Initially NICMOS and STIS imaging was obtained
with the goal to identify targets for spectroscopic followup with
FORS (to search for \lya\ emission), and ISAAC (to search for
H$\alpha$, H$\beta$, [\ion{O}{ii}], and [\ion{O}{iii}] emission).
The final spectroscopy was not evenly
distributed on all QSOs; instead slits were placed on the QSOs using
PAs to include the candidate galaxies resulting in up to four slit
positions on individual QSOs during the followup.
In order to assess the efficiency of the two strategies, we now ask the question:
how many of the DLAs in the original sample would we have detected with
the current X-shooter strategy?

In order to make the comparison meaningful, given our current
knowledge, we here only consider the unbiased, intervening
DLAs, i.e., we do not include sub-DLAs, DLAs which, at the
time, had already been detected in emission, nor proximate 
($z_{\rm abs} \approx z_{\rm em}$) DLAs which may be in an altered physical
environment (\citealt*{Moller1998a}; \citealt{Ellison2010}).
Also, for a few of the targeted systems the metallicity has still not
been determined, and therefore we cannot include them. From the target
list of \citet{Warren2001}, we then have a complete and unbiased list
of 15 DLAs which represents a random DLA sample.

\begin{figure}
	\includegraphics[width=0.48\textwidth]{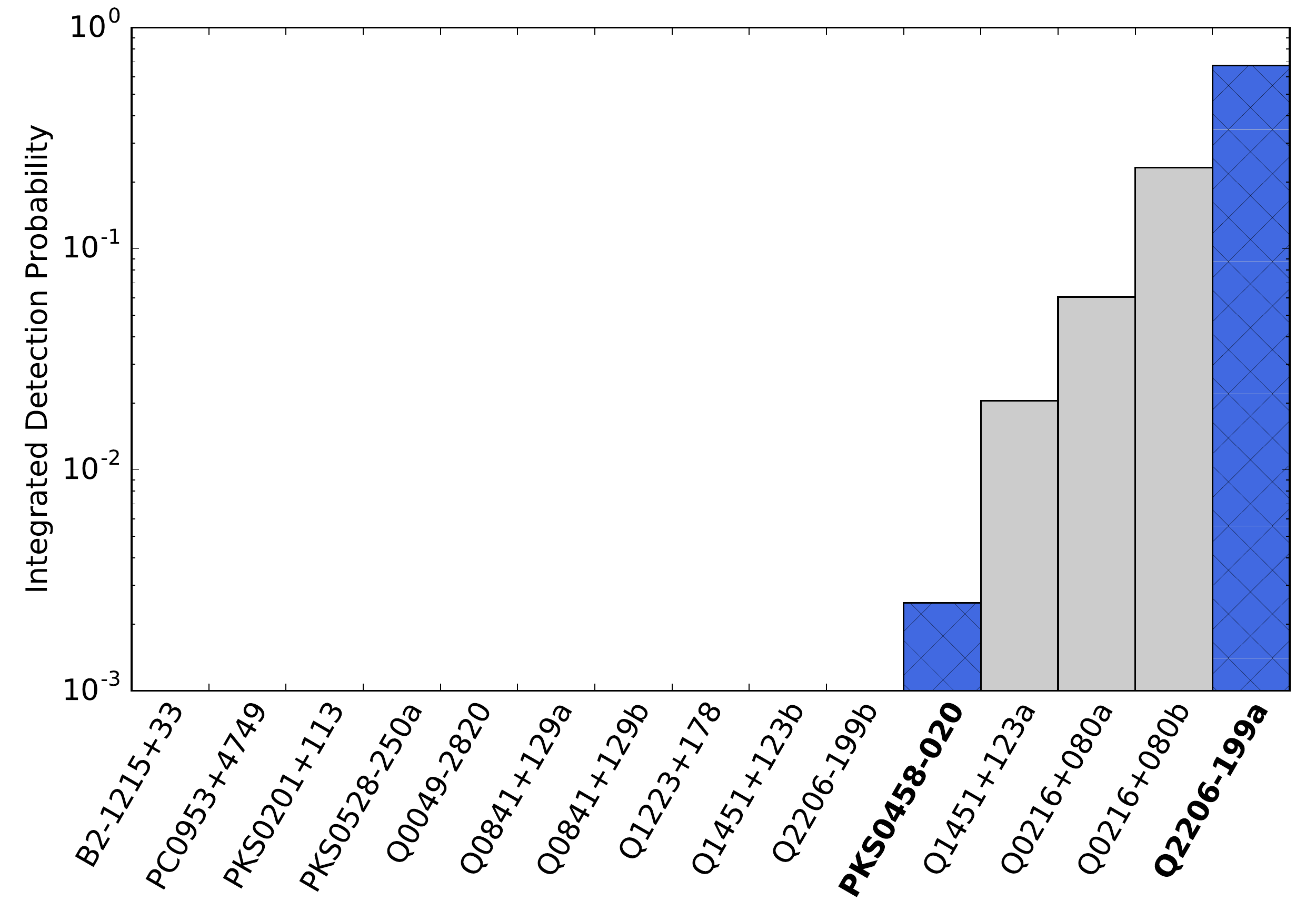}
	\caption{Individual detection probabilities for the unbiased,
	intervening DLA sample from the {\it HST} study by \citet{Warren2001}.
	For each target, the detection probability is indicated by the height
	of the bar. The targets are sorted by ascending probability going from
	left to right. For targets where no bar is shown, the modelled detection
	probability is $P=0$.
	The two objects that were detected from this sample
	(PKS0458--020 and Q2206--199a) are highlighted by blue, hatched bars
	and their labels are set in boldface.\label{fig:HST_sim}}
\end{figure}

Applying the exact same analysis of Section~\ref{model} to the sample defined
above, we obtain individual detection probabilities for each DLA
as shown in Fig.~\ref{fig:HST_sim}, 
and a total predicted number of detections of
$1.0 \pm 0.7$. This is fully consistent with
what we would predict for a random sample ($15 \times 0.064 = 1.0$)
given the detection probability of 6.4\% inferred from our control sample,
and also in good agreement with the two
detections that resulted from the survey. The conclusion is
therefore that either of the two observing strategies will provide the
same number of detections for the same sample. The real difference lies
in the strategy for the sample definition. Surveys are still today
being conducted on randomly selected samples, their success rates for
detection can easily be predicted from the examples presented in this
paper.

Interestingly, the two successful detections of the {\it HST}/VLT survey,
Q2206--199a and PKS0458--020, are the most likely by far and the fifth
most likely, respectively. This mirrors the detection distribution 
in the X-shooter sample. In short, we also here see that our model
predicts well the expected number of detections in a sample, but it is
not able to predict precisely which will be detected on a one-to-one
basis. This reflects the fact that effectively the prediction is based
on the underlying mass--metallicity relation, which has a substantial
scatter of 0.38 dex in metallicity \citep{Moller2013}.

\subsection{Integral Field Spectroscopy with SINFONI}
\label{comparison:peroux}

\citet[][and references therein]{Peroux2012} carry out a survey for DLA (and sub-DLA) counterparts by looking for nebular lines in the near-infrared using the integral field spectrograph SINFONI mounted at the VLT. Specifically for the high-redshift part of their sample, they search for H$\alpha$ emission in the $K$-band. In the following, we will compare our model expectations to the \citet{Peroux2012} study. We restrict our comparison to the 11 $z>2$ DLAs (i.e., only targets with $\log\NHI>20.3$ are considered) in their sample and we use their quoted [Si/H] as the metallicity indicator, since [Zn/H] is not available for the entire sample, and in cases where both Zn and Si are provided, these are consistent.

By assuming that the star formation rate inferred from the UV continuum flux directly traces the star formation rate inferred from H$\alpha$, we are able to estimate the expected H$\alpha$ flux from the model. We here assume that the flux is a point source broadened only by the Gaussian seeing, since the H$\alpha$ emission is much less extended spatially than \lya.
We then perform a similar set of realizations as described in Sect.~\ref{model}, except we convert the continuum flux directly to H$\alpha$ flux following \citet{Kennicutt1998}.
We model the flux within an 8$\times$8~arcsec$^2$ field of view to mimic the SINFONI setting used by \citet{Peroux2012}. As in Sect.~\ref{model}, we compare the modelled fluxes to the detection limits stated by the authors\footnote{Note that we use 3-$\sigma$ limits in this work, and subsequently convert the 2.5-$\sigma$ limits stated by \citet{Peroux2012}.}.
Based on 2000 realizations, we find an average number of H$\alpha$ detections of $\langle N_{\rm H\alpha} \rangle = 0.6$ out of 11 targets with more than 90\% of the realizations resulting in either zero or one detection. This is in perfect agreement with the one detection reported by \citet{Peroux2012}.

The direct conversion from UV luminosity to H$\alpha$ luminosity relies on the assumption that the two different calibrations probe star formation activity on the same scales (both in space and time). While this undoubtedly introduces further scatter in the modelling results, we have tested that the assumption is consistent for other systems where we have the data available.
Two DLAs in our X-shooter campaign have precise measurements of H$\alpha$ via deep ($\sim$10 hr) follow-up observations: namely 2222--0946 and 0918+1636-2. By applying the same modelling approach as described above for the H$\alpha$ flux, we find an expected flux ranging from $3.1$ to $10\times10^{-17}$~erg~s$^{-1}$~cm$^{-2}$ (1$\sigma$) for DLA 2222--0946 in perfect agreement with the $5.7\times10^{-17}$~erg~s$^{-1}$~cm$^{-2}$ observed \citep{Peroux2012, Krogager2013}.
For DLA 0918+1636-2, we find an expected flux of H$\alpha$ ranging from $8.4$ to $31\times10^{-17}$~erg~s$^{-1}$~cm$^{-2}$ (1$\sigma$) perfectly consistent with the observed flux of $27\times10^{-17}$~erg~s$^{-1}$~cm$^{-2}$ \citep{Fynbo2013b}.
We therefore find that the direct conversion provides reasonable agreement for the available data.

\subsection{Direct Imaging below the Lyman Limit}
\label{comparison:fumagalli}

By observing bluewards of the Lyman-limit of higher redshift Lyman-limit systems, \citet{Fumagalli2015} have carefully searched for UV continuum emission from the lower redshift DLAs along the same sight-lines. This way the higher redshift absorber effectively serves as a `blocking filter' which removes the quasar light and allows the emission from the DLA to be observed free of quasar contamination \citep[see][]{OMeara2006, Christensen2009, Fumagalli2010}. \citeauthor{Fumagalli2015} find a very low {\it in situ} star formation rate, i.e., only star formation right at the absorber location is considered. The survey is carried out in two parts: one from ground-based imaging using Keck I and another from space-based imaging using {\it HST}. From the ground-based survey, three DLA counterpart candidates are identified; however, one of these is shown to have a non-negligible probability of being an interloper and another is very likely contaminated by quasar leakage due to the optical depth at the Lyman-limit being insufficient to completely block the background quasar.
Neither of the three candidates have spectroscopic confirmation.
On average, an upper limit on the in situ star-formation of $\dot\psi \lesssim 0.65$~M$_{\odot}$~yr$^{-1}$ is inferred within an aperture of 1\farcs5 diameter. From the space-based survey, no detections are reported and an average upper limit of $\dot\psi \lesssim 0.38$~M$_{\odot}$~yr$^{-1}$ is inferred within an aperture of 0\farcs25 diameter.

With the modelling approach presented in Sect.~\ref{model}, we can now compare the results of \citet{Fumagalli2015} to the expectations from our model. We only have to make minor adjustments to the modelling approach. Instead of estimating the \lya\ flux, we use the continuum luminosity at 1700~\AA\ to infer the star formation rate, $\dot\psi$ \citep{Kennicutt1998}. We calculate a spatial profile using a 10 times smaller scale-length than for the \lya\ emission. This agrees well with observations of continuum emission associated to \lya\ emitters which typically has a scale-length smaller than 1~kpc. The spatial profile is convolved with the appropriate seeing FWHM and the profile within the assumed aperture is integrated to give an `observed star formation rate', $\psi_{\rm obs}$. If $\psi_{\rm obs}$ is larger than the reported upper-limit for a given target, the given realization is marked as detected. For each target in the sample\footnote{Only targets with available metallicity measurements have been included. Sample properties have been taken from \citet{Fumagalli2014}.}, we create 2000 realizations to calculate the distribution of expected detections assuming the same cosmological parameters as in the work by \citeauthor{Fumagalli2015}.

From the modelling of the ground-based sample (12 DLAs), we infer an average number of detections of $\langle N \rangle = 1.6\pm1.0$. The most probable number of detections for the ground-based sample is 1--2 with equal probabilities, in perfect agreement with the observations (0--3 detections).
For the {\it HST} sample (of 14 DLAs), we find that no detections are expected in 95\% of the cases, thus fully consistent with the non-detections reported by \citet{Fumagalli2015}.

We note that what the authors deem the most likely emission counterpart candidate from the ground-based sample (5:G5) has a very low detection probability in our modelling. Nonetheless, the non-detections reported by \cite{Fumagalli2015} and low star formation rates inferred for average DLAs is fully consistent with an underlying metallicity--luminosity relation as implemented in the F08 model.

The concept of `in situ SFR' measured in small apertures centred at the
position of the background quasar means, as correctly pointed out by
the authors,
that only a fraction of the SFR of the host galaxy is included; consequently,
if the impact parameter is large, that fraction could be very small as only
the outskirts of the host is considered.  Effectively this means that for
each host there is an impact-parameter dependent aperture correction which
must be applied to convert the reported upper limits (on in situ UV-flux or SFR)
to a total upper limit.  \citeauthor{Fumagalli2015} have not computed this in their analysis,
however, based on our model we are able to obtain an aperture correction for each target.
The actual flux fractions within the apertures range from 3\% to 50\% (HST data; 0\farcs25 aperture)
and $\gtrsim$40\% (ground based data; 1\farcs5 aperture).
We shall use those flux aperture corrections in Sect.~\ref{discussion}
where we discuss the luminosity relation.
For our computation of detection probabilities the aperture
correction is inherently included as described above.

\subsection{SDSS Spectral Stacking}
\label{comparison:stacking}

\citet{Rahmani2010} use the SDSS spectra from DR7 to constrain the average \lya\ flux from DLAs by stacking 341 spectra of DLAs with $\log\NHI \geq 20.62$. The authors do not detect any flux in the bottom of the DLA absorption trough and place an upper limit on the line flux of $F_{\rm Ly\alpha}<3.0\times10^{-18}\ {\rm erg\ s^{-1}\ cm^{-2}}$ (using a clipped mean) and $F_{\rm Ly\alpha}<3.9\times10^{-18}\ {\rm erg\ s^{-1}\ cm^{-2}}$ (using a regular mean). We compare the inferred flux limit with expectations from our model by simulating a sample of 341 DLAs at an average redshift of $z_{\rm abs}=2.86$ following the method laid out in Sect.~\ref{model}. Instead of calculating the flux observed through the three X-shooter slits, we calculate the flux observed in a circular fiber with a diameter of 3\arcsec\ centred on the quasar position. Also, we assume an average seeing of 1\farcs5 to match the site conditions for the SDSS observatory \citep{York2000}. This way we generate 200 stacks each containing a set of 341 DLAs with randomly assigned metallicities. Based on the modelled stacks, we observe a mean \lya\ flux of $F_{\rm Ly\alpha} = 2.6\times10^{-18}\ {\rm erg\ s^{-1}\ cm^{-2}}$. Hence, the expected \lya\ emission given the F08-CG model is in perfect agreement with the non-detection reported by \citet{Rahmani2010}. In 25\% of the realizations, we observe a stacked flux above the reported 3$\sigma$ limit (for the clipped mean), while this is only observed in 13\% of the realizations assuming the limit calculated with a regular mean.

Furthermore, in two recent studies, \citet{Noterdaeme2014} and \citet{Joshi2017} have stacked a large number of DLAs from the BOSS spectrograph (as part of the SDSS--III) to look for \lya\ emission.
\citet{Noterdaeme2014} report a positive detection of \lya\ $\left( L_{\rm Ly\alpha} = 6 \pm 2 \times 10^{41}~{\rm erg~s^{-1}} \right)$ for DLAs with $\log\NHI > 21.7$.
Including more DLAs with lower $\NHI$ ($\log\NHI > 21$), \citet{Joshi2017} report a tentative detection of \lya\ and infer a luminosity of $L_{\rm Ly\alpha} = 5.2\pm3.3 \times10^{40}\ {\rm erg\ s^{-1}}$.

Taking into account the smaller fiber width of the BOSS instrument (2\arcsec), we model as before the expected \lya\ luminosity for a stack of 95 DLAs \citep{Noterdaeme2014} using the DLA redshifts as stated in their paper to generate the sample. Since the metallicities are not available, we draw them randomly from the overall DLA metallicity distribution as in Sect.~\ref{model:control_sample}. Similarly, we model the stack of 704 DLAs \citep{Joshi2017} using random absorber redshifts in the range $2.3 < z_{\rm DLA} < 3.4$ and random metallicities. We have adopted the same cosmology for our modelling as assumed by these authors.
Since our model does not take into account the column density of \HI, we find similar expected luminosities for the two stacks.
Based on 200 realizations of each of the stacks, we infer an expected  luminosity of $\log ( L_{\rm Ly\alpha} / {\rm erg\ s^{-1}}) = 41.1\pm0.3$.
This value is in between the two results reported by \citet{Noterdaeme2014} and \citet{Joshi2017}, but consistent at the 2.3-$\sigma$ and 1.3-$\sigma$ levels, respectively. The differences between the two studies can be understood as a result of the anti-correlation between impact parameter and column density, leading to smaller average impact parameters for the higher column density DLAs in the study by \citet{Noterdaeme2014}. The smaller average impact parameter means that a larger fraction of the DLA counterpart emission fits within the central 2\arcsec\ as probed by the BOSS fiber. A similar correlation between luminosity and $\NHI$ is observed in the \citet{Joshi2017} study. 
Hence, the expected luminosity will be underpredicted as too much light is placed outside the fiber.
Since we do not have an exact parametrization of the impact parameter distribution for a given $\NHI$ (like we do for metallicity and impact parameter), we are at present not able to take this effect into account in our modelling.
Interestingly, \citet{Joshi2017} also find that the luminosity correlates with the equivalent width of \SiII\,$\lambda\,1526$ in agreement with the results of our X-shooter campaign.

\begin{table}
\caption{Summary of model comparison results. \label{tab:comparison}}
\begin{center}
\begin{tabular}{lccc}
\hline
Sample                       		&    Detected    &   Predicted           & \# DLAs$^{(a)}$ \\  
\hline
X-shooter campaign      			&        5       &   $5.7\pm1.1$         & 11  \\[2mm]  
\citet{Warren2001}                 &        2       &   $1.0\pm0.7$         & 15  \\[2mm]  
\citet{Fumagalli2015} (ground)     &      0--3      &   $1.6\pm1.0$         & 12  \\[2mm]  
\citet{Fumagalli2015} ({\it HST})  &        0       &   $<1\ \ (2\sigma)$   & 14  \\[2mm]  
\citet{Peroux2012}                 &        1       &   $\le1\ \ (2\sigma)$ & 11  \\
\hline
\citet{Rahmani2010}$^{(b)}$        &      $<$ 3.9   &  2.6                  & 341 \\[2mm] 
\citet{Noterdaeme2014}$^{(c)}$     &  60 $\pm$ 20   &  13$^{+12}_{-7}$      & 95  \\[2mm] 
\citet{Joshi2017}$^{(c)}$        &  5.2 $\pm$ 3.3 &  13$^{+12}_{-7}$      & 704 \\
\hline
\end{tabular}
\end{center}

$^{(a)}$ Number of DLAs in each sample.\\[1mm]
$^{(b)}$ \lya\ flux in units of $10^{-18}$~erg~s$^{-1}$~cm$^{-2}$.\\[1mm]
$^{(c)}$ \lya\ luminosity in units of $10^{40}$~erg~s$^{-1}$.

\end{table}

\section{Discussion}
\label{discussion}

We have compared all major surveys for emission counterparts (both
continuum and line emission) of high redshift DLAs to a model based on
the framework of \citet{Fynbo2008} with a constant metallicity
gradient applied \citep{Christensen2014}. The results, in terms of
detection rates, of this comparison are summarised in
Table~\ref{tab:comparison} and overall we observe a very good
agreement.
In particular, our detailed analysis of this large
body of DLA samples (1203 DLAs in total, distributed on 8 different
surveys which were observed using a wide variety of methods)
clarifies the reason for the very low detection success rates in
blind surveys versus the high success rate in our targeted X-shooter
survey.

While the model provides very good agreement on average, we note that the
predictions for individual targets are limited by uncertainties due to the
scatter in the underlying scaling relations.
In the following section, we will discuss these limitations further in the
context of the X-shooter campaign, for which we have the most direct constraints.

Firstly, our analysis highlights that \lya\ is a notoriously difficult line to
interpret since its emission profile (as well as its mere detection) depends
heavily on the geometry and dust distribution of the interstellar medium in the
galaxy \citep[e.g.,][]{Neufeld1991, Laursen2009b}. Thus the detection
(or non-detection) of other emission lines is important in order to obtain more
robust conclusions about the counterpart properties of DLAs as these lines will
be less prone to complex scattering which complicates the \lya\ escape. However,
for the analysis of the X-shooter campaign in this work, we only consider the
\lya\ line. The detailed analysis of the nebular lines in the NIR data for the
X-shooter campaign will be presented in a forthcoming paper
(Fynbo et al. in preparation).

\subsection{The X-shooter Campaign}
\label{discussion:individual}

\begin{figure}
	\includegraphics[width=0.48\textwidth]{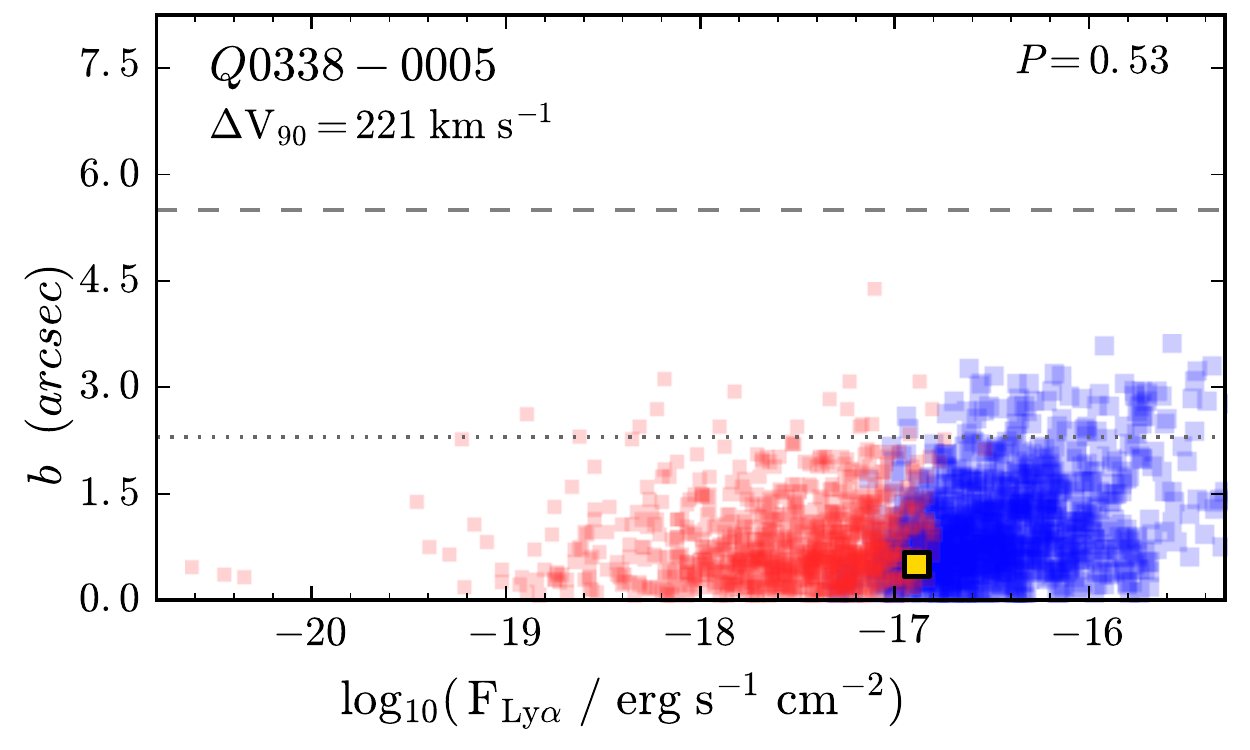}
	\caption{Same as Fig.~\ref{fig:individual_sim} for Q0338--0005 based on $\Delta V_{90}$ rather than metallicity, assuming the relation of \citet{Ledoux2006}. \label{fig:0338_test_sim}}
\end{figure}

The first thing we will address is the one outlier in our modelling: the detection of \lya\ flux from the DLA towards Q0338--0005. This DLA showed a high equivalent width of \ion{Si}{ii}$\,\lambda1526$ in the SDSS spectrum, from which the sample selection was performed. However, with the higher resolution data, we observe strong blending of this line with \lya\ forest absorption. This explains the low observed metallicity of this object ([Zn/H] $= -1.4$), which in turn explains the low predicted flux. So why do we then detect the counterpart for this DLA? The most plausible explanation for this target is that the DLA traces metal-poor gas in the vicinity of a brighter galaxy, most likely inflowing gas. Indeed, if we look at the velocity-width of the low-ionization absorption lines we observe a high value of $\Delta V_{90}=221$~km~s$^{-1}$, significantly higher than what is typically observed for such low metallicities. If we assume the relation between [M/H] and $\Delta V_{90}$ of \citet{Ledoux2006}, we can calculate an expected metallicity given the observed velocity width, [M/H]$_{\rm L06} \approx -0.6$. For this metallicity, we find that the expected flux from our model is consistent with the data, see Fig.~\ref{fig:0338_test_sim}.
We therefore conclude that this DLA most likely traces metal-poor (possibly inflowing) gas in the halo of a much more metal-rich galaxy, whose stronger gravitational potential dominates the velocity width of the gas observed in absorption. By using the combined constraints from both metallicity and absorption kinematics, we can thus improve our model predictions.

\subsubsection{The Effect of Dust}
\label{discussion:dust}
While dust attenuation is generally not considered important for DLAs (due to the low average metallicities), at the high metallicities probed in our sample, the amount of dust is expected to be non-negligible \citep[e.g.,][]{Ledoux2003, DeCia2016}. Indeed, the depletion of iron relative to zinc observed in our sample is consistent with some degree of dust in the absorbing medium (an average [Fe/Zn] ratio of $-0.7$ is observed in our statistical sample). 
Moreover, the \lya\ equivalent width is strongly affected by resonant scattering, which is difficult to quantify without the help of detailed radiative transfer simulations \citep{Laursen2009b, Noterdaeme2012a, Krogager2013}.
The attenuating effect of dust absorption and multiple scatterings has indirectly been taken into account since the observationally determined equivalent width distribution, that we use in our model, already has this effect imprinted on it.
However, given the way LAEs are selected in large surveys, the average dust extinction is most likely underestimated. We have tested the influence of an additional dust correction on our metal-rich sample of DLAs.
We incorporate a prescription for dust attenuation in absorbers from \citet{Zafar2013} who quantify visual extinction in terms of total metal column, i.e., $\log \NHI + [{\rm M/H}]$. The authors infer a metals-to-dust ratio of $\log \left(\kappa\, /\, {\rm cm^{2}}~A_V^{-1} \right)= 21.2 \pm 0.3$ \citep[see also][]{Vladilo2005}. Using this expression, we then calculated the visual extinction, $A_V$, for each DLA as:
\begin{equation}
	\log\,A_V = \log\,\NHI + [{\rm M/H}] - \log\,\kappa \ .
\end{equation}

\begin{figure}
	\includegraphics[width=0.48\textwidth]{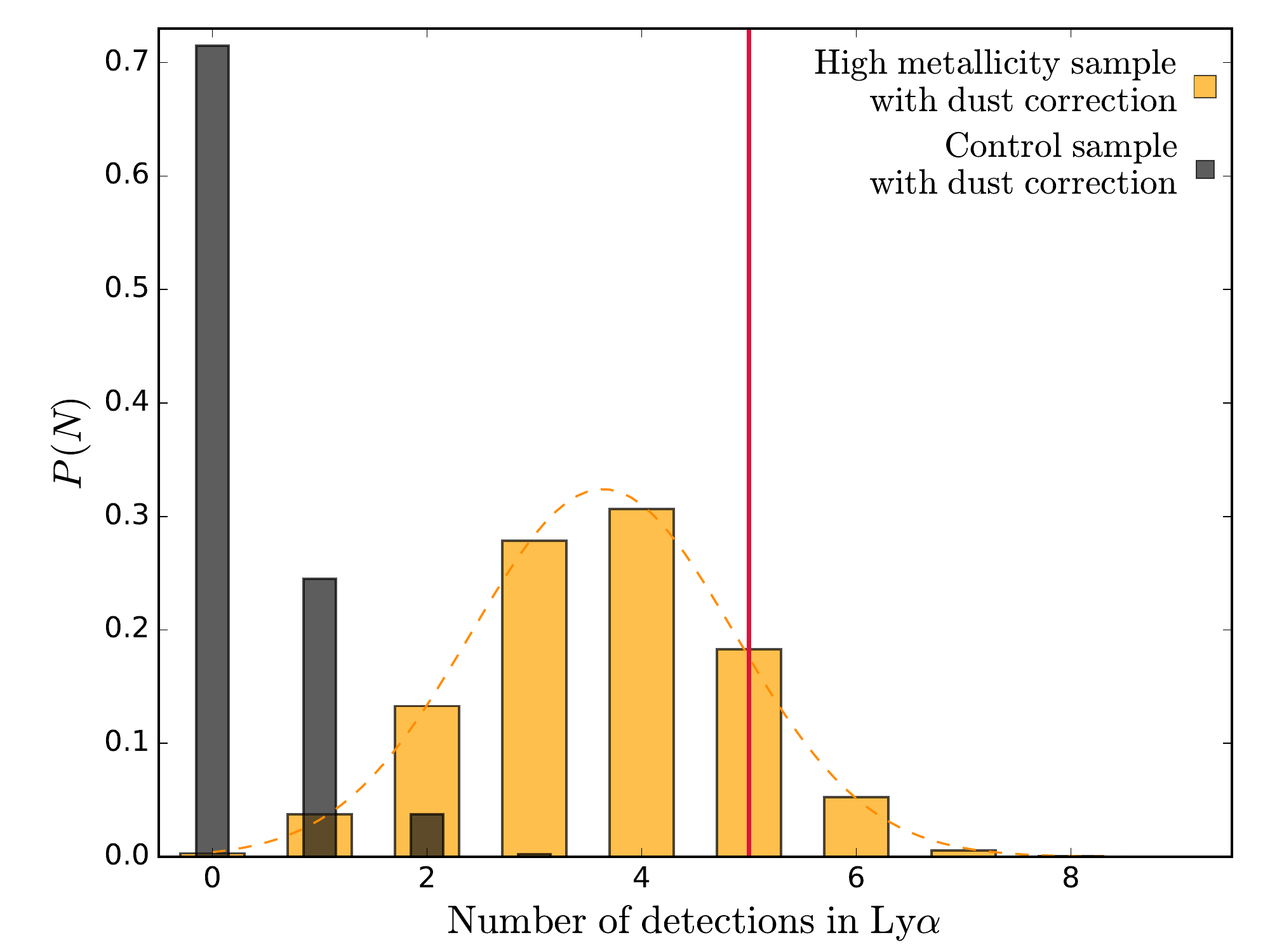}
	\caption{Number of detected \lya\ emission lines out of our statistical sample of 11 DLAs with dust correction. The yellow distribution shows the results from the simulated datasets given the model described in Sect.~\ref{model} with the additional dust correction described in the text. The distribution is well approximated by a Gaussian distribution (dashed line). The red, vertical line marks the actual number of detections in our dataset (5 out of 11). The grey distribution with narrow bins shows the results from the simulation of 11 DLAs drawn randomly from the overall metallicity distribution for $z\approx2$ DLAs.\label{fig:simulations_dust}}
\end{figure}

\noindent
The extinction of the \lya\ flux was then calculated assuming SMC type extinction.
The expected number of detections when taking dust into account is $N_{\rm Ly\alpha} = 3.6\pm1.2$, still consistent with the observed number of detections at 1.1$\sigma$, see Fig.~\ref{fig:simulations_dust}. This correction should be considered an upper limit on the effect of dust since an average correction is already included, as mentioned above. This is in good agreement with our actual observations lying in between of the two cases.

It is, however, not known exactly how the dust along the absorption line-of-sight relates to the dust affecting the emission counterpart \citep[see][]{Wiseman2017}. Nonetheless, we can gauge the impact of differences in the amount of dust at the absorber and the counterpart by looking at the detailed analysis of the DLA 0918+1636-2 \citep{Fynbo2013b} and DLA 2222--0946 \citep{Krogager2013}. For the DLA 2222--0946, the amount of dust inferred from the SED of the emission counterpart ($A_V=0.08_{-0.07}^{+0.29}$~mag) is fully consistent with the extinction derived from the depletion observed in absorption ($A_V=0.10\pm0.03$~mag). 
However, for the DLA 0918+1636-2, the $A_V$ inferred for the emission counterpart (1.54 mag) is much higher than the $A_V$ inferred from the absorber ($\sim 0.2$~mag). The impact parameter for DLA 0918+1636-2 is also much larger ($\sim16$~kpc) than for DLA 2222--0946 ($\sim6$~kpc), hence the variations might depend on impact parameter.
\citet{Wiseman2017} find similar results: While the presence of significant extinction at the emission counterpart seems to be related to the presence of dust in the absorber, the relation is far from one-to-one.

Such variations in the dust extinction can explain some of the tension we observe for the individual realizations of high-metallicity absorbers where no \lya\ emission is observed. Particularly for the DLA 0918+1636-2, \citet{Fynbo2013b} report detections of [\ion{O}{ii}], [\ion{O}{iii}], H$\beta$, and H$\alpha$, but no emission from \lya\ is observed. When including the amount of extinction reported by \citet{Fynbo2013b} in our modelling for this target, the non-detection is fully consistent with our model.

Two other DLAs with high $P$-values but no detections (Q0845+2008, and Q1057+0629) are likely affected by a similar effect as for Q0918+1636-2, given their very similar metallicities ([Zn/H] $\approx -0.1$) and depletion patterns. For these two targets, their counterparts are thus very likely detectable in the near-infrared from nebular lines such as H$\alpha$. The near-infrared constraints from our X-shooter data will be presented in a forthcoming paper (Fynbo et al. in preparation).

\subsubsection{The effect of $\log\NHI$}
\label{discussion:logNHI}
\begin{figure}
	\includegraphics[width=0.48\textwidth]{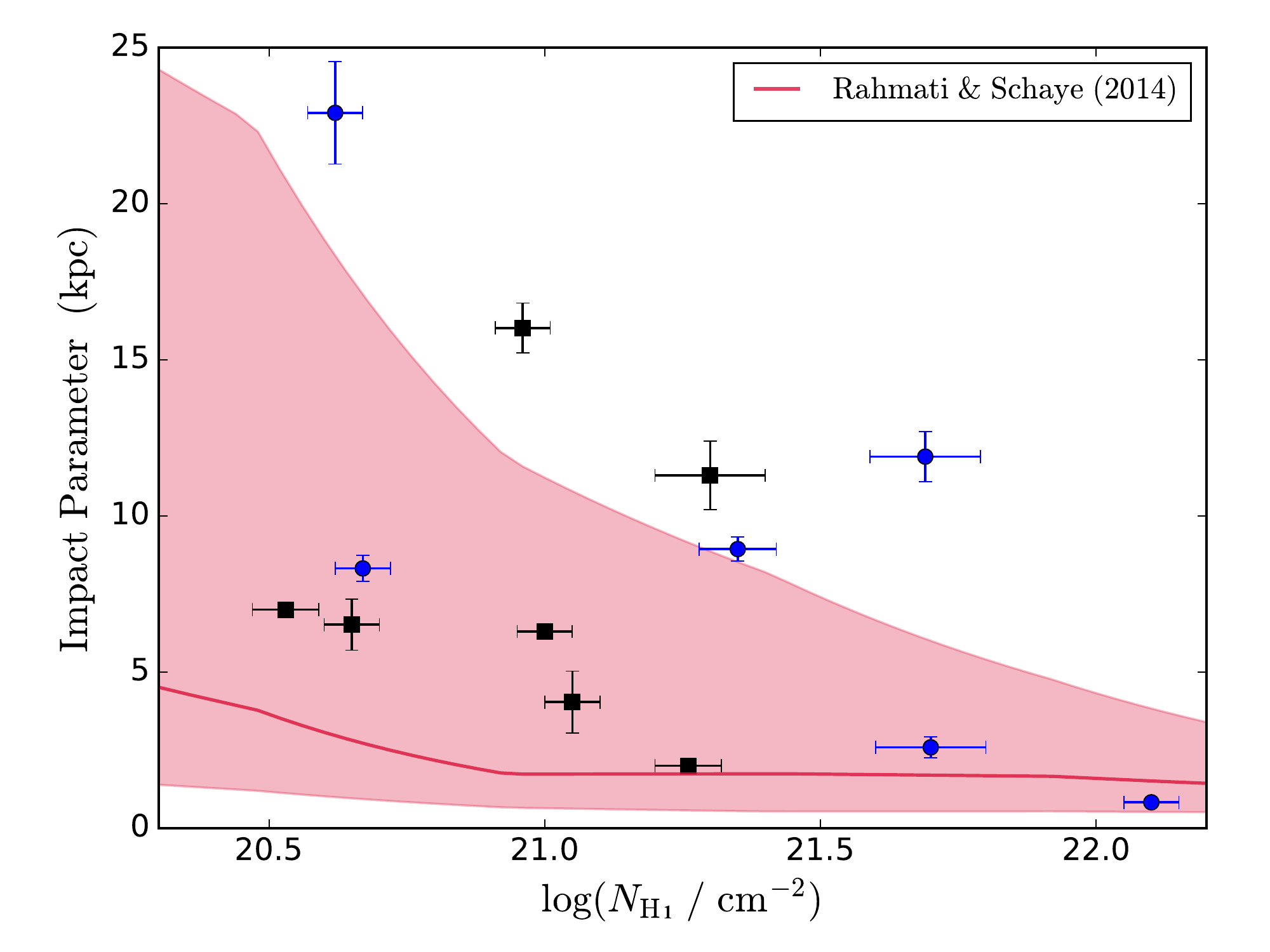}
	\caption{Column density of neutral hydrogen versus impact parameter. Black squares and blue circles mark the detections from our X-shooter campaign and the literature sample, respectively, see Table~\ref{tab:sample}. The red line  shows the median of the distribution of impact parameters from the simulations by \citet{Rahmati2014}, and the red shaded region encompasses 68\% of this impact parameter distribution.
	\label{fig:b_NHI}}
\end{figure}

An additional effect, which is not included in the model is the anti-correlation between impact parameter and $\log\NHI$.  In Fig.~\ref{fig:b_NHI}, we show the observed impact parameters as a function of the neutral hydrogen column density for all spectroscopically confirmed, $z \gtrsim 2$ DLA counterparts (see Table~\ref{tab:sample}). From the data, we observe an anti-correlation between $\log(b)$ and $\log(\NHI)$. Using a Pearson correlation test, we find $r = -0.59$ significant at the 0.035 level. A similar anti-correlation has been reported by several studies in the literature \citep{Moller1998b, Zwaan2005, Monier2009, Peroux2011a, Rubin2015}.
Three of the DLAs from the X-shooter campaign with no \lya\ detection (Q0845+2008, Q1057+0629, and Q1435+0354) have the three lowest values of $\NHI$ in our sample $\left( \log \NHI = 20.4 - 20.5 \right)$.
The counterparts for these three DLAs are therefore more likely to have large impact parameters.
Given our observational setup using three slits centred on the quasar, we are consequently more likely to miss these targets. As mentioned in Sect.~\ref{comparison:stacking}, the anti-correlation between impact parameter and $\log(\NHI)$ has not been taken into account in the modelling, and the detection probabilities for the individual targets with low $\NHI$ might therefore be over-estimated. As discussed in Sect.~\ref{comparison:stacking}, the anti-correlation between $\NHI$ and impact parameter would furthermore improve the model predictions for the stacking results.

\begin{table*}
\begin{center}
\caption{DLA emission counterparts. \label{tab:sample}}
\begin{tabular}{lcccccc}
\hline
Target       &  $\zDLA$ &   $\logNHI$    &      [M/H]      &     $b$       &      $b$        &  References \\
             &          &                &                 &  (arcsec)     &     (kpc)       &            \\  
\hline
Q2206--199a    &   1.92   & $20.67\pm0.05$ & $-0.54\pm0.05$  & $0.99\pm0.05$ &  $ 8.32\pm0.42$ & Abs.: [5]; Em.: [2, 3]  \\  
Q1135--0010   &   2.21   & $22.10\pm0.05$ & $-1.10\pm0.08$  & $0.10\pm0.01$ &  $ 0.83\pm0.08$ & [11] \\  
HE2243--60    &   2.33   & $20.62\pm0.05$ & $-0.72\pm0.05$  & $2.80\pm0.20$ &  $22.91\pm1.64$ & [12] \\  
PKS0528--250b &   2.81   & $21.35\pm0.07$ & $-0.91\pm0.07$  & $1.14\pm0.05$ &  $ 8.94\pm0.39$ & Abs.: [5]; Em.: [1] \\  
J2358+0149    &   2.98   & $21.69\pm0.10$ & $-1.83\pm0.18$  & $1.5\pm0.1$ &  $ 11.9\pm0.8$ & [15] \\ 
\hline
Q0338--0005   &   2.23   & $21.12\pm0.05$ & $-1.36\pm0.07$  & $0.49\pm0.12$ &    $3.7\pm1.0$  & Abs.: [16]; Em.: [10, 16] \\ 
PKS0458--020  &   2.04   & $21.70\pm0.10$ & $-1.22\pm0.08$  & $0.31\pm0.04$ &    $2.7\pm0.3$  & Abs.: [5]; Em.: [4, 16]\\ 
Q0918+1636-1  &   2.41   & $21.26\pm0.06$ &  $-0.6\pm0.2$   &     $<0.3$    &        $<2$     & [9] \\ 
Q0918+1636-2  &   2.58   & $20.96\pm0.05$ & $-0.12\pm0.05$  &  $2.0\pm0.1$  &   $16.2\pm0.8$  & [8] \\ 
Q1313+1441    &   1.79   & $21.3 \pm0.1 $ &  $-0.7\pm0.1$   &  $1.3\pm0.1$  &   $11.3\pm1.1$  & [16] \\ 
Q2059--0528   &   2.21   & $21.00\pm0.05$ & $-0.96\pm0.06$  &     $<0.8$    &        $<6.3$   & [14] \\ 
Q2222--0946   &   2.35   & $20.65\pm0.05$ & $-0.38\pm0.05$  &  $0.8\pm0.1$  &    $6.3\pm0.3$  & [7, 13] \\ 
Q2348--011-1  &   2.43   & $20.53\pm0.06$ & $-0.33\pm0.08$  &  $0.8\pm0.2$  &    $6.7\pm1.6$  & Abs.: [6, 16]; Em.: [16] \\ 
\hline
\end{tabular}
\end{center}

{References:
[1] \citet{Moller1993};
[2] \citet{Warren2001};
[3] \citet{Moller2002};
[4] \citet{Moller2004};
[5] \citet{Ledoux2006};\\

[6] \citet{Noterdaeme2007};
[7] \citet{Fynbo2010};
[8] \citet{Fynbo2011};
[9] \citet{Fynbo2013b};
[10] \citet{Krogager2012};\\

[11] \citet{Noterdaeme2012a};
[12] \citet{Bouche2013};
[13] \citet{Krogager2013};
[14] \citet{Hartoog2015};
[15] \citet{Srianand2016};
[16] This work.
}\\[1mm]

{\sc Note} --- We do not consider the two counterparts, Q0139--0824 and Q0953+47, included in \citet{Krogager2012}, since no further peer-reviewed results for these two targets have been published since then. We do not include sub-DLAs nor proximate DLAs ($z_{\rm abs}\approx z_{\rm em}$) only the DLA towards PKS0528--250b which has been shown to be unrelated to the quasar (\citealt*{Moller1998a}; \citealt{Ellison2010}). Lastly, we only consider spectroscopically confirmed counterparts.
\end{table*}

\subsection{Impact Parameter--Metallicity Relation}
\label{discussion:impact_parameter}

\begin{figure}
	\includegraphics[width=0.48\textwidth]{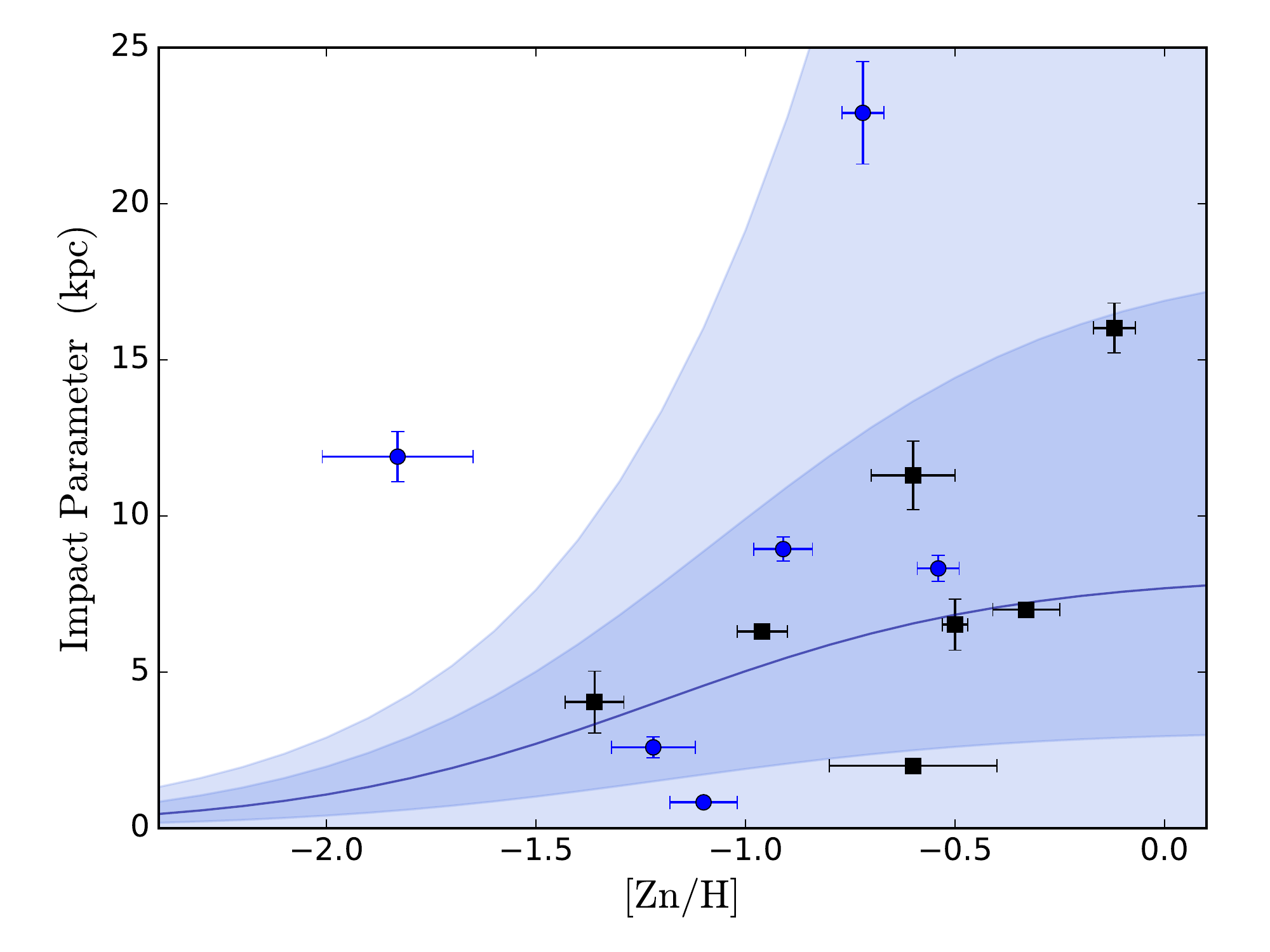}
	\caption{Metallicity versus impact parameter. Black squares refer to detections from the X-shooter campaign presented in this work. Blue circles indicate detections from the literature, see Table~\ref{tab:sample}. The light blue shaded region shows the extent of DLAs as a function of metallicity from the model of \citetalias{Fynbo2008}-CG. The solid blue line marks the median impact parameter expected from the \citetalias{Fynbo2008}-CG model and the darker shaded region encompasses 68\% of the impact parameter distribution.
	\label{fig:b_M}}
\end{figure}

In Fig.~\ref{fig:b_M}, we show the distribution of impact parameter versus metallicity for the entire sample of DLA emission counterparts with spectroscopic confirmation at $z \gtrsim 2$, summarised in Table~\ref{tab:sample}. The expected relation between impact parameter and metallicity from the model by \citet{Fynbo2008} is shown as the blue shaded area in this figure.
We observe a very good agreement between the allowed impact parameters of the model and the data points. Only one point is not in agreement, namely the DLA galaxy reported by \citet{Srianand2016}. If we further look at the region containing the 68\% most probable impact parameters (dark shaded region; from the 16th and 84th percentiles), we find that 9 out of 13 targets (i.e., 69\%) lie within this region. Thus consistent with the model expectation.
If we consider the number of points above and below the median of the impact parameter distribution (solid blue curve), we find that $62^{+30}_{-21}$\% lie above the median and $38^{+27}_{-16}$\% below, consistent with the expectation.

The impact parameter distribution in our model is based on the simple assumption that the DLA gas is arranged in a thin, circular disk. In reality we do not expect such a simple approximation to represent the true physical distribution of gas in and around galaxies; nonetheless, our modelled impact parameters are consistent with state-of-the-art simulations \citep{Bird2013, Rahmati2014}. This indicates that our simplified prescription of DLA cross-sections provides a reasonable, analytical description. A more detailed handling of the \HI\ distribution around galaxies is beyond the scope of this article, and we highlight that even in simulations the exact DLA cross-section varies significantly depending on the resolution and the technique applied \citep[for a comparison between smoothed particle hydrodynamics and moving mesh simulations, see][]{Bird2013}.

\subsection{How do DLAs Trace Galaxies?}
\label{discussion:DLAgalaxies}
\begin{figure}
	\includegraphics[width=0.48\textwidth]{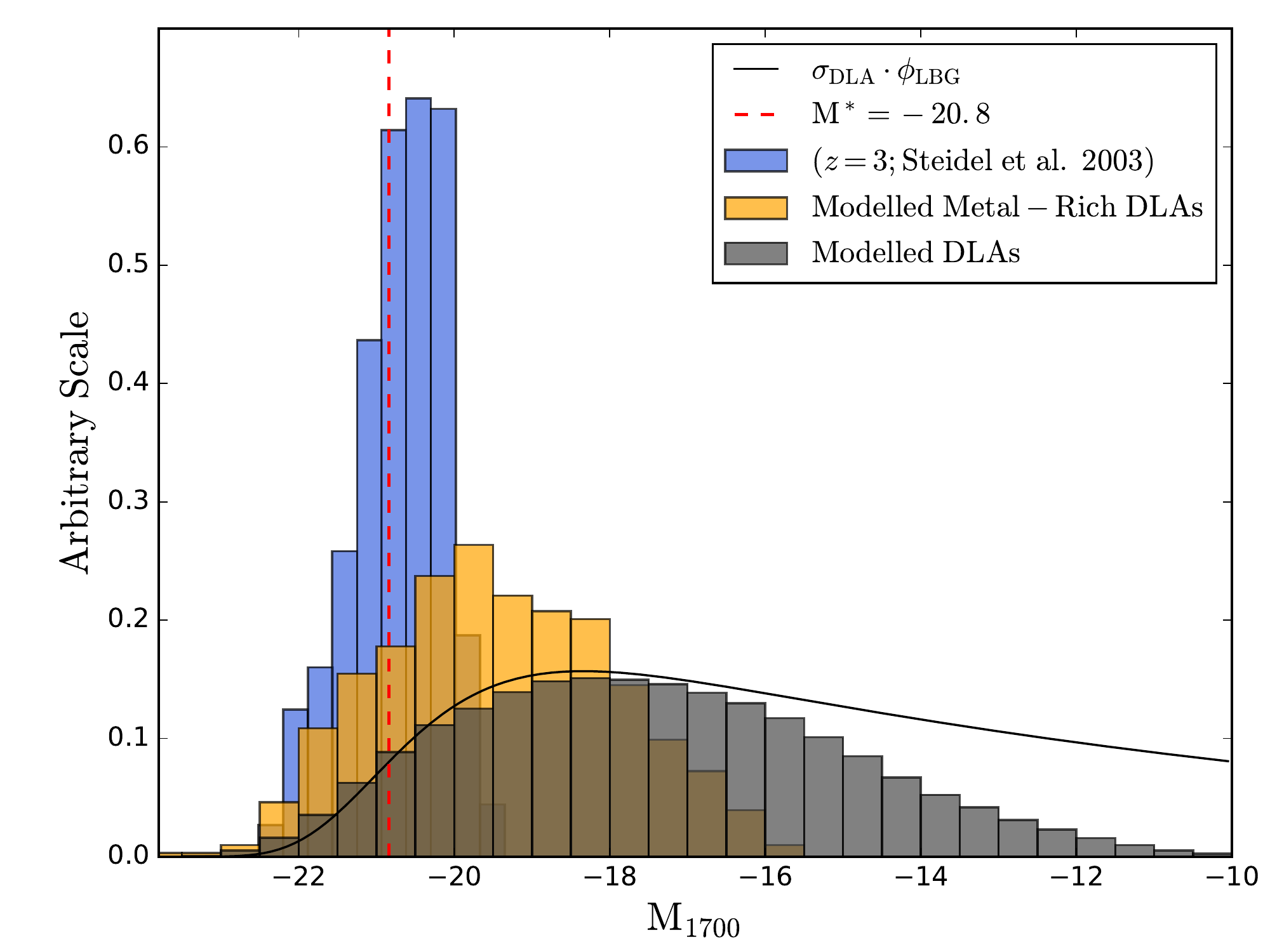}
	\caption{Distribution of rest-frame UV luminosities. The solid black line marks the LBG luminosity function weighted by the DLA cross-section from the \citetalias{Fynbo2008} model.
	\label{fig:lumdist}}
\end{figure}

The underlying metallicity--luminosity relation has important implications for the way DLAs sample the galaxy population. The main consequence is that DLAs sample galaxies over most of the luminosity function -- not simply the faintest galaxies. To illustrate this effect, we have estimated luminosities from our model based on the observed metallicity distribution of DLAs. For each absorption metallicity, we trace back the distribution of galaxies with various impact parameters that contribute to that given observed metallicity. This results in a skewed distribution of central galaxy metallicities depending on the metallicity gradient in place. The central galaxy metallicities are then converted to luminosities given the relation described in Sect.~\ref{model}. The resulting distribution is shown as the gray distribution in Fig.~\ref{fig:lumdist}. For comparison, we show the UV luminosity function at redshift $z\approx3$ \citep{Reddy09} weighted by the luminosity-dependent cross-section for DLAs as the black curve.
In the same figure, we show the distribution of luminosities for a high-metallicity sample similar to our X-shooter campaign, and a spectroscopic, flux-limited sample of LBGs from \citet{Steidel2003}. It is seen that the overall DLA population thus samples the luminosity function over a much larger range than what is accessible in emission studies of LBGs, i.e., DLAs sample galaxies down to $\sim$8 magnitudes fainter.
Moreover, the overlap between the high-metallicity distribution and the LBG distribution is in perfect agreement with detailed studies of DLA counterparts at high redshift where multi-wavelength continuum and line emission detections allow a precise determination of physical quantities such as stellar mass, star formation rate and global metallicity \citep{Bouche2013, Krogager2013, Fynbo2013b, Christensen2014}.

The scenario, in which DLAs trace a large range of stellar masses and star-formation rates is supported by the recent numerical modelling results by \citet{Berry2016}. From their models, \citeauthor{Berry2016} find that galaxies with stellar masses ranging from $10^6$~M$_{\odot}$ to $10^{11}$~M$_{\odot}$ contribute to the cross-section of DLAs. On average the stellar-mass range probed by DLAs is $\sim10^8$~M$_{\odot}$. This agrees well with the estimated average stellar mass of DLAs of $10^{8.5}$~M$_{\odot}$ from the mass--metallicity relation by \citet{Moller2013}.

Similar results are found by \citet{Rahmati2014} in their simulations of \HI\ absorbers in a cosmological context. In Fig.~\ref{fig:b_NHI}, we show the distribution of impact parameters from \citet{Rahmati2014} together with the sample of DLA counterparts. There is generally good agreement between the simulated distribution and the observed impact parameter distribution as a function of $\log\NHI$, although the data on average show larger impact parameters than the median predictions from Rahmati \& Schaye.
These authors explain this apparent offset by the so-called `identification bias', in which observers identify the brightest nearby galaxy to a given absorber, although a smaller and fainter galaxy might be the `true' counterpart. While this effect is difficult to rule out in observations, we find it unlikely to be a major effect for the high metallicity DLAs where the metallicity--luminosity relation predicts a bright counterpart with star formation rates in agreement with the authors' numerical results (SFR $\sim 10^0$ to $10^1$~M$_{\odot}$~yr$^{-1}$). The discrepancy might be explained by the fact that \citeauthor{Rahmati2014} show impact parameters for all their simulated galaxies irrespective of the stellar mass, whereas the observed data presented in Fig.~\ref{fig:b_NHI} predominantly traces metal-rich DLAs, which given our model traces only the brightest galaxies. This is consistent with the correlation between impact parameter and star-formation rate (and stellar mass) shown by \citeauthor{Rahmati2014}.

At lower metallicities, the identification bias will be more severe, as the counterparts will be very faint, and hence, a bright, unrelated counterpart is more likely to be associated. This is also reflected in the impact parameter--metallicity distribution in Fig.~\ref{fig:b_M}, where metal-poor DLAs are expected to have small counterparts. 
Indeed, the DLA counterpart presented by \citet{Srianand2016} might be the case of such an identification bias, given its low metallicity and very high column density of $\NHI$.

\subsection{The Metallicity--Luminosity Relation Revisited}
\label{discussion:ZLrelation}

\begin{figure}
	\includegraphics[width=0.48\textwidth]{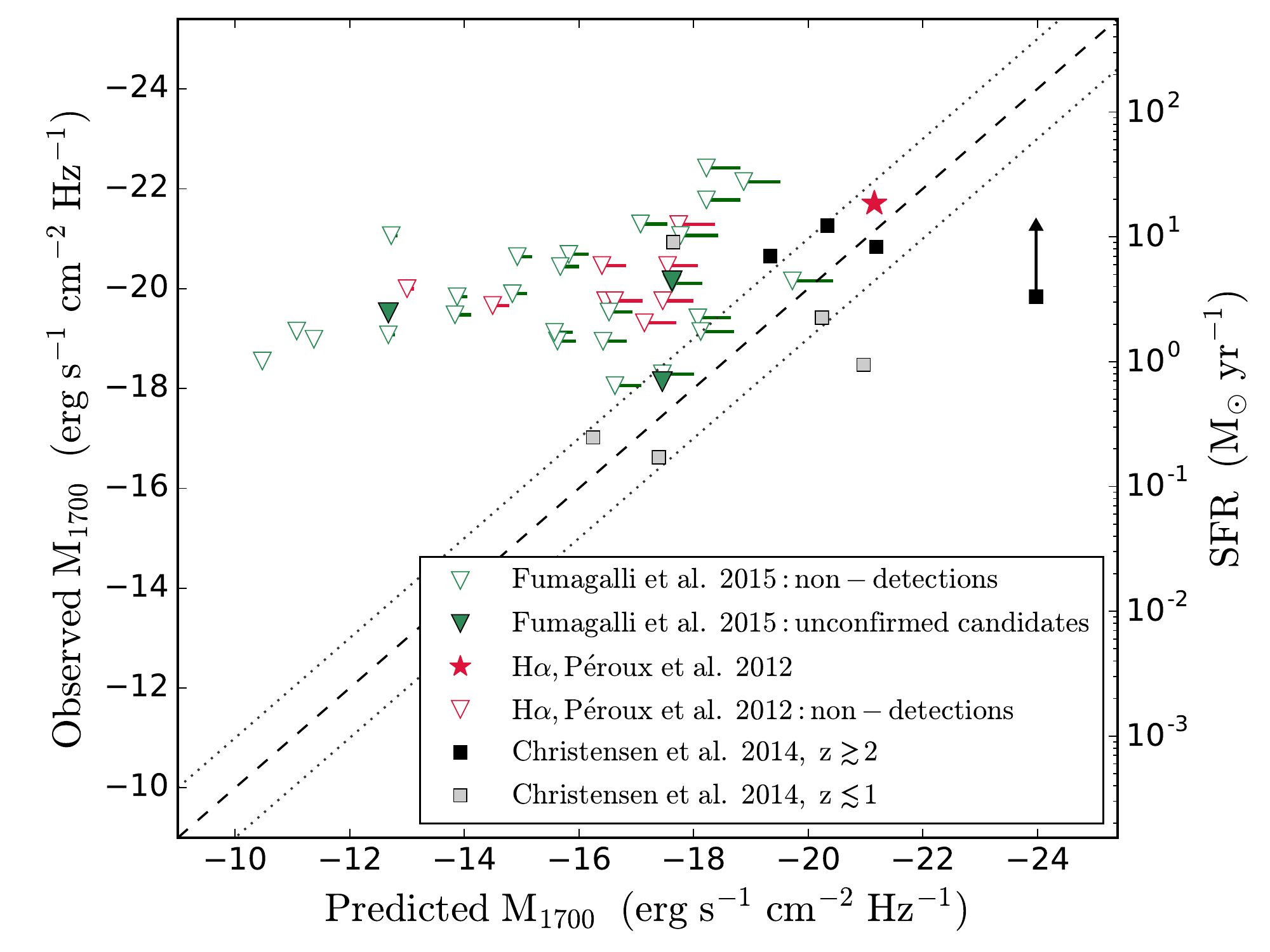}
	\caption{Observed versus predicted absolute magnitude at rest-frame 1700~\AA\ given our model (Sect.~\ref{model}). The rest-frame UV continuum measurements by \citet{Fumagalli2015} and \citet{Christensen2014} should be read on the left hand axis, whereas the converted line flux measurements from \citet{Peroux2012} should be read on the right hand axis. The dashed line marks the one-to-one relation. For comparison, the dotted lines indicate a scatter of $\pm$0.4~dex in luminosity, corresponding to the observed scatter in the mass--metallicity relation for DLAs \citep{Moller2013}. The horizontal line segments for the non-detections indicate the extent of the median correction for metallicity gradients given the modelled impact parameter distribution for each target. The upward arrow on one black point marks the correction for dust as measured by \citet{Fynbo2013b}. Since measurements of $A_V$ are not available for the remainder of the \citet{Christensen2014} sample, we have not corrected for dust. \label{fig:predicted_flux}}
\end{figure}

Our employed model links the measured absorption metallicities to a probability
distribution of continuum luminosities, which can further be converted to
star formation rates using the relation by \citet{Kennicutt1998}. The tight agreement
with the observed detection rates confirms that the model relations provide
a good description of the true underlying relations, nevertheless it is
useful to visualize this agreement directly. A large fraction of the surveys
analysed here were based purely on the search for Ly$\alpha$,
which due to the large scatter in emission properties makes it very uninformative to visualize the required parameters.
However, a few high redshift DLA galaxies have been directly imaged in the continuum,
thereby providing a measured luminosity and impact parameter as
recently provided in the compilation of \citet{Christensen2014}.
In Fig.~\ref{fig:predicted_flux} we plot the observed luminosity (left
axis) of those galaxies versus their model-predicted luminosity (solid black
squares). It is seen that the galaxies on average follow the one-to-one
relation between observed and predicted luminosity (dashed line), albeit
with some scatter. The survey by \citet{Fumagalli2015}
was a UV continuum imaging survey resulting in upper limits (shown as
open green triangles) and three unconfirmed candidates (solid green
triangles). As detailed in Sect.~\ref{comparison:fumagalli}, the
`in situ' limits provided in that work require aperture correction
factors in the range of 2 to 33 in order to obtain actual upper
limits on the total host luminosities.
In Figure 6, we show the observed versus predicted {\it total} luminosity, i.e., the corrections
have been applied to both the observed and modelled `in situ' aperture fluxes.
The corrections therefore simply shift the green points along a diagonal and
do not change the probability for detection as calculated in Section 5.3.
It is seen that most upper limits and candidates from this survey are well
above the predicted luminosities (owing primarily to the low metallicity of their sample),
consistent with the reported 0--3 detections.

The H$\alpha$ IFU survey by \citet{Peroux2012} can be converted to star-formation rates
(right axis) as described in Sect.~\ref{comparison:peroux}. One
detection (red star) and several upper limits (open red triangles)
from this survey are plotted. Their detection (Q2222--0946) was also
reported as a continuum detection by \citet{Krogager2013},
and in the figure the two points are seen both to fall close to the predicted relation;
i.e., both the continuum and the H$\alpha$ detection follow the predicted relation
with only a small difference.
For all detections and upper limits, we have corrected the points to a fiducial
redshift of $z=2.3$ (the median of our statistical sample)
using the redshift evolution derived in \citet{Moller2013}. In the
high redshift regime those corrections are all very small.

It is seen from the figure that both continuum and emission-line based
blind high redshift surveys have reached roughly the same detection
limits in terms of the expected total luminosity of the DLA galaxy.
Also, it is evident that all currently known high redshift
detections are found exactly where this limit intersects the one-to-one relation,
i.e., where our model predicts luminosities that are bright enough to
be detected. The fact that our model predicts the correct fluxes
for the known detections shows that we have the correct zero-point for
the metallicity--luminosity relation, but at present the detections do not span enough
range in luminosity to simultaneously confirm the slope, which is effectively
adopted from emission selected samples.  Some indication may be
obtained from considering also lower redshift DLA galaxies.
The compilation of photometric data for DLA galaxies \citep{Christensen2014}
also contains a number of low redshift galaxies. Again we correct those
detections to the fiducial redshift of $z=2.3$ and plot them as grey squares.
Including those low redshift objects we see that they, together with the high-redshift detections,
follow the predicted relation over 2 orders of magnitude in luminosity,
i.e., we have a good indication that the slope is correct.
Nonetheless this should be confirmed in the two redshift regimes independently.

One of the high redshift detections falls significantly
below the prediction. This object is the heavily dust obscured
DLA galaxy 0918+1636-2 (discussed in Section~\ref{discussion:dust}) with an $A_V = 1.54$~mag.
The correction for this obscuration is shown by the arrow which brings
it closer to the prediction. Dust obscuration is known to correlate with
metallicity, and is therefore likely to be relevant only in the far right
side of Fig 6.

In passing it is interesting also to note that one of the unconfirmed
candidates from the UV samples is very close to the predicted
flux. It would be worthwhile to attempt a confirmation of this
candidate.

\section{Summary and Conclusions}
\label{conclusion}

In this work, we have presented the analysis of the last part of the survey initiated by \citet{Fynbo2010} targeting metal-rich DLAs, as well as a final discussion of the results from the complete
survey. In particular we have tested the hypothesis \citep{Fynbo2008} that a simple model, in which the luminosity of DLA galaxies scales with metallicity, is consistent with the detections both in this and in all
other major surveys for emission counterparts of DLAs during the past two and a half decades. Our main results can be summarised as follows:

\begin{itemize}

	\item In our X-shooter campaign targeting metal-rich DLAs, we find a detection rate of 64\%. This is significantly higher than previous blind surveys ($\sim$10\%), and confirms our hypothesis that luminosity correlates with metallicity.

	\item Combining the X-shooter campaign with 7 major past surveys, we find that they all -- individually -- produce detection rates (or stacked fluxes) in close agreement with our simple model based on the F08 model, modified to use the metallicity gradient found for DLAs by \citet{Christensen2014}.

	\item For surveys with individual \lya\ detections, we find a significant scatter in the detection probabilities for the individual targets; i.e., even though the {\it total} predicted number of detections matches the actual detections, it is not always the most likely targets which are detected. We ascribe this to be a natural result of the scatter in the underlying scaling relations as well as the obscuring effect of dust, see last point below.

	\item Having shown that the model is in close agreement with results from all major surveys, we then use the metallicity--luminosity relation to compute the expected luminosity distribution of DLA galaxies and compare this to the observed luminosity distribution of a spectroscopic sample of LBGs. We show that both samples cover the bright end, however, the DLA host sample traces galaxies down to $\sim$8 magnitudes fainter. 

	\item We investigate the relation of metallicity and impact parameter and find that this relation is consistent with our model expectation. Moreover, we study the distribution of \HI\ column density as a function of impact parameter and conclude that recent numerical simulations successfully reproduce the observed anti-correlation between impact parameter and $\NHI$. 

	\item We show that dust attenuation, both in emission and in the absorption sightline, is important for high metallicity DLAs, and argue that this is likely one of the important contributions to the scatter in detection probability reported above.

\end{itemize}


The steep dependence of luminosity on metallicity together with the low average
metallicity of DLAs imply that the vast majority of DLAs are much too
faint for direct detection with current facilities. Hence, we propose that
the time for blind surveys of DLA counterparts is over.
At the lowest metallicities, current detection limits are $\sim$3.5 orders of
magnitude away from reaching the required luminosity/SFR limit to test
the model (Fig.~\ref{fig:predicted_flux}).

With an understanding of the basic selection of DLAs in place, we can now
start to address specific issues, e.g., the exact scaling of cross-section
with metallicity and $\NHI$, how these evolve with redshift, and the
impact of metallicity gradients.
Having better observational constraints on such quantities will allow the
use of DLAs to study the faint-end slope of the luminosity function.
Moreover, such observations will serve as direct constraints for future
numerical simulations.
A more detailed modelling approach, which self-consistently produces observables
over a larger part of the spectral energy distribution in a cosmological context, would also improve our ability to compare predictions and observations.
Lastly, we stress that while our model provides a good overall description of the data, it is not unique and other models might equally well reproduce the observations.

For now, the main source of uncertainty in our modelling
is due to the intrinsic scatter in the fundamental relations. An important
step for future work is therefore to explore the sources of this scatter
\citep[e.g.,][]{Christensen2014} and to understand the nature of outliers
as well as the details of dust correction \citep[e.g.,][]{DeCia2016, Wiseman2017}.
By focusing on those aspects, new insight can be achieved before the advent of
30 meter class telescopes.

\section*{Acknowledgments}
We thank the anonymous referee for the constructive feedback which helped improve the
manuscript. We also wish to thank Ali Rahmati for sharing his model results with us.
J.-K. Krogager acknowledges financial support from the Danish Council for Independent
Research (EU-FP7 under the Marie-Curie grant agreement no. 600207) with reference
DFF-MOBILEX--5051-00115.
The research leading to these results has received funding from the European Research
Council under the European Union's Seventh Framework Program (FP7/2007-2013)/ERC Grant
agreement no. EGGS-278202.
The work is based on observations carried out at the ESO Paranal Observatory, Chile,
under programmes 084.A-0303, 084.A-0524, 086.A-0074, 088.A-0601, and 089.A-0068.

\def\aj{AJ}
\def\araa{ARA\&A}
\def\apj{ApJ}
\def\apjl{ApJ}
\def\apjs{ApJS}
\def\apss{Ap\&SS}
\def\aap{A\&A}
\def\aapr{A\&A~Rev.}
\def\aaps{A\&AS}
\def\mnras{MNRAS}
\def\memras{MmRAS}
\def\nat{Nature}
\def\pasp{PASP}
\def\aplett{Astrophys.~Lett.}

\bibliographystyle{apj}
\bibliography{/Users/krogager/Documents/Papers/bibtex/bibliography.bib}

\appendix

\section{Best Fit Absorption Profiles}
\label{app:fits}

\begin{figure}
	\includegraphics[width=0.48\textwidth]{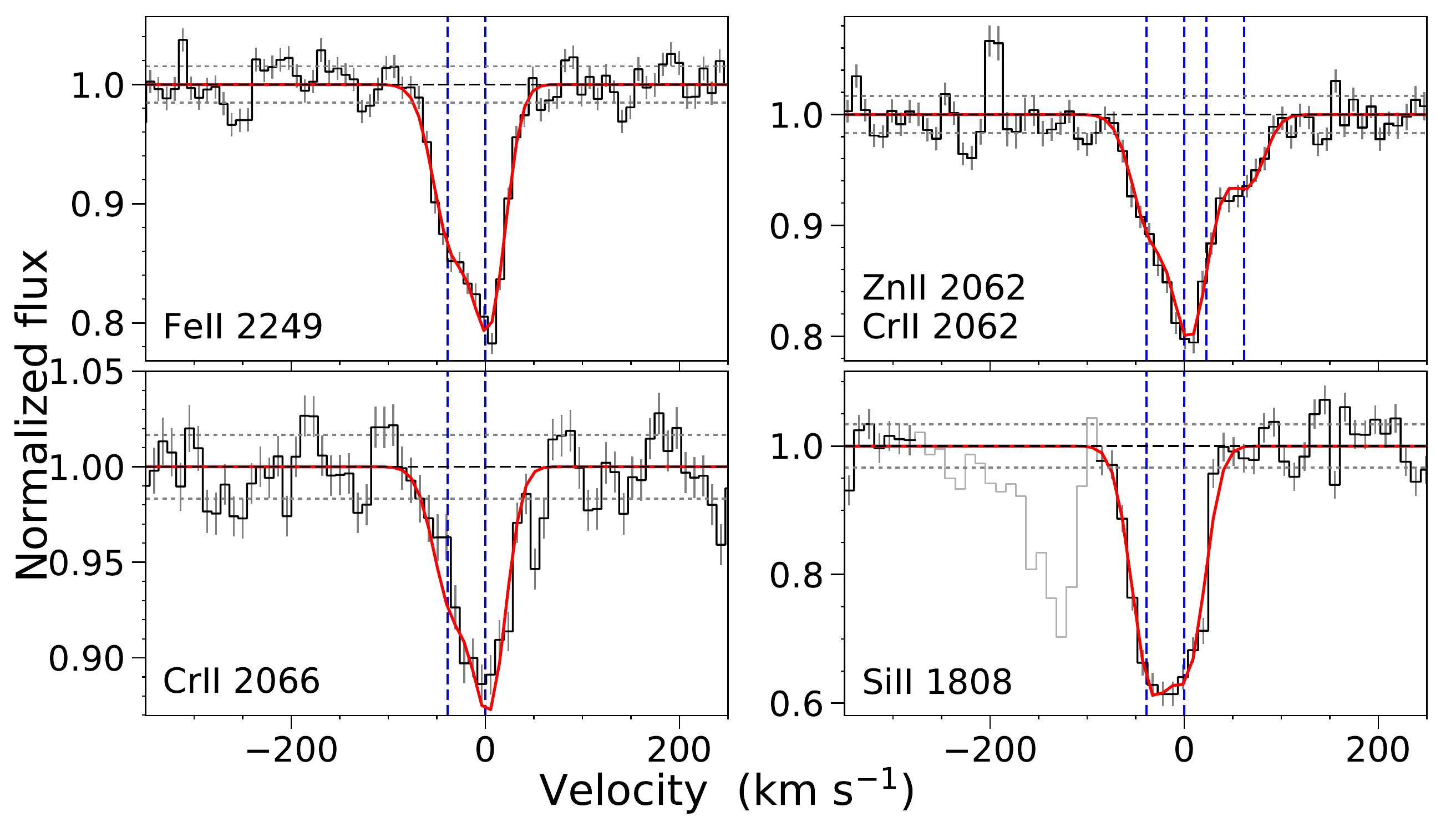}
	\caption{Fitted metal lines for the $\zDLA=2.179$ DLA toward Q0316+0040.
			 The X-shooter data are shown as black data with grey errorbars indicating
			 the 1$\sigma$ uncertainty from the pipeline error spectrum. Regions that were
			 masked in the fits are shown as a thin grey line with no errorbars. These are
			 either line blends of other species or other absorption systems, or from telluric
			 absorption features. The fitted components are shown as blue, dashed, vertical
			 lines. The fitted transitions in each frame are shown in the lower left corner.
	\label{fig:Q0316_bestfit}}
\end{figure}

\begin{figure}
	\includegraphics[width=0.48\textwidth]{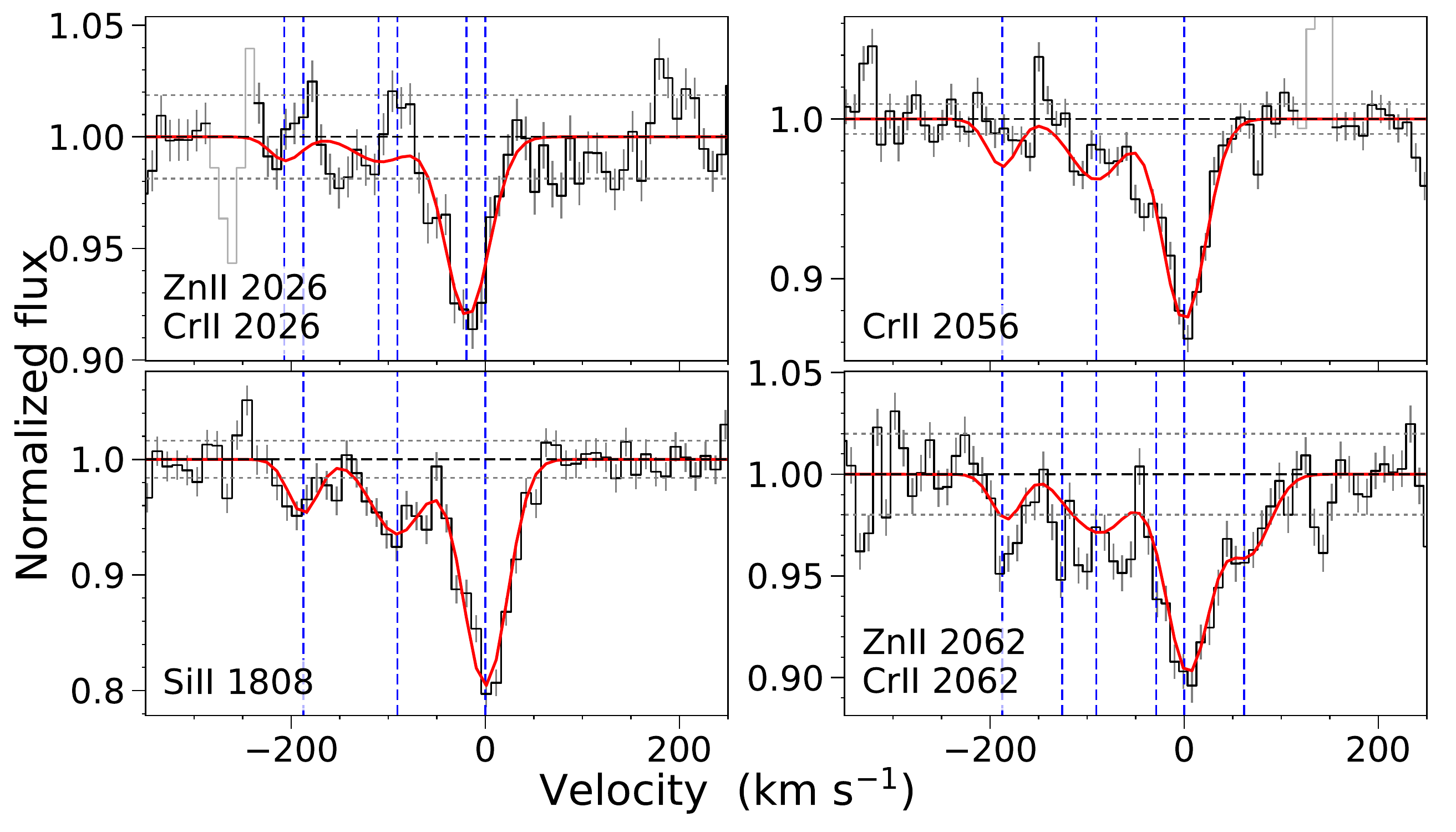}
	\caption{Fitted metal lines for the $\zDLA=2.229$ DLA toward Q0338$-$0005.
			 Same as Figure~\ref{fig:Q0316_bestfit}.
	\label{fig:Q0338_bestfit}}
\end{figure}

\begin{figure}
	\includegraphics[width=0.48\textwidth]{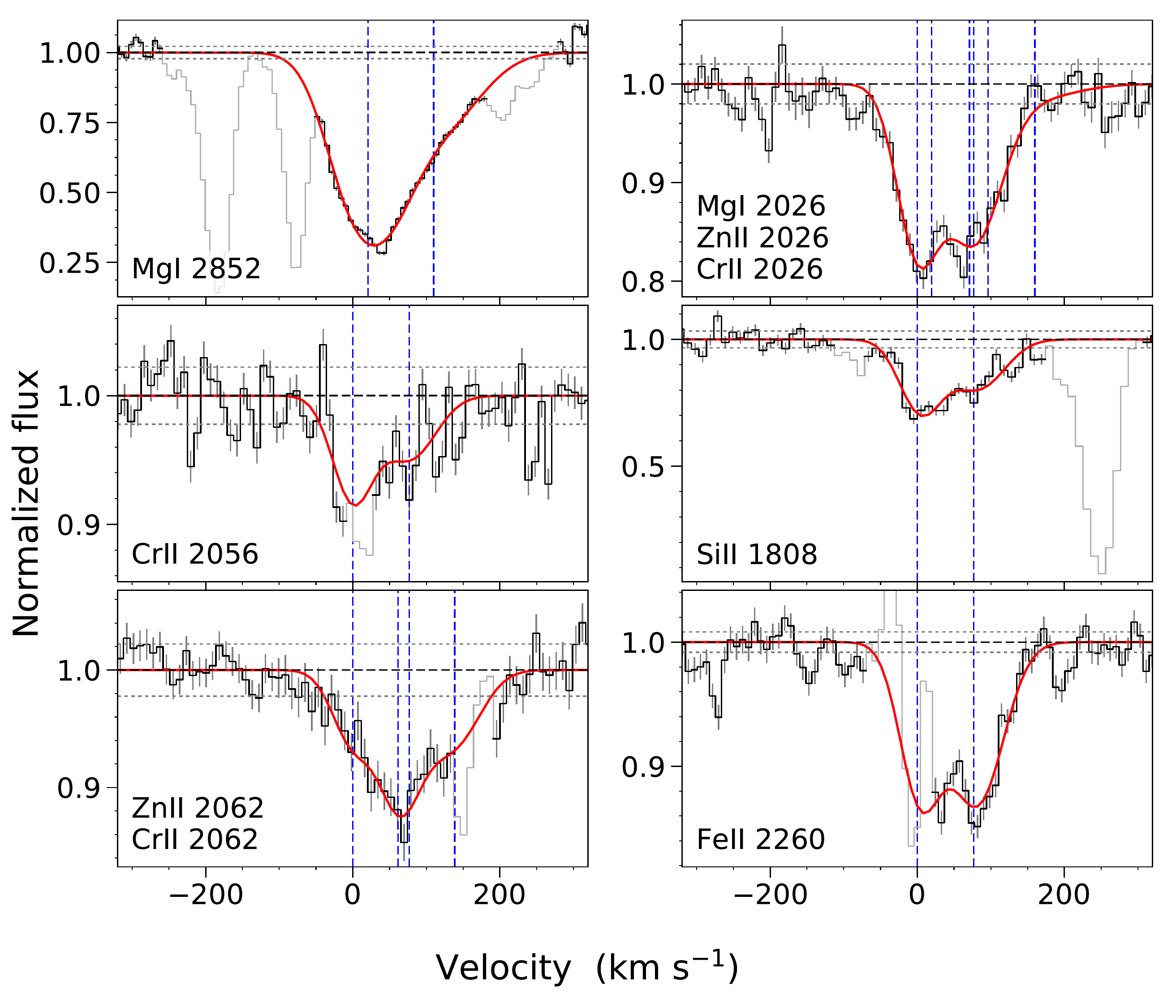}
	\caption{Fitted metal lines for the $\zDLA=2.237$ DLA toward Q0845+2008.
			 Same as Figure~\ref{fig:Q0316_bestfit}.
	\label{fig:Q0845_bestfit}}
\end{figure}

\begin{figure}
	\includegraphics[width=0.48\textwidth]{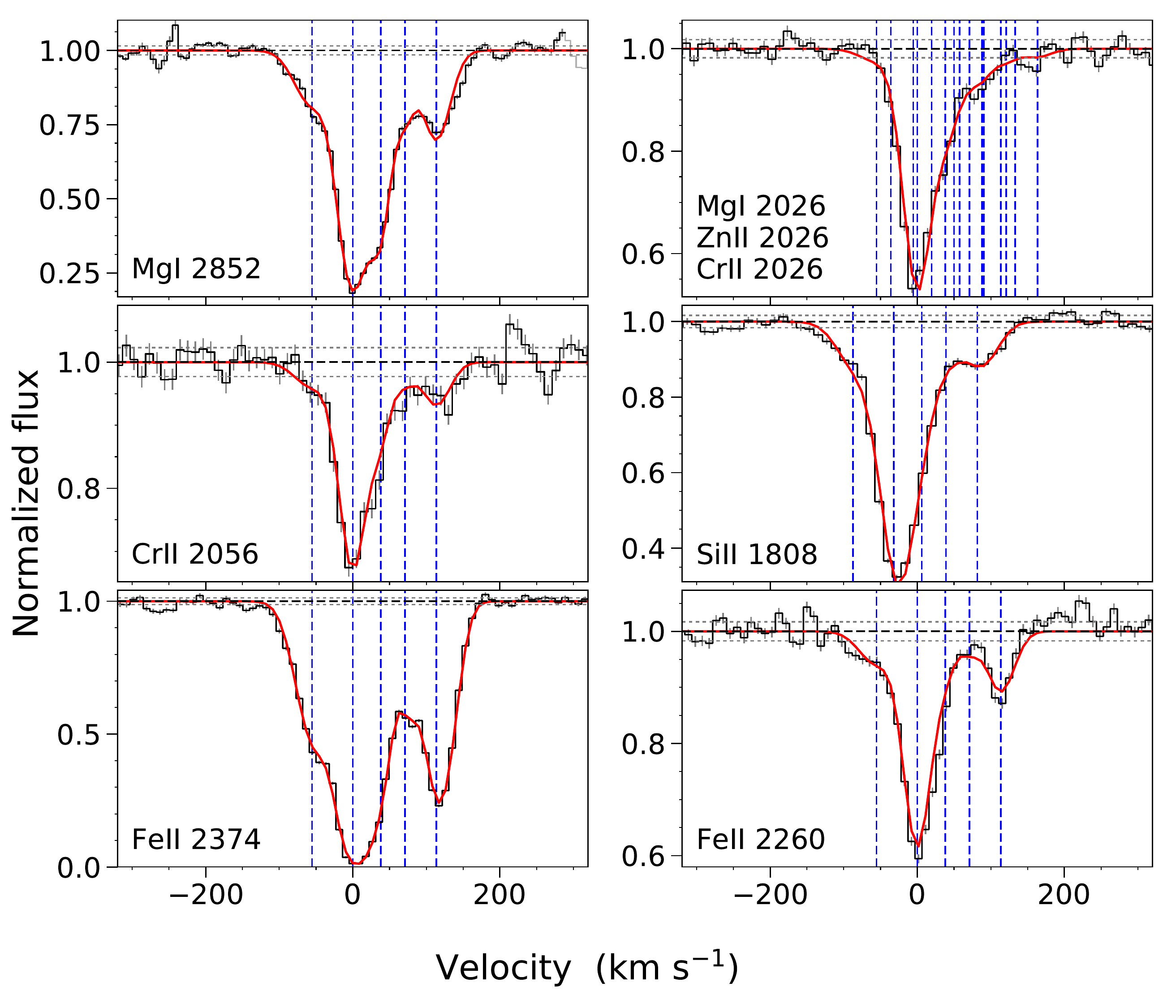}
	\caption{Fitted metal lines for the $\zDLA=1.794$ DLA toward Q1313+1441.
			 Same as Figure~\ref{fig:Q0316_bestfit}.
	\label{fig:Q1313_bestfit}}
\end{figure}

\begin{figure}
	\includegraphics[width=0.48\textwidth]{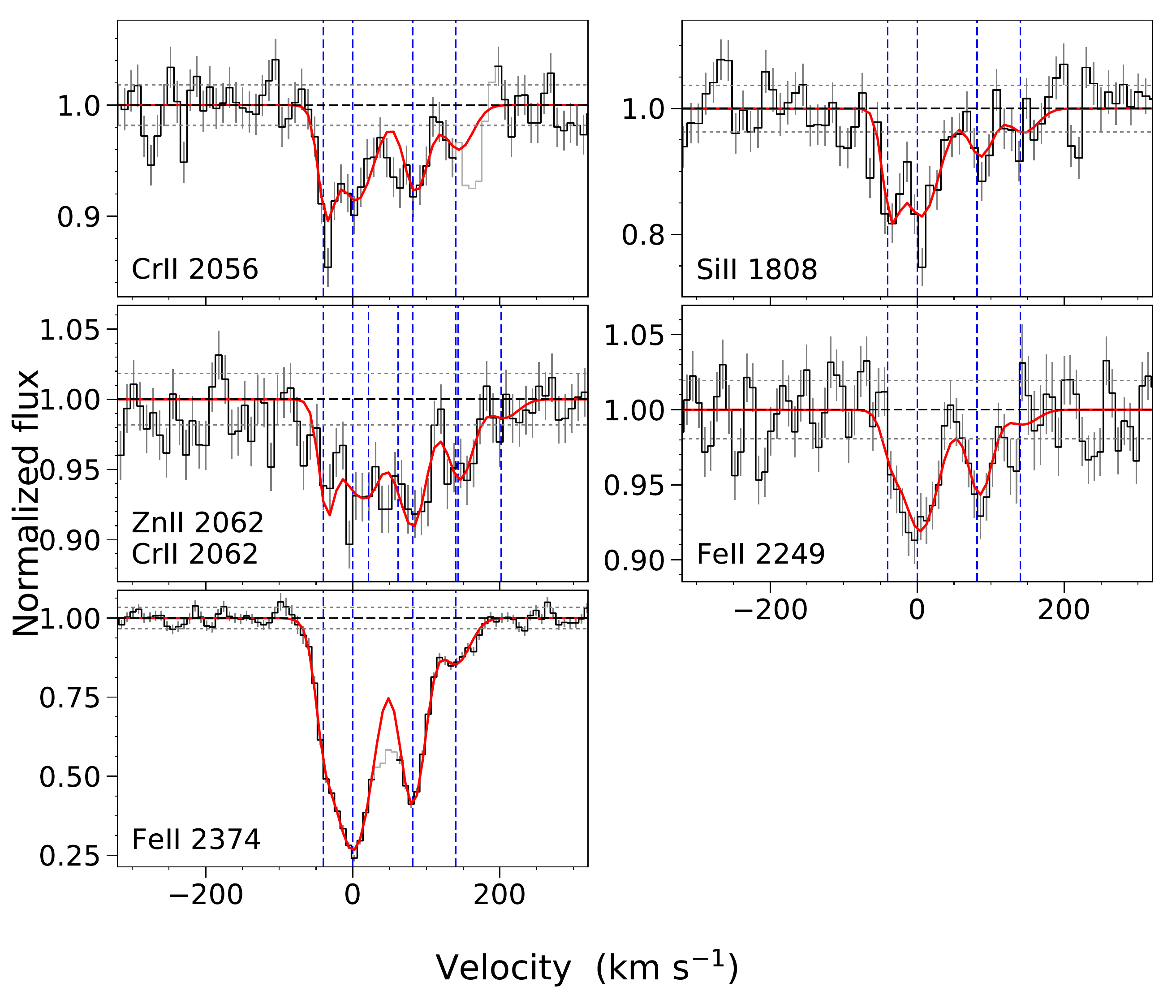}
	\caption{Fitted metal lines for the $\zDLA=2.269$ DLA toward Q1435+0354.
			 Same as Figure~\ref{fig:Q0316_bestfit}.
	\label{fig:Q1435_bestfit}}
\end{figure}

\begin{figure}
	\includegraphics[width=0.48\textwidth]{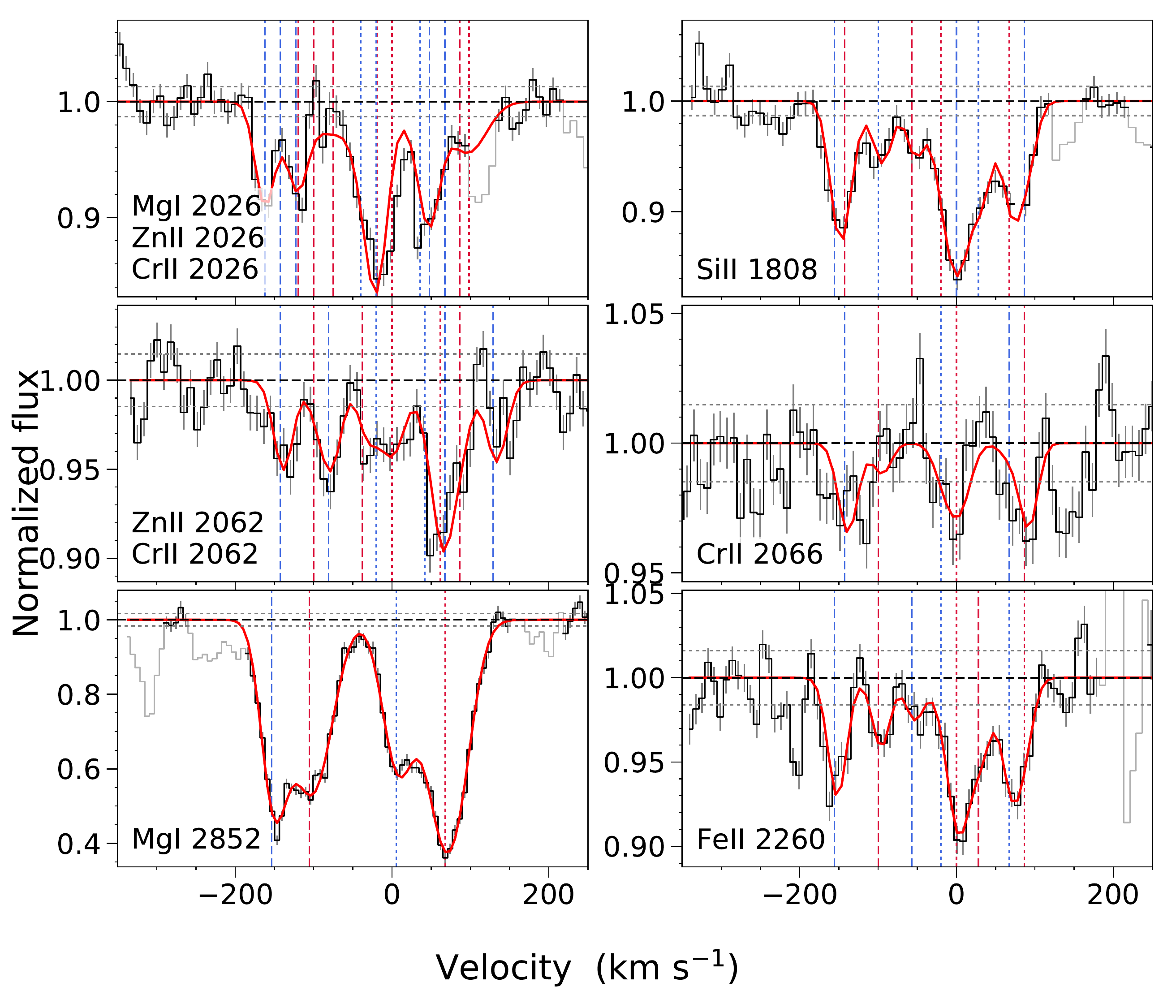}
	\caption{Fitted metal lines for the $\zDLA=2.425$ DLA toward Q2348$-$011.
			 Same as Figure~\ref{fig:Q0316_bestfit}.
	\label{fig:Q2348-1_bestfit}}
\end{figure}

\begin{figure}
	\includegraphics[width=0.48\textwidth]{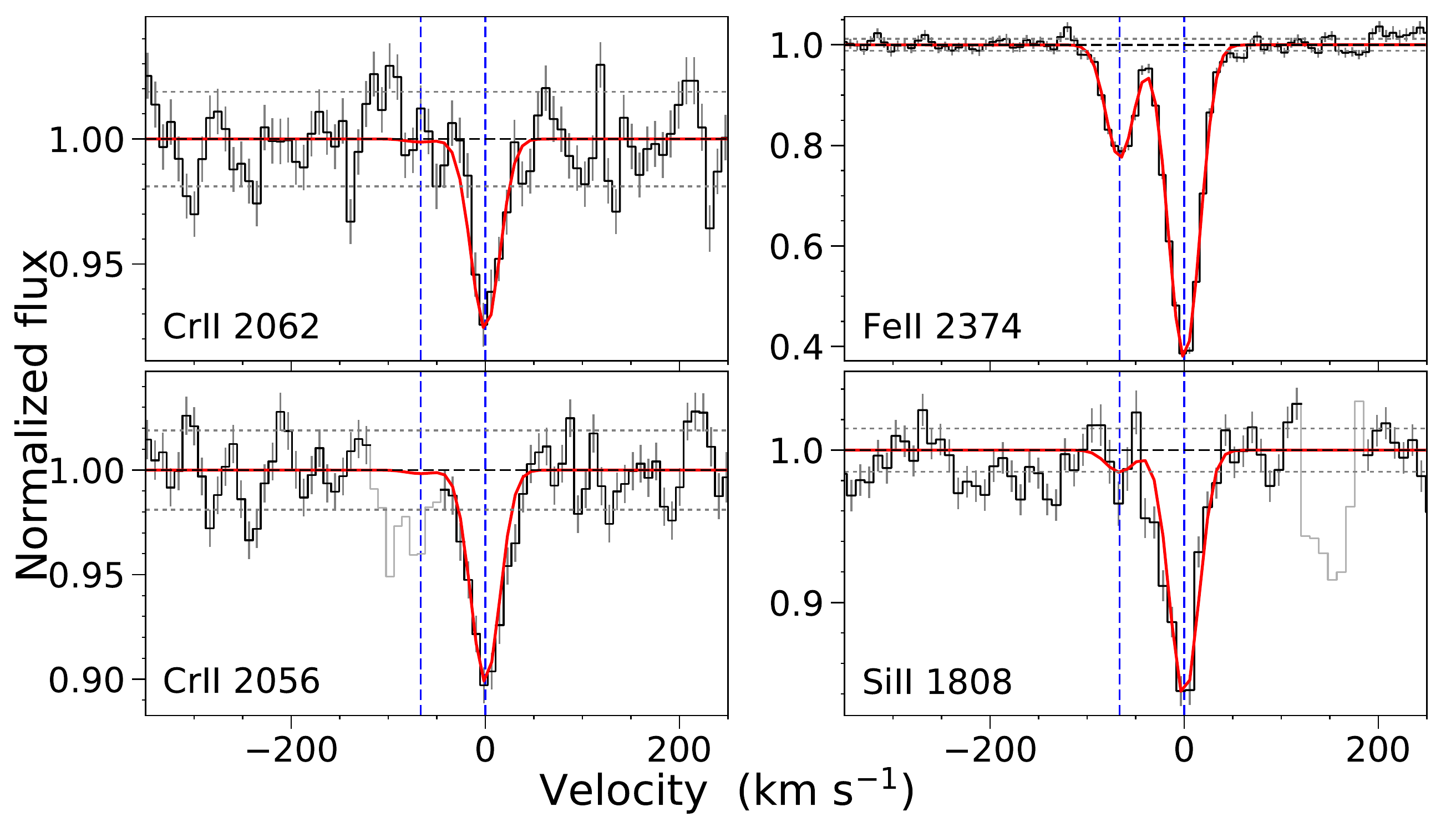}
	\caption{Fitted metal lines for the $\zDLA=2.614$ DLA toward Q2348$-$011.
			 Same as Figure~\ref{fig:Q0316_bestfit}.
	\label{fig:Q2348-2_bestfit}}
\end{figure}

The best-fit absorption line profiles are shown in Figs.~\ref{fig:Q0316_bestfit}--\ref{fig:Q2348-2_bestfit} for each quasar analysed in this work. For each quasar, we show all the low-ionization lines that were fitted. The measured spectral resolution for each spectrum is summarized in Table~\ref{tab:spectral_res} and the best-fit parameters are summarized in Tables~\ref{tab:fit_Q0316}--\ref{tab:fit_Q2348-2}.

\begin{table}
\caption{Spectral resolving power measured at 7600~\AA. \label{tab:spectral_res}}
\begin{center}
\begin{tabular}{lc}
\hline
Target  &  $\mathcal{R}$   \\
\hline
Q0316+0040	 		&	13100 \\
Q0338--0005	 		&	13000 \\
Q0845+2008	 		&	11100 \\
Q1313+1441$^{a}$	&	11800 \\
Q1435+0354	 		&	11700 \\
Q2348--011   		&	11600 \\
\hline
$^{a}$ In the UVB arm, we infer a resolution of $\mathcal{R} = $8000.
\end{tabular}
\end{center}
\end{table}

\begin{table}
\caption{Best fit parameters for Q0316+0040. \label{tab:fit_Q0316}}
\begin{center}
\begin{tabular}{lcccc}
\hline
\# & Velocity$^{a}$  &  $b$          &  Ion  &  $\log(N)$   \\
   &  (km~s$^{-1}$)  & (km~s$^{-1}$) &       &   \\
\hline
1  &  $-38.9$  &  $20.9$  &  \ion{Fe}{ii}  & $14.71 \pm 0.09$ \\
    &               & ''   &   \ion{Cr}{ii}  &  $13.10 \pm 0.10$  \\
    &               & ''   &   \ion{Si}{ii}  &  $15.34 \pm 0.07$  \\
    &               & ''   &   \ion{Zn}{ii}  &  $12.25 \pm 0.14$  \\[1mm]
2  &  $ 0.0$  &  $16.8$  &  \ion{Fe}{ii}  & $14.84 \pm 0.07$ \\
    &               & ''   &   \ion{Cr}{ii}  &  $13.28 \pm 0.06$  \\
    &               & ''   &   \ion{Si}{ii}  &  $15.21 \pm 0.10$  \\
    &               & ''   &   \ion{Zn}{ii}  &  $12.29 \pm 0.11$  \\
\hline
\end{tabular}
\end{center}
$^{a}$ Relative to the systemic redshift $z_{\rm sys}=2.1797$.
\end{table}

\begin{table}
\caption{Best fit parameters for Q0338--0005. \label{tab:fit_Q0338}}
\begin{center}
\begin{tabular}{lcccc}
\hline
\# & Velocity$^{a}$  &  $b$          &  Ion  &  $\log(N)$   \\
   &  (km~s$^{-1}$)  & (km~s$^{-1}$) &       &   \\
\hline

1  &  $-187.5$  &  $17.0$  &  \ion{Cr}{ii}  & $12.32 \pm 0.13$ \\
    &               & ''   &   \ion{Si}{ii}  &  $14.27 \pm 0.16$  \\
    &               & ''   &   \ion{Zn}{ii}  &  $<11.16$  \\[1mm]
2  &  $-90.6$  &  $34.2$  &  \ion{Cr}{ii}  & $12.65 \pm 0.08$ \\
    &               & ''   &   \ion{Si}{ii}  &  $14.64 \pm 0.09$  \\
    &               & ''   &   \ion{Zn}{ii}  &  $11.38 \pm 0.27$  \\[1mm]
3  &  $ 0.0$  &  $27.9$  &  \ion{Cr}{ii}  & $13.12 \pm 0.02$ \\
    &               & ''   &   \ion{Si}{ii}  &  $15.08 \pm 0.03$  \\
    &               & ''   &   \ion{Zn}{ii}  &  $12.21 \pm 0.05$  \\
\hline
\end{tabular}
\end{center}
$^{a}$ Relative to the systemic redshift $z_{\rm sys}=2.2289$.
\end{table}

\begin{table}
\caption{Best fit parameters for Q0845+2008. \label{tab:fit_Q0845}}
\begin{center}
\begin{tabular}{lcccc}
\hline
\# & Velocity$^{a}$  &  $b$          &  Ion  &  $\log(N)$   \\
   &  (km~s$^{-1}$)  & (km~s$^{-1}$) &       &   \\
\hline

1  &  $ 0.0$  &  $34.9$  &  \ion{Fe}{ii}  & $<14.76$ \\
    &               & ''   &   \ion{Zn}{ii}  &  $12.65 \pm 0.06$  \\
    &               & ''   &   \ion{Si}{ii}  &  $15.37 \pm 0.06$  \\
    &               & ''   &   \ion{Cr}{ii}  &  $13.01 \pm 0.09$  \\[1mm]
2  &  $76.9$  &  $46.9$  &  \ion{Fe}{ii}  & $14.89 \pm 0.06$ \\
    &               & ''   &   \ion{Zn}{ii}  &  $12.60 \pm 0.07$  \\
    &               & ''   &   \ion{Si}{ii}  &  $15.28 \pm 0.08$  \\
    &               & ''   &   \ion{Cr}{ii}  &  $12.90 \pm 0.10$  \\
\hline
\end{tabular}
\end{center}
$^{a}$ Relative to the systemic redshift $z_{\rm sys}=2.2360$.
\end{table}

\begin{table}
\caption{Best fit parameters for Q1313+1441. \label{tab:fit_Q1313}}
\begin{center}
\begin{tabular}{lcccc}
\hline
\# & Velocity$^{a}$  &  $b$          &  Ion  &  $\log(N)$   \\
   &  (km~s$^{-1}$)  & (km~s$^{-1}$) &       &   \\
\hline

1  &  $-55.6$  &  $28.5$  &  \ion{Fe}{ii}  & $14.34 \pm 0.04$ \\
    &               & ''   &   \ion{Mg}{i}  &  $11.93 \pm 0.05$  \\
    &               & ''   &   \ion{Cr}{ii}  &  $12.61 \pm 0.29$  \\
    &               & ''   &   \ion{Zn}{ii}  &  $<11.69$  \\
    &               & ''   &   \ion{Si}{ii}  &  $14.94 \pm 0.03$  \\[1mm]
2  &  $ 0.0$  &  $20.1$  &  \ion{Fe}{ii}  & $15.14 \pm 0.02$ \\
    &               & ''   &   \ion{Mg}{i}  &  $12.75 \pm 0.02$  \\
    &               & ''   &   \ion{Cr}{ii}  &  $13.51 \pm 0.04$  \\
    &               & ''   &   \ion{Zn}{ii}  &  $13.02 \pm 0.03$  \\
    &               & ''   &   \ion{Si}{ii}  &  $15.86 \pm 0.01$  \\[1mm]
3  &  $38.2$  &  $11.2$  &  \ion{Fe}{ii}  & $14.17 \pm 0.08$ \\
    &               & ''   &   \ion{Mg}{i}  &  $12.40 \pm 0.03$  \\
    &               & ''   &   \ion{Cr}{ii}  &  $12.84 \pm 0.15$  \\
    &               & ''   &   \ion{Zn}{ii}  &  $12.08 \pm 0.17$  \\
    &               & ''   &   \ion{Si}{ii}  &  $14.95 \pm 0.04$  \\[1mm]
4  &  $70.9$  &  $8.6$  &  \ion{Fe}{ii}  & $13.86 \pm 0.07$ \\
    &               & ''   &   \ion{Mg}{i}  &  $11.75 \pm 0.08$  \\
    &               & ''   &   \ion{Cr}{ii}  &  $<12.26$  \\
    &               & ''   &   \ion{Zn}{ii}  &  $<11.33$  \\
    &               & ''   &   \ion{Si}{ii}  &  $14.53 \pm 0.08$  \\[1mm]
5  &  $113.7$  &  $20.7$  &  \ion{Fe}{ii}  & $14.52 \pm 0.01$ \\
    &               & ''   &   \ion{Mg}{i}  &  $12.08 \pm 0.03$  \\
    &               & ''   &   \ion{Cr}{ii}  &  $12.76 \pm 0.18$  \\
    &               & ''   &   \ion{Zn}{ii}  &  $<11.37$  \\
    &               & ''   &   \ion{Si}{ii}  &  $14.83 \pm 0.04$  \\
\hline
\end{tabular}
\end{center}
$^{a}$ Relative to the systemic redshift $z_{\rm sys}=1.7941$.
\end{table}

\begin{table}
\caption{Best fit parameters for Q1435+0354. \label{tab:fit_Q1435}}
\begin{center}
\begin{tabular}{lcccc}
\hline
\# & Velocity$^{a}$  &  $b$          &  Ion  &  $\log(N)$   \\
   &  (km~s$^{-1}$)  & (km~s$^{-1}$) &       &   \\
\hline

1  &  $ 0.0$  &  $5.5$  &  \ion{Fe}{ii}  & $13.63 \pm 0.07$ \\
    &               & ''   &   \ion{Cr}{ii}  &  $12.65 \pm 0.08$  \\
    &               & ''   &   \ion{Si}{ii}  &  $14.70 \pm 0.16$  \\
    &               & ''   &   \ion{Zn}{ii}  &  $<11.42$  \\[1mm]
2  &  $40.3$  &  $31.1$  &  \ion{Fe}{ii}  & $14.59 \pm 0.01$ \\
    &               & ''   &   \ion{Cr}{ii}  &  $12.99 \pm 0.05$  \\
    &               & ''   &   \ion{Si}{ii}  &  $15.06 \pm 0.06$  \\
    &               & ''   &   \ion{Zn}{ii}  &  $12.31 \pm 0.13$  \\[1mm]
3  &  $121.6$  &  $18.2$  &  \ion{Fe}{ii}  & $14.26 \pm 0.01$ \\
    &               & ''   &   \ion{Cr}{ii}  &  $12.78 \pm 0.07$  \\
    &               & ''   &   \ion{Si}{ii}  &  $14.51 \pm 0.16$  \\
    &               & ''   &   \ion{Zn}{ii}  &  $11.94 \pm 0.33$  \\[1mm]
4  &  $180.3$  &  $22.4$  &  \ion{Fe}{ii}  & $13.56 \pm 0.04$ \\
    &               & ''   &   \ion{Cr}{ii}  &  $12.54 \pm 0.23$  \\
    &               & ''   &   \ion{Si}{ii}  &  $14.27 \pm 0.29$  \\
    &               & ''   &   \ion{Zn}{ii}  &  $11.67 \pm 0.41$  \\
\hline
\end{tabular}
\end{center}
$^{a}$ Relative to the systemic redshift $z_{\rm sys}=2.2685$.
\end{table}

\begin{table}
\caption{Best fit parameters for Q2348--011-1. \label{tab:fit_Q2348-1}}
\begin{center}
\begin{tabular}{lcccc}
\hline
\# & Velocity$^{a}$  &  $b$          &  Ion  &  $\log(N)$   \\
   &  (km~s$^{-1}$)  & (km~s$^{-1}$) &       &   \\
\hline

1  &  $-155.8$  &  $6.6$  &  \ion{Fe}{ii}  & $14.16 \pm 0.06$ \\
    &               & ''   &   \ion{Si}{ii}  &  $14.52 \pm 0.10$  \\[1mm]
2  &  $-142.7$  &  $8.1$  &  \ion{Zn}{ii}  & $12.04 \pm 0.06$ \\
    &               & ''   &   \ion{Si}{ii}  &  $14.16 \pm 0.20$  \\
    &               & ''   &   \ion{Cr}{ii}  &  $12.59 \pm 0.08$  \\[1mm]
3  &  $-99.8$  &  $5.1$  &  \ion{Fe}{ii}  & $13.89 \pm 0.11$ \\
    &               & ''   &   \ion{Zn}{ii}  &  $11.73 \pm 0.12$  \\
    &               & ''   &   \ion{Si}{ii}  &  $14.31 \pm 0.11$  \\
    &               & ''   &   \ion{Cr}{ii}  &  $12.09 \pm 0.26$  \\[1mm]
4  &  $-56.8$  &  $5.7$  &  \ion{Fe}{ii}  & $13.69 \pm 0.17$ \\
    &               & ''   &   \ion{Si}{ii}  &  $14.15 \pm 0.13$  \\[1mm]
5  &  $-20.1$  &  $9.8$  &  \ion{Fe}{ii}  & $<13.41$ \\
    &               & ''   &   \ion{Zn}{ii}  &  $11.60 \pm 0.18$  \\
    &               & ''   &   \ion{Si}{ii}  &  $14.39 \pm 0.10$  \\
    &               & ''   &   \ion{Cr}{ii}  &  $12.17 \pm 0.28$  \\[1mm]
6  &  $ 0.0$  &  $11.9$  &  \ion{Fe}{ii}  & $14.31 \pm 0.06$ \\
    &               & ''   &   \ion{Zn}{ii}  &  $12.34 \pm 0.04$  \\
    &               & ''   &   \ion{Si}{ii}  &  $14.69 \pm 0.06$  \\
    &               & ''   &   \ion{Cr}{ii}  &  $12.48 \pm 0.14$  \\[1mm]
7  &  $28.0$  &  $7.4$  &  \ion{Fe}{ii}  & $13.92 \pm 0.11$ \\
    &               & ''   &   \ion{Si}{ii}  &  $14.42 \pm 0.08$  \\[1mm]
8  &  $67.4$  &  $10.5$  &  \ion{Fe}{ii}  & $14.18 \pm 0.07$ \\
    &               & ''   &   \ion{Mg}{i}  &  $12.61 \pm 0.01$  \\
    &               & ''   &   \ion{Zn}{ii}  &  $12.07 \pm 0.05$  \\
    &               & ''   &   \ion{Si}{ii}  &  $14.57 \pm 0.07$  \\
    &               & ''   &   \ion{Cr}{ii}  &  $<11.90$  \\[1mm]
9  &  $86.6$  &  $3.4$  &  \ion{Fe}{ii}  & $13.53 \pm 0.24$ \\
    &               & ''   &   \ion{Si}{ii}  &  $13.95 \pm 0.18$  \\
    &               & ''   &   \ion{Cr}{ii}  &  $12.50 \pm 0.15$  \\
\hline
1 & $-153$ & $15$  &  \ion{Mg}{i}  &  $12.31 \pm 0.05$  \\
2 & $-106$ & $30$  &  \ion{Mg}{i}  &  $12.43 \pm 0.04$  \\
3 & $6$    & $22$  &  \ion{Mg}{i}  &  $12.26 \pm 0.02$  \\
4 & $68$   & $28$  &  \ion{Mg}{i}  &  $12.61 \pm 0.01$  \\
\hline
\end{tabular}
\end{center}
$^{a}$ Relative to the systemic redshift $z_{\rm sys}=2.42630$.
\end{table}

\begin{table}
\caption{Best fit parameters for Q2348--011-2. \label{tab:fit_Q2348-2}}
\begin{center}
\begin{tabular}{lcccc}
\hline
\# & Velocity$^{a}$  &  $b$          &  Ion  &  $\log(N)$   \\
   &  (km~s$^{-1}$)  & (km~s$^{-1}$) &       &   \\
\hline

1  &  $-66.4$  &  $12.4$  &  \ion{Fe}{ii}  & $13.65 \pm 0.02$ \\
    &               & ''   &   \ion{Cr}{ii}  &  $<11.03$  \\
    &               & ''   &   \ion{Si}{ii}  &  $<13.71$  \\[1mm]
2  &  $ 0.0$  &  $14.0$  &  \ion{Fe}{ii}  & $14.29 \pm 0.01$ \\
    &               & ''   &   \ion{Cr}{ii}  &  $12.85 \pm 0.03$  \\
    &               & ''   &   \ion{Si}{ii}  &  $14.83 \pm 0.03$  \\
\hline
\end{tabular}
\end{center}
$^{a}$ Relative to the systemic redshift $z_{\rm sys}=2.6138$.
\end{table}

\clearpage

\section{2D Spectra}
\label{app:lya}

\begin{figure}
	\includegraphics[width=0.49\textwidth]{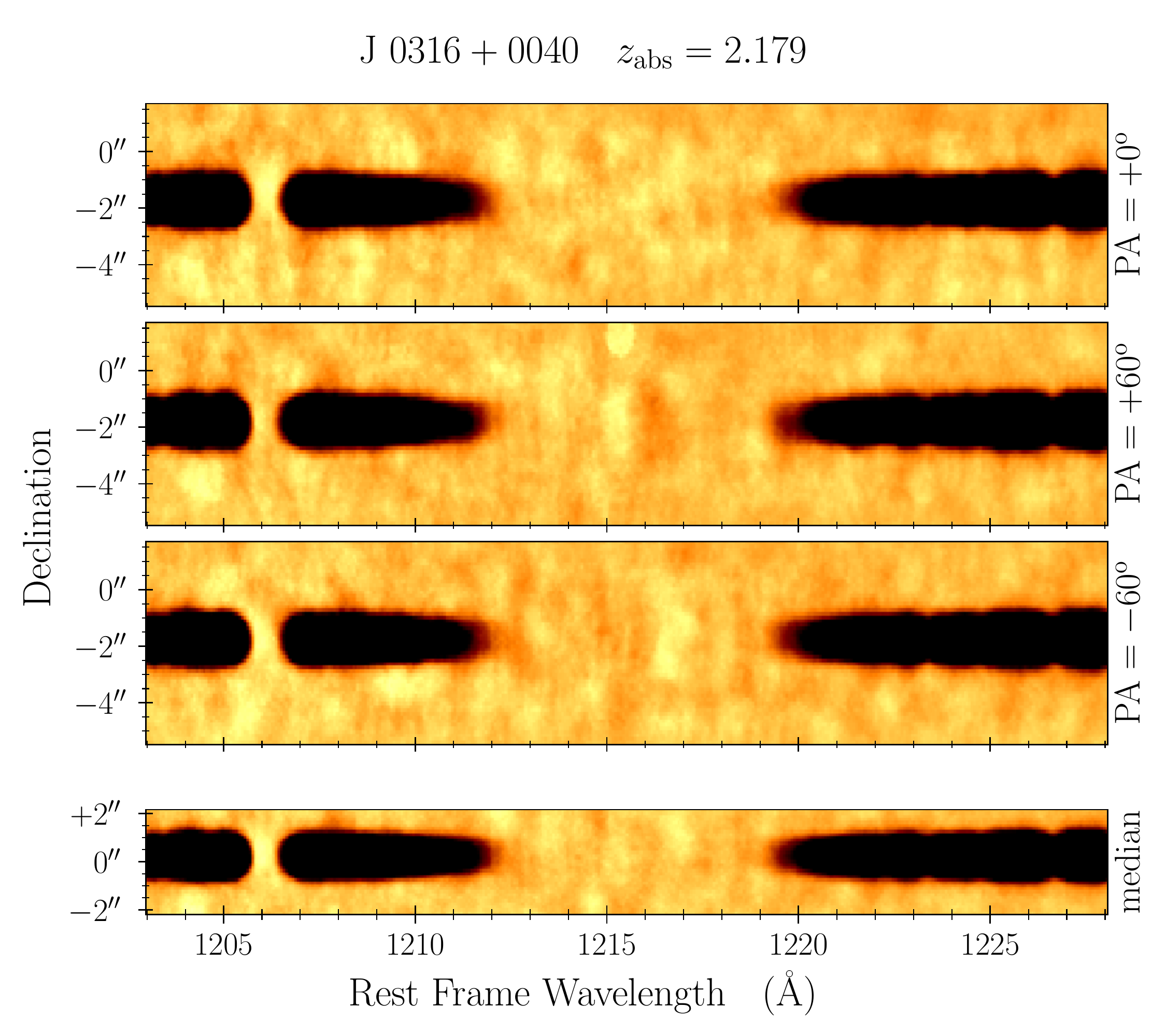}
	\caption{Cutout of the 2D-spectrum (UVB) around the $\zDLA=2.179$ DLA toward Q0316+0040. The top three panels show the individual position angles (PA1=$0$\degr, PA2=$+60$\degr, and PA3=$-60$\degr east of north), while the bottom panel shows the median combination of all three PAs in order to search for emission at very small impact parameters. All spectra have been smoothed by a $5\times5$ top hat filter for visual purposes.
	\label{fig:Q0316_lya}}
\end{figure}

\begin{figure}
	\includegraphics[width=0.49\textwidth]{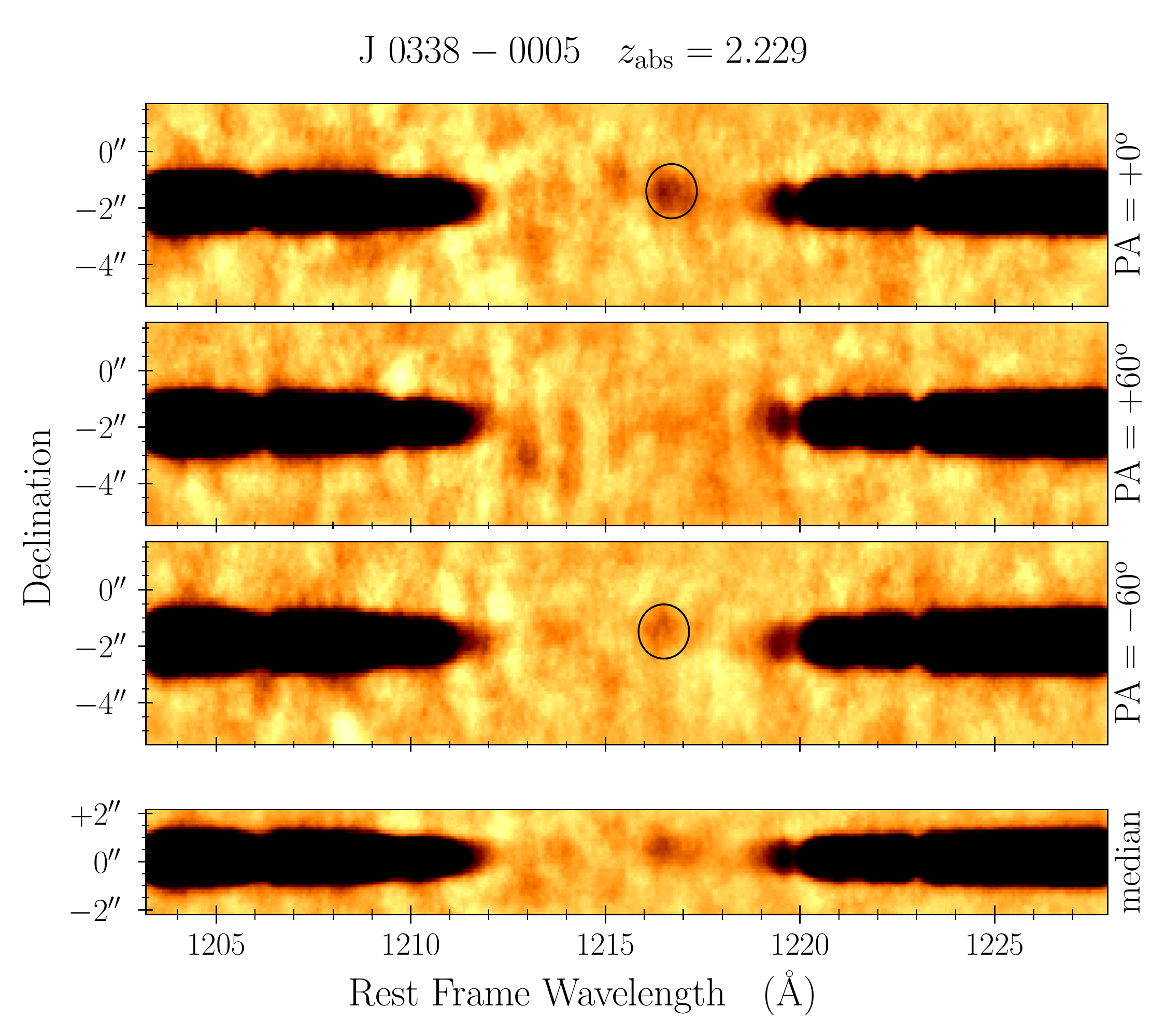}
	\caption{Same as Figure~\ref{fig:Q0316_lya} for the $\zDLA=2.229$ DLA toward
			 Q0338$-$0005. Emission is detected in PA1 and PA3 (though less significant in PA3).
	\label{fig:Q0338_lya}}
\end{figure}

\begin{figure}
	\includegraphics[width=0.49\textwidth]{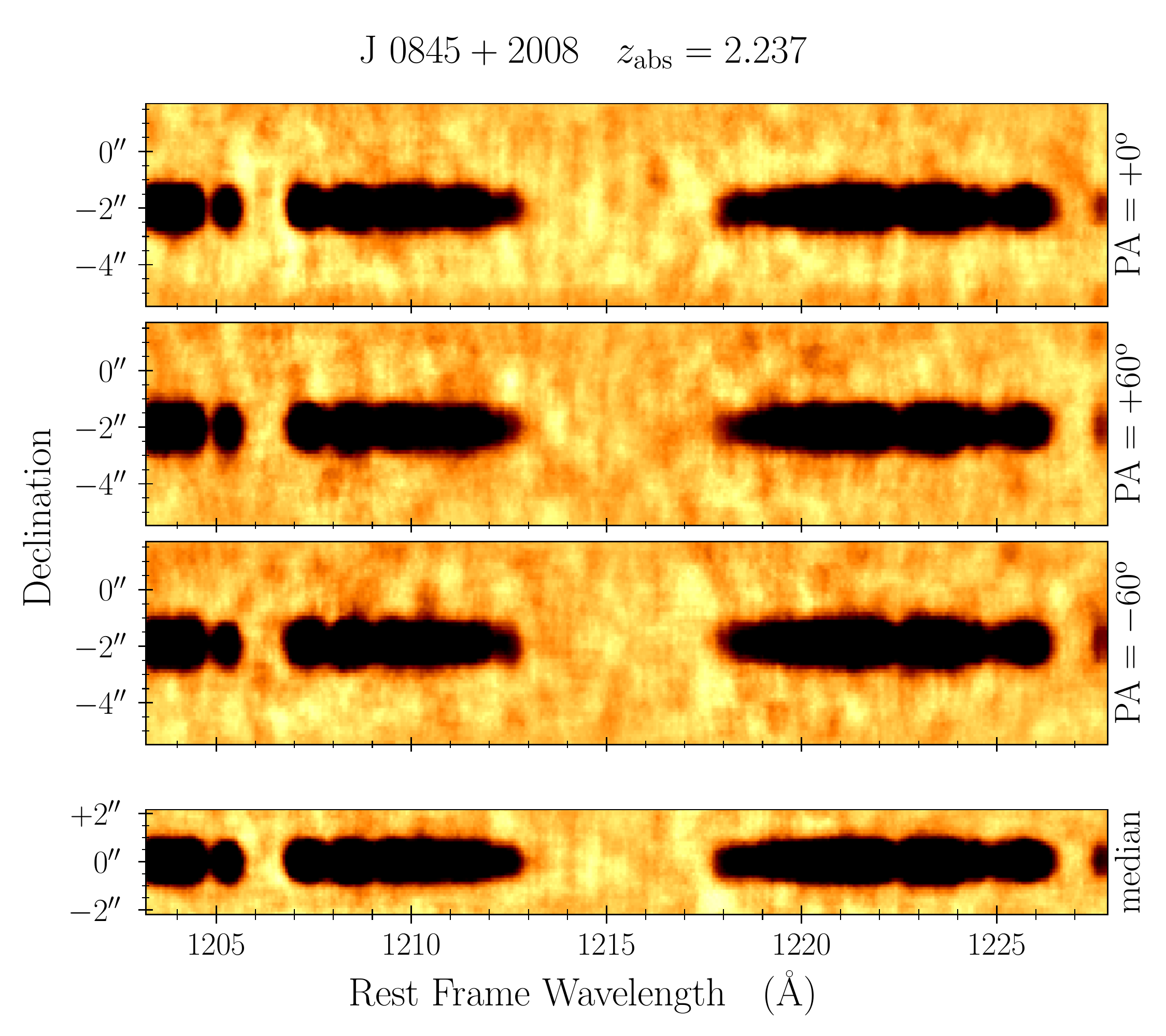}
	\caption{Same as Figure~\ref{fig:Q0316_lya} for the $\zDLA=2.237$ DLA toward Q0845+2008.
	\label{fig:Q0845_lya}}
\end{figure}

\begin{figure}
	\includegraphics[width=0.49\textwidth]{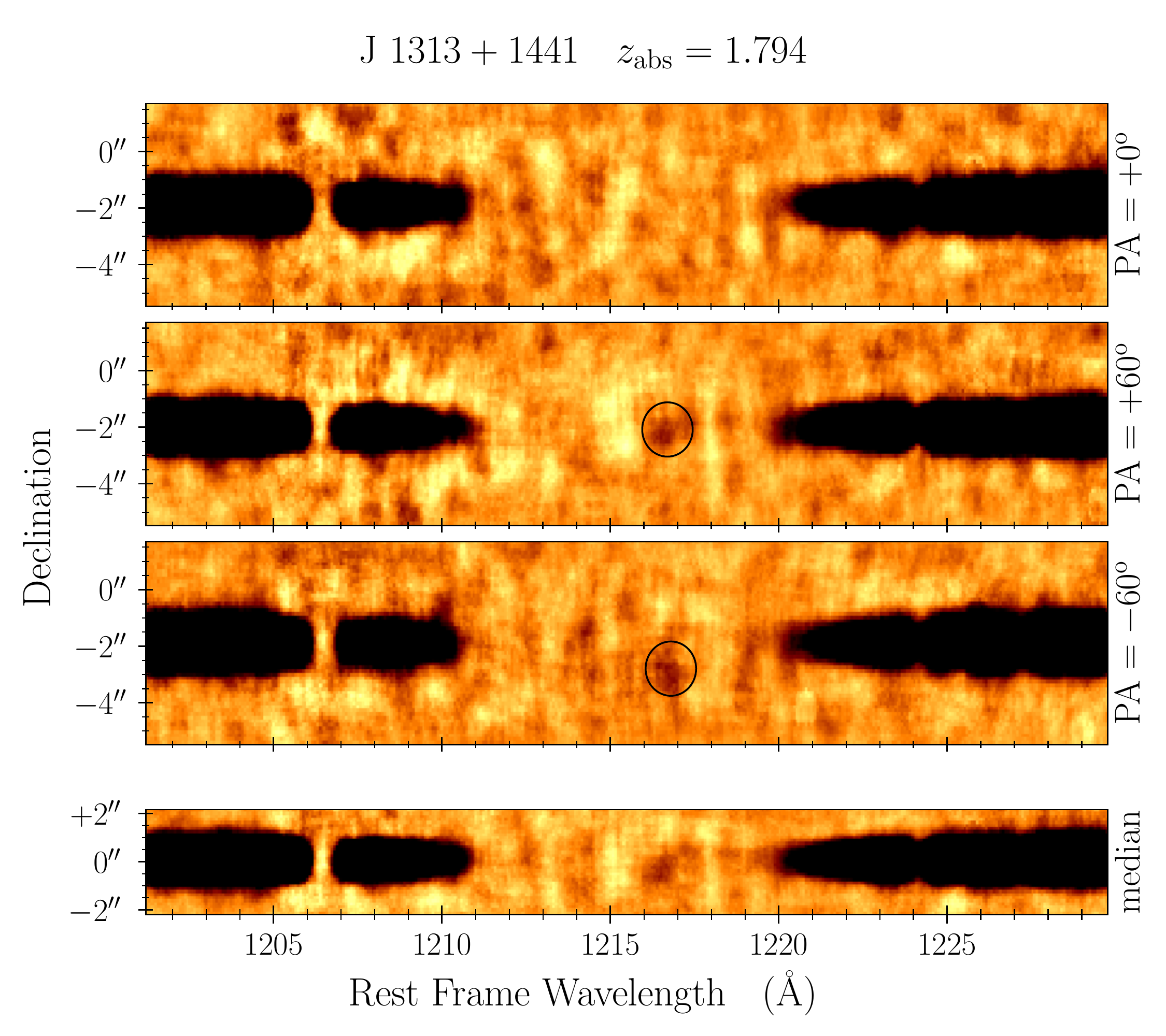}
	\caption{Same as Figure~\ref{fig:Q0316_lya} for the $\zDLA=1.794$ DLA toward Q1313+1441.
			 Emission is detected in PA2 and PA3. Due to the lower redshift of this DLA, the
			 spectrum is farther in the blue where the sky background noise is larger.
	\label{fig:Q1313_lya}}
\end{figure}

\begin{figure}
	\includegraphics[width=0.49\textwidth]{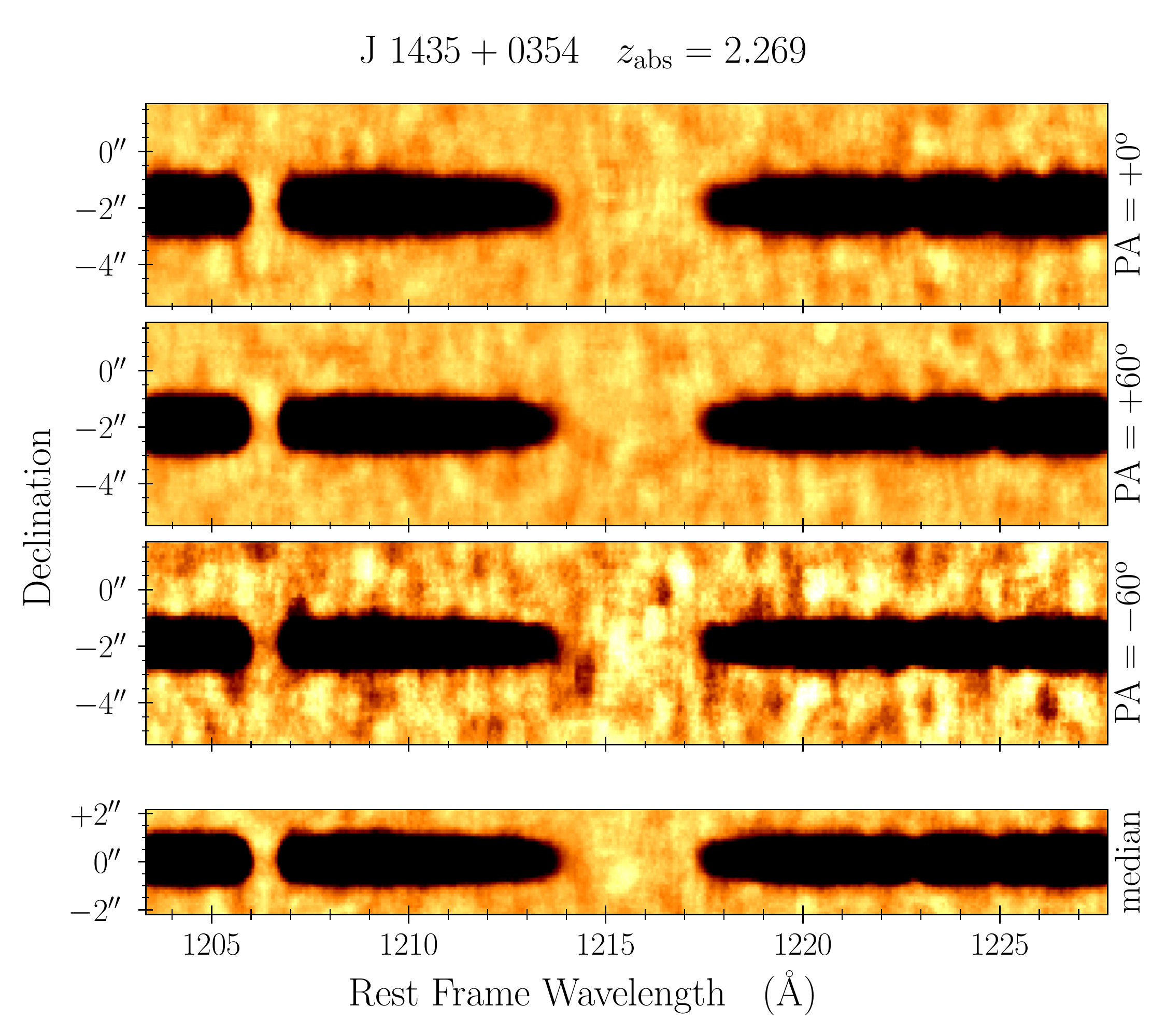}
	\caption{Same as Figure~\ref{fig:Q0316_lya} for the $\zDLA=2.269$ DLA toward Q1435+0354.
			 The spectrum for PA3 is more noisy than the two remaining PAs due to the reduced
			 exposure time for this particular PA, see Table~\ref{tab:log}.
	\label{fig:Q1435_lya}}
\end{figure}

\begin{figure}
	\includegraphics[width=0.49\textwidth]{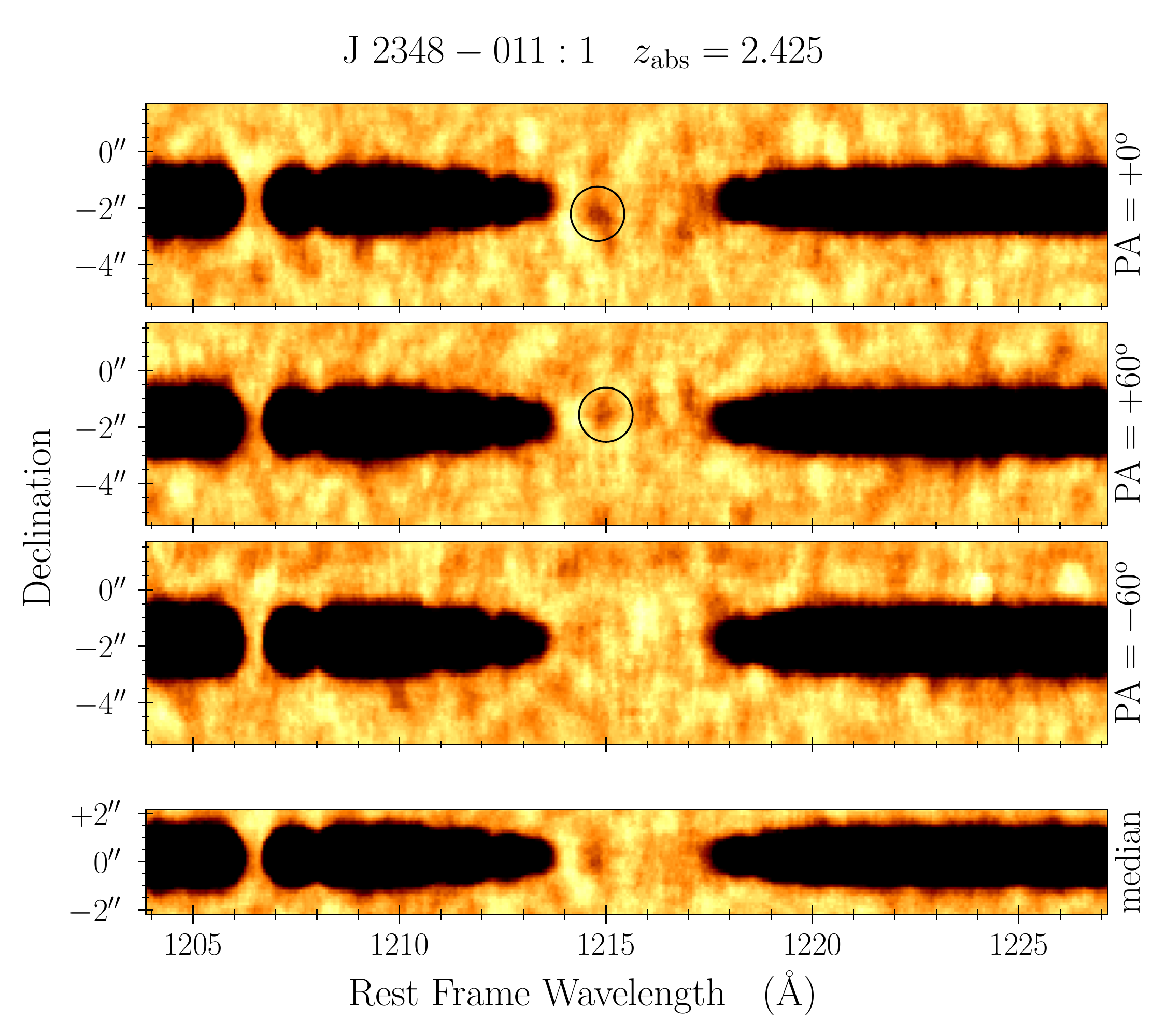}
	\caption{Same as Figure~\ref{fig:Q0316_lya} for the $\zDLA=2.425$ DLA toward
			 Q2348$-$011. Emission is detected in PA1, and tentatively in PA2.
	\label{fig:Q2348-1_lya}}
\end{figure}

\begin{figure}
	\includegraphics[width=0.49\textwidth]{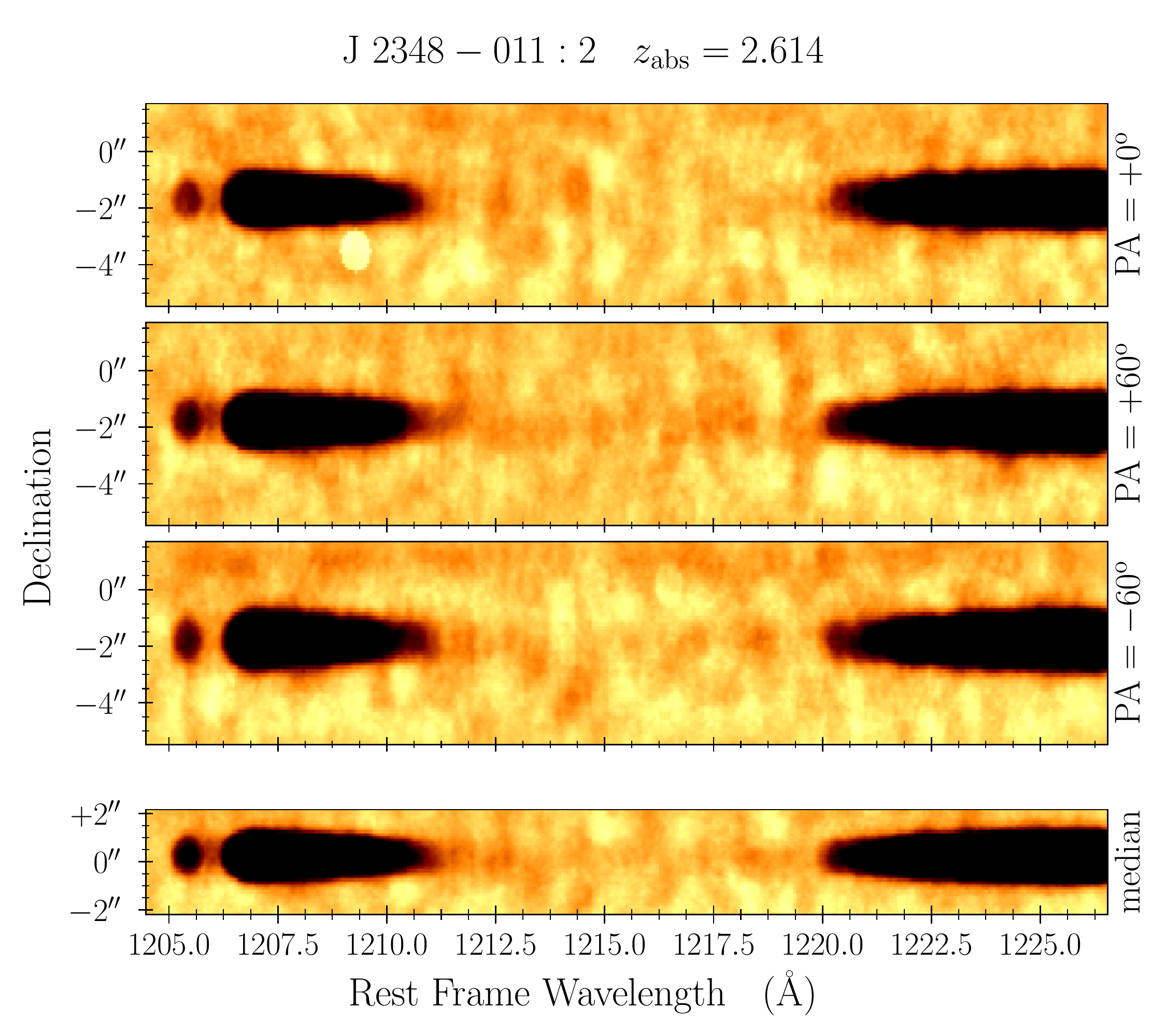}
	\caption{Same as Figure~\ref{fig:Q0316_lya} for the $\zDLA=2.614$ DLA toward
			 Q2348$-$011.
	\label{fig:Q2348-2_lya}}
\end{figure}

The individual spectra around the \lya\ line of the DLA for all the targets analysed in this work are shown in Figs.~\ref{fig:Q0316_lya}--\ref{fig:Q2348-2_lya}.

\newpage
\clearpage

\section{Estimating Impact Parameter Distribution}
\label{app:PDF}

\begin{figure}
	\includegraphics[width=0.49\textwidth]{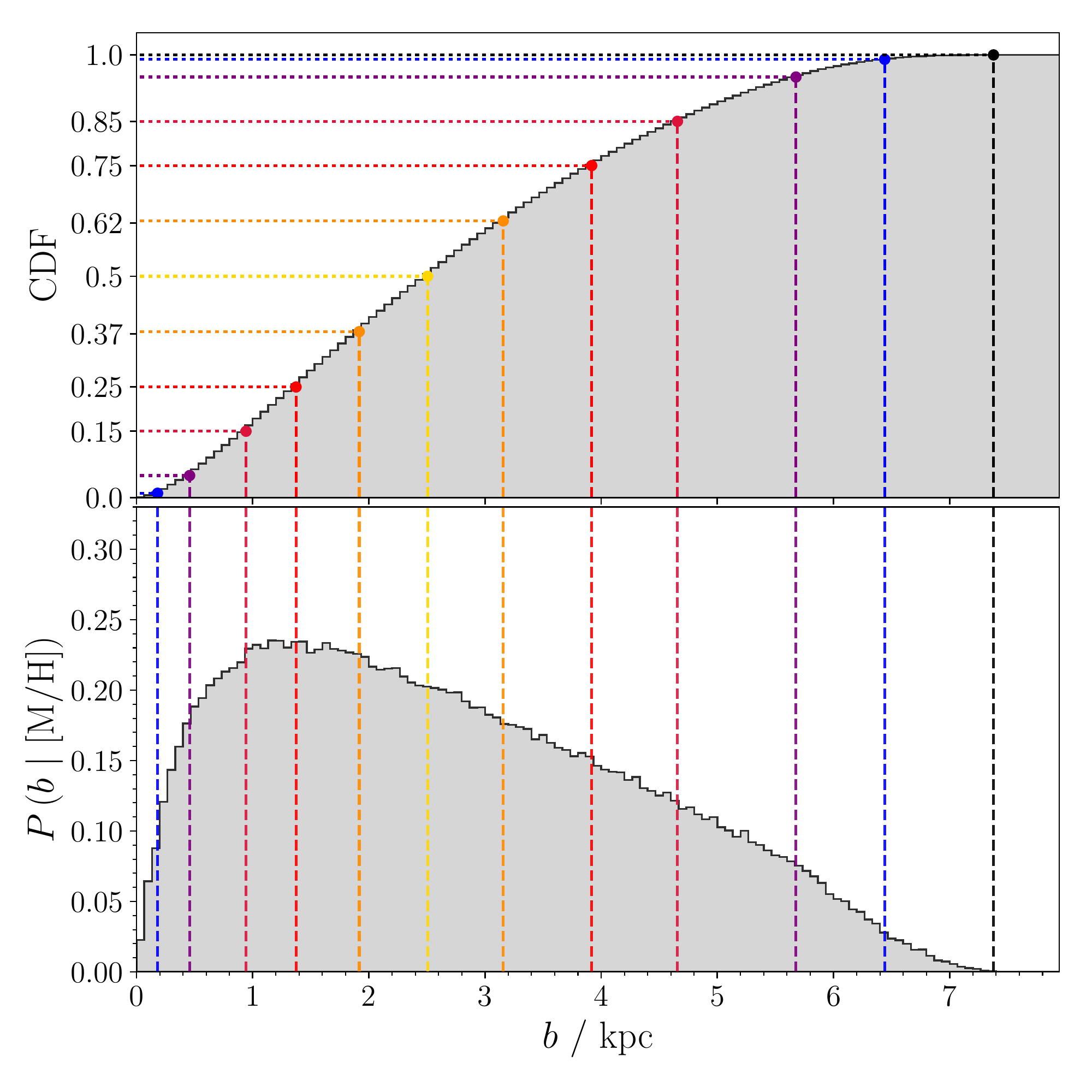}
	\caption{Cumulative distribution function (top) and probability distribution function (bottom) of impact parameters for a metallicity of [M/H] = $-$1.5 from the model of \citet{Fynbo2008}. The dotted lines correspond to the percentiles: 1, 5, 15, 25, 37.5, 50, 62.5, 75, 85, 95, 99, 100. These are indicated on the y-axis of the top panel, except for the first two and the last two, which would be spaced too closely together to be visible.
	\label{fig:ex_pdf}}
\end{figure}

\begin{figure}
	\includegraphics[width=0.49\textwidth]{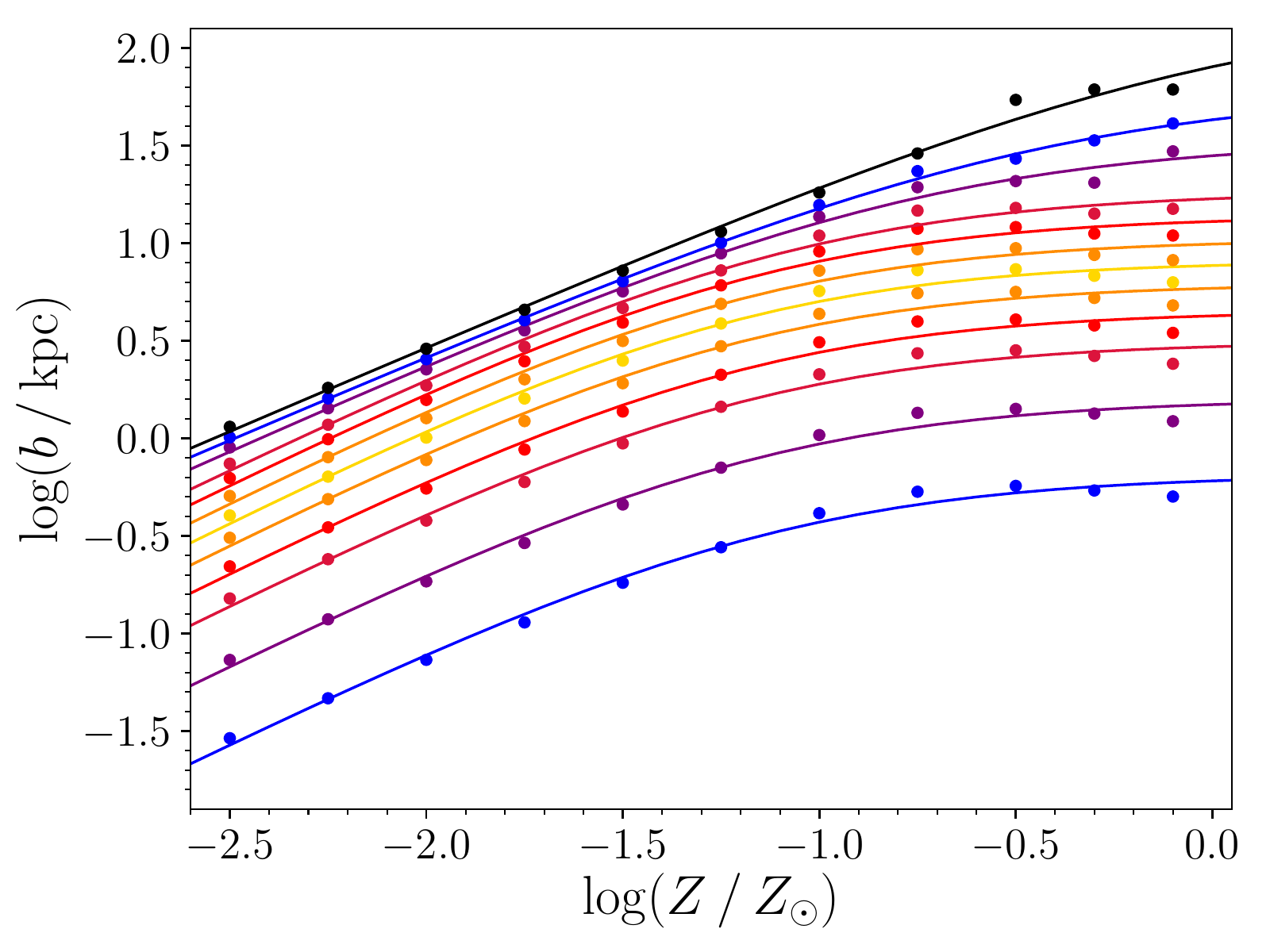}
	\caption{Impact parameter percentiles as function of metallicity. The points indicate (from bottom to top) the 1, 5, 15, 25, 37.5, 50, 62.5, 75, 85, 95, 99, and 100 percentiles, and the lines show the fitted functional form as a function of metallicity. The percentiles shown correspond to the same percentiles as in Fig.~\ref{fig:ex_pdf}.
	\label{fig:pdf_model}}
\end{figure}

We estimate the probability density function (PDF) of impact parameter for a given metallicity,
$P\,(b\,|\,[{\rm M/H}])$, based on the model realizations from the model of \citet{Fynbo2008}.
We parametrize the PDF as a function of metallicity in order to speed up the evaluation of the
PDF for the many model realizations in our analysis, and in order to increase the precision for high-metallicity, where the model realizations from \citet{Fynbo2008} are scarcely sampled.
Instead of fitting the PDF with an assumed functional form for different metallicities,
we evaluate the percentiles of the impact parameter distribution in various metallicity bins and
quantify the metallicity evolution of these percentiles. In order to recover the distribution
with high precision, we evaluate 12 percentiles of the distribution (see Fig.~\ref{fig:ex_pdf})
and fit those with a fixed functional form as a function of metallicity.
We observe that the percentiles as function of metallicity follow a power-law for low metallicities, whereas the evolution flattens for higher metallicities.
In order to reproduce the observed behaviour of the percentiles, we fit the 12 percentiles of the
impact parameter distribution by the following function of metallicity:

\begin{equation}
	\log(b_i / {\rm kpc}) = \alpha_i - \log(\beta_i + 10^{\gamma_i {\rm [M/H]}})\ \ ,
\end{equation}

\noindent where the subscript $i$ denotes the $i$-th percentile: 1, 5, 15, 25, 37.5, 50, 62.5, 75, 85, 95, 99, 100. The best-fit evolution of the percentiles as functions of metallicity is shown in
Fig.~\ref{fig:pdf_model}.
During the modelling (described in Sect.~\ref{model}) we are then able to evaluate the percentiles
of the distribution for any given metallicity. This allows us to reconstruct the cumulative
probability function (CDF) from which we can draw random samples by use of inverse transform sampling.

\label{lastpage}

\end{document}